\documentclass{aa}  

\usepackage{graphicx}
\usepackage{txfonts}
\usepackage{amssymb,amsfonts,amsmath,bm}
\usepackage{epstopdf,epsfig,subfig,float}
\usepackage{hyperref}
\usepackage{color}
\usepackage[dvipsnames]{xcolor}
\usepackage{natbib}
\usepackage{soul}
\usepackage[english]{babel}
\usepackage{placeins}
\usepackage{tcolorbox}
\usepackage{multirow}

\definecolor{myblue}{rgb}{0.1,0.1,0.66}
\definecolor{kulblue}{HTML}{116E8A}
\hypersetup{colorlinks,breaklinks,
            linkcolor=myblue,urlcolor=myblue,
            anchorcolor=myblue,citecolor=myblue}
\definecolor{kulblue}{HTML}{116E8A}

\newcommand\bb[1]{\boldsymbol{#1}}

\def\aap{Astron. Astrophys.}

\def\aapr{Astron. Astrophys. Rev.}
\def\araa{Annu. Rev. Astronom. Astrophys.}
\def\apj{Astrophys. J.}
\def\apjl{Astrophys. J. Lett.}

\def\asr{Adv. Space Res.}

\def\pof{Phys. Fluids}
\def\pop{Phys. Plasmas}
\def\ppcf{Plasma Phys. Control. Fusion}

\def\prv{Phys. Rev.}
\def\mnras{Mon. Not. R. Astron. Soc.}

\def\apss{Ap\&SS}               

\def\ssr{Space Science Reviews}

\def\frass{Front. Astron. Space Sci.}
\def\ijmpd{Int. J. Mod. Phys. D}
\def\lrsp{Living Rev. Sol. Phys.}
\def\ssr{Space Sci. Rev.}
\def\jpp{J. Plasma Phys.}

\begin{document}

   \title{Revisiting the role of cosmic-ray driven Alfv\'en waves in pre-existing magnetohydrodynamic turbulence}

   \subtitle{I. Turbulent damping rates and feedback on background fluctuations}

   \author{Silvio Sergio Cerri}

   \institute{Universit\'{e} C\^{o}te d'Azur, Observatoire de la C\^{o}te d'Azur, CNRS, Laboratoire Lagrange, France\\ \email{silvio.cerri@oca.eu}}


 
  \abstract
   {Alfv\'en waves (AWs) excited by the cosmic-ray (CR) streaming instability (CRSI) are a fundamental ingredient for CR confinement. The effectiveness of such self-confinement relies on a balance between the CRSI growth rate and the damping mechanisms acting on quasi-parallel AWs excited by CRs. One relevant mechanism is called turbulent damping, in which an AW packet injected in pre-existing turbulence undergoes a cascade process due to its nonlinear interaction with fluctuations of the background.}
   {The turbulent damping of an AW packet in pre-existing magnetohydrodynamic (MHD) turbulence is re-examined, revised, and extended to include the most recent theories of MHD turbulence that account for dynamic alignment and reconnection-mediated regimes. The case in which the role of feedback of CR-driven AWs on pre-existing turbulence is important is also discussed.}
   {The  Els\"asser formalism is employed. Particular attention is given to the role of a nonlinearity parameter $\chi^w$ that estimates the strength of the nonlinear interaction between CR-driven AW packets and the background fluctuations. We point out the difference between $\chi^w$ and the parameter $\chi^z$ that instead describes the intrinsic strength of nonlinear interactions between pre-existing fluctuations. Turbulent damping rates of quasi-parallel AW packets and cosmic-ray feedback (CRF) are derived within this formalism.
   }
   {When the strength of the nonlinear interaction is properly taken into account, we find that (i) the turbulent damping rate of quasi-parallel AWs in sub-Alfv\'enic turbulence depends on the background-fluctuation amplitude to the third power, and hence is strongly suppressed; (ii) the dependence on the AW's wavelength (and thus on the CR gyro-radius from which it is excited) is different from what has been previously obtained; and  (iii) when dynamic alignment of cascading fluctuations and the possibility of a reconnection-mediated range is included in the picture, the turbulent damping rate exhibits novel regimes and breaks. Finally, a criterion for CRF is derived and a simple phenomenological model of CR-modified scaling of background fluctuations is provided.}
   {}

   \keywords{turbulence -- magnetohydrodynamics -- cosmic rays}

   \maketitle
%

\section{Introduction}\label{sec:intro}

Turbulent magnetized plasmas permeate a wide range of space and astrophysical environments~\citep[e.g.,][]{QuataertGruzinovAPJ1999,SchekochihinCowleyPOP2006,BrandenburgLazarianSSRv2013,BrunoCarboneLRSP2013,FerrierePPCF2020}.
Understanding the properties of the turbulent cascade, and  how the fluctuation energy is transferred from injection to dissipation scales, thus heating the plasma and also producing nonthermal particles in the process,  is a relevant task in itself since it can elucidate the role that turbulence plays in the dynamics and thermodynamics of several astrophysical systems.
Inspired by the seminal work of \citet{Kolmogorov1941} in hydrodynamics, turbulence in magnetized plasmas has been the object of several theoretical efforts aimed at obtaining universal scaling for its fluctuations on large (fluid) magnetohydrodynamic (MHD) scales~\citep[e.g.,][]{IroshnikovAZ1963,KraichnanPOF1965,GoldreichSridharAPJ1995,NgBhattacharjeePOP1997,GaltierJPP2000,ChoLazarianPRL2002,BoldyrevPRL2006,LazarianSSRv2012,ChandranAPJ2015,MalletMNRAS2015,BoldyrevLoureiroAPJ2017,MalletMNRAS2017,CerriAPJ2022,SchekochihinJPP2022}.
At the same time, these astrophysical environments are also populated with cosmic rays (CRs), which are  charged particles with supra-thermal (relativistic) energies that pervade  the interstellar, intergalactic, and intracluster media~\citep[e.g.,][]{BrunettiJonesIJMPD2014,AmatoBlasiASR2018,FaucherOhARAA2023,RuszkowskiPfrommerAARv2023} and get scattered by magnetic-field fluctuations~\citep{GinzburgSyrovatskiiBOOK1964,BerezinskyBOOK1990}.
While cosmic-ray transport partly depends upon the properties of pre-existing turbulence~\citep[e.g.,][]{SchlickeiserMillerAPJ1998,ChandranPRL2000,LercheSchlickeiserAA2001,YanLazarianPRL2002,YanLazarianAPJ2008,TeufelAA2003,ShalchiSchlickeiserAA2004,FornieriMNRAS2021,LazarianXuAPJ2021,LemoineJPP2023,KempskiMNRAS2023}, CRs can also generate their own scattering fluctuations through streaming instabilities~\citep[e.g.,][]{KulsrudPearceAPJ1969,LeeAPJ1972,SkillingMNRAS1975,GaryBOOK1993,BellMNRAS2004,AmatoMSAI2011,WeidlAPJ2019a,WeidlAPJ2019b,MarcowithPOP2021}.
The level at which self-generated fluctuations saturate depends on a balance between the instability growth and the damping mechanisms that these waves are subjected to. Depending on the Galactic environment, the damping processes that were originally considered are the ion-neutral (IN) damping~\citep{KulsrudPearceAPJ1969} and the nonlinear Landau (NLL) damping~\citep{LeeVoelkApSS1973}.
These cosmic-ray driven Alfv\'en-wave (AW) packets, however,   also interact with pre-existing fluctuations of the turbulent environment in which they are generated. This interaction has been suggested to represent another source of damping, the process called turbulent damping, for which a CR-generated AW packet is cascaded to dissipation by its nonlinear interaction with background fluctuations.
This damping mechanism was originally proposed by \citet{FarmerGoldreichAPJ2004}, and subsequently extended by \citet{LazarianAPJ2016} to account for different regimes of background turbulence. 
However, an important parameter that has not been taken into account in these previous works is the strength of the nonlinear interaction (usually referred to as the  nonlinearity parameter $\chi$) between the AW packet and pre-existing turbulent fluctuations. This parameter is indeed different from (and typically much smaller than) the nonlinear parameter describing the regime of background turbulence, which also needs to be taken into account~\citep[as done by][]{LazarianAPJ2016}. We  show here that taking this difference into account completely changes the estimate of the turbulent damping rate, which is  almost always much lower than any rate derived previously.  
Moreover, the damping rate and its scaling strongly depend upon the properties of background turbulence. In the previous literature, only what we can call
the classic theories of MHD turbulence have been taken into account:  isotropic ``Kolmogorov-like'' turbulence, hereafter the K41 regime ~\citep[][]{Kolmogorov1941}; a weak cascade of Alfv\'enic fluctuations, hereafter the W0 regime~\citep[][]{NgBhattacharjeePOP1997,GaltierJPP2000};
and a critically balanced Alfv\'enic cascade, hereafter the  GS95 regime ~\citep{GoldreichSridharAPJ1995}. 
However, advanced theories of MHD turbulence that extend the above classic picture have been formulated in the past years.
This is the case, for instance, of a theory that incorporates dynamic (i.e., scale-dependent) alignment of fluctuations in a critically balanced Alfv\'enic cascade, hereafter the B06 regime~\citep[][]{BoldyrevPRL2006}, which can intrinsically lead at even smaller scales to a regime usually referred to as tearing-mediated turbulence, hereafter the TMT regime~\citep[i.e., a regime where magnetic reconnection mediates the generation of smaller-scale fluctuations,][]{BoldyrevLoureiroAPJ2017,MalletMNRAS2017}.
It is also worth mentioning that the conditions under which critical balance and the associated cascades develop (i.e., GS95, B06, and TMT regimes) may not cover all the possible scenarios in MHD turbulence~\citep[e.g., see discussion in][]{OughtonMatthaeusAPJ2020}. However, analytical (phenomenological) scaling of turbulent fluctuations and their anisotropy can be only derived for these cases. Moreover, several  numerical simulations and in situ measurements in the solar wind have provided solid evidences for these regimes~\citep[e.g.,][and references therein]{ChenJPP2016,SahraouiRMPP2020,SchekochihinJPP2022}.
Therefore, it is of interest to derive turbulent damping rates for all these theories. 
The results obtained here have indeed implications for the   cosmic-ray self-confinement, since its effectiveness for CR scattering is the result of a competition between different damping mechanisms and a balance between the most-relevant damping rate and the growth rate of the CR streaming instability~\citep[e.g.,][]{FarmerGoldreichAPJ2004,BlasiPRL2012,LazarianAPJ2016,KempskiQuataertMNRAS2022,XuLazarianAPJ2022}.
For instance, by adopting the rates obtained in \citet{FarmerGoldreichAPJ2004} and \cite{LazarianAPJ2016},  turbulent damping can compete with or even dominate over the IN and NLL damping processes, depending on the properties of the Galactic environment under consideration and on the CR energy~\citep[see, e.g.,][and references therein]{NavaMNRAS2019,KempskiQuataertMNRAS2022,RecchiaAA2022,XuLazarianAPJ2022}; this picture can be significantly challenged by the new turbulent damping rates obtained here, and will be addressed in detail in a following work (hereafter Paper II).

This paper is organized as follows. In Section~\ref{sec:Elsasser_formalism} the Els\"asser formalism and the definitions of the timescales and nonlinear parameter are introduced. In Section~\ref{sec:general_damping} the formalism is employed to derive general expressions for the turbulent damping rates, whose scaling are then explicitly derived in Section~\ref{sec:explicit_scalings} for various turbulence regimes and within different theories of MHD turbulence.
Additionally, some considerations about the feedback of CR-driven AWs on pre-existing fluctuations and possible phenomenological models for the CR-modified background turbulence spectrum are discussed in Section~\ref{sec:AW_feedback}.
Finally, a summary and discussion of the results is provided in Section~\ref{sec:conclusions}. 
It is worth noting  that this work is meant to primarily provide a rigorous, general formalism for deriving the turbulent damping rates of CR-driven AW packets, as well as some criteria for the possible relevance of CR feedback on pre-existing fluctuations. A more extensive discussion about different damping rates, the role of coherent structures and compressible turbulence, as well as the implications for specific astrophysical systems will be the focus of Paper II.


\section{Setting the stage: The Els\"asser formalism}\label{sec:Elsasser_formalism}

The magnetohydrodynamic (MHD) equations for an incompressible plasma with mass density $\rho_0$, viscosity $\nu$, and resistivity $\eta$,  can be conveniently expressed in terms of the Els\"asser variables $\bb{z}^\pm = \bb{u} \pm \bb{B}/\sqrt{4\pi\rho_0} = \bb{u}\pm\bb{v}_{\rm A}$~\citep{ElsasserPR1950}, where $\bb{u}$ is the fluid velocity, $\bb{B}$ is the magnetic field, and $\bb{v}_{\rm A}$ denotes the Alfv\'en-speed vector associated  with $\bb{B}$. The incompressible MHD equations in terms of $\bb{z}^\pm$ read as
\begin{equation}\label{eq:MHD_Elsasser_pm}
    \frac{\partial\bb{z}^\pm}{\partial t}
    +
    (\bb{z}^\mp\cdot\bb{\nabla})\,\bb{z}^\pm
    =
    -\frac{\bb{\nabla}P_{\rm tot}}{\rho_0}
    + 
    \mu_{+}\,\nabla^2\bb{z}^\pm
    +
    \mu_{-}\,\nabla^2\bb{z}^\mp\,,
\end{equation}
\begin{equation}\label{eq:MHD_Elsasser_incompr}
    \bb{\nabla}\cdot\bb{z}^\pm=0\,,
\end{equation}
where $P_{\rm tot} = P_{\rm th} + B^2/8\pi$ is the sum of the thermal and magnetic pressure, and $\mu_\pm=(\nu\pm\eta)/2$.
Here we assume $\nu=\eta$ for simplicity, so that $\mu_{+}=\eta$ and $\mu_{-}=0$. By splitting the variables into a background quantity (denoted by a ``0'' in subscript\footnote{The subscript ``0'' formally implies a  large-scale average procedure $\bb{B}_0=\langle\bb{B}\rangle_{L}$. In the latter, $L\sim\ell_0$ will be the injection scale of turbulence.}
) and purely transverse fluctuations; in other words, $\bb{z}^\pm=\bb{z}_0^\pm+\delta\bb{z}_\perp^\pm$, where $\bb{z}_0^\pm=\pm\bb{B}_0/\sqrt{4\pi\rho_0}=\pm\bb{v}_{\rm A,0}$ is the Alfv\'en speed associated with the background magnetic field $\bb{B}_0$, and $\delta\bb{z}_\perp^\pm=\delta\bb{u}_\perp\pm\delta\bb{B}_\perp/\sqrt{4\pi\rho_0}$ the fluctuating Els\"asser fields, equation \eqref{eq:MHD_Elsasser_pm} rewrites as
\begin{equation}\label{eq:MHD_Elsasser_pm_fluct}
    \bigg(\frac{\partial}{\partial t}\,
    \mp\,
    \underbrace{v_{\rm A,0}\,\nabla_\|}_{\omega_{\rm lin}^\pm\,\sim\,k_\|^\pm\,v_{\rm A,0}}\,
    +\,
    \underbrace{\delta\bb{z}_\perp^\mp\cdot\bb{\nabla}_\perp}_{\omega_{\rm nl}^\pm\,\sim\,k_\perp^\pm\,\delta z_\perp^\mp}\,
    -\underbrace{\eta\,\nabla^2}_{\omega_{\rm diss}^{\pm}\sim\, \eta\,k^\pm{}^2 }\bigg)\,\delta\bb{z}_\perp^\pm
    =
    -\frac{\bb{\nabla}\delta P_{\rm tot}}{\rho_0}\,,
\end{equation}
where the parallel and perpendicular directions are defined with respect to $\bb{B}_0$ for the global equations (but are later defined with respect to a scale-dependent mean field $\langle\bb{B}\rangle_{k}$ in a turbulent environment); we also mention that the term $\bb{\nabla}\delta P_{\rm tot}/\rho_0$ in practice contributes just as a multiplicative factor (in Fourier space) to the nonlinear term (and associated timescale) on the left-hand side.\footnote{By taking the divergence of \eqref{eq:MHD_Elsasser_pm_fluct} and using the incompressiblity condition $\bb{\nabla}\cdot\delta\bb{z}_\perp^\pm=0$, we find that pressure fluctuations satisfy the condition $\bb{\nabla}\cdot[\bb{\nabla}\delta P_{\rm tot}] = \nabla^2 \delta P_{\rm tot}=\rho_0\bb{\nabla}_\perp\cdot[(\delta\bb{z}_\perp^\mp\cdot\bb{\nabla}_\perp)\,\delta\bb{z}_\perp^\pm]$ \citep{SchekochihinJPP2022}.} 
One important feature of the formulation in \eqref{eq:MHD_Elsasser_pm_fluct} is that it explicitly shows that the nonlinear term $(\delta\bb{z}_\perp^\mp\cdot\bb{\nabla}_\perp)\delta\bb{z}_\perp^\pm$ is due only to the interaction of counter-propagating Alfv\'en-wave packets, $\delta\bb{z}_\perp^{+}$ being transverse fluctuations propagating at the Alfv\'en speed $v_{\rm A,0}$ in the direction of $\bb{B}_0$, while $\delta\bb{z}_\perp^{-}$ are fluctuations propagating at the same speed in the direction of $-\bb{B}_0$.

From \eqref{eq:MHD_Elsasser_pm_fluct} one can define a nonlinear parameter $\chi$, which measures the strength of nonlinear effects with respect to the linear propagation term, namely
\begin{equation}\label{eq:MHD_Elsasser_chi_pm}
    \chi^\pm
    \sim
    \frac{|(\delta\bb{z}_\perp^\mp\cdot\bb{\nabla}_\perp)\,\delta\bb{z}_\perp^\pm|}{|(v_{\rm A,0}\,\nabla_\|)\,\delta\bb{z}_\perp^\pm|}
    \sim
    \frac{\omega_{\rm nl}^\pm}{\omega_{\rm lin}^\pm}
    \sim
    \frac{\tau_{\rm A}^\pm}{\tau_{\rm nl}^\pm}
    \sim
    \frac{k_\perp^\pm\, \delta z_k^\mp}{k_\|^\pm\, v_{\rm A,0}}\,,
\end{equation}
and involves the wave-vector components $(k_\|^\pm,k_\perp^\pm)$ of the evolving fluctuation $\delta\bb{z}_\perp^\pm$ and the amplitude $\delta\bb{z}_\perp^\mp$ of the counter-propagating fluctuation that induces the nonlinearities on $\delta\bb{z}_\perp^\pm$.
In the following this parameter plays a central role to estimate the nonlinear cascade rate (or turbulent damping) of an Alfv\'en-wave packet interacting with background fluctuations.
In particular, to obtain the correct turbulent damping rate it is necessary to make a careful distinction between the nonlinear parameter $\chi^z$, which  characterizes counter-propagating pre-existing fluctuations, and the nonlinear parameter $\chi^w$, which  describes the interaction between the AW packet and background turbulence.

\section{Turbulent damping of an Alfv\'en-wave packet}\label{sec:general_damping}

We consider here an Alfv\'en-wave packet that is injected in an environment filled with pre-existing Alfv\'enic turbulence. We let $\delta\bb{w}$ be the initial Els\"asser variable of the packet, and $\lambda_\perp^w$ and $\lambda_\|^w$ respectively its wavelength perpendicular to and parallel to a mean-magnetic field $\langle\bb{B}\rangle_{\lambda^w}$ at such scales,\footnote{One can think of each component $i$ of this mean field at scale $\lambda$ as defined, for instance, by $\langle B_i\rangle_\lambda\sim\left(\int_{1/\ell_0}^{k'< 1/\lambda} B_{i,\bb{k'}}^2{\rm d}\bb{k'}\right)^{1/2}$, i.e., a magnetic field that is the result of the contribution from the background field $\bb{B}_0$ plus all the fluctuations $\delta\bb{B}_{\lambda'}$ at scales $\lambda'>\lambda$, such that the nonlinear timescale $\tau_{{\rm nl},\lambda'}$ over which the associated  turbulent eddy evolves is much longer than the nonlinear evolution timescale $\tau_{{\rm nl},\lambda}$ of fluctuations at the scale $\lambda$, i.e., $\tau_{{\rm nl},\lambda'}\gg\tau_{{\rm nl},\lambda}$, so that turbulent eddies on scales $\lambda'$ appear to be   frozen over the turnover time of turbulent eddies at scale $\lambda$. Operationally, this mean field can be defined in different ways \citep[e.g.,][]{ChoVishniacAPJ2000,ChoLazarianAPJL2004,HorburyPRL2008,WicksMNRAS2010,ChenMNRAS2011,MatthaeusAPJ2012,MalletMNRAS2016,CerriFSPAS2019}, but what is the most appropriate operational definition is still a matter of debate \citep[see, e.g.,][]{OughtonMatthaeusAPJ2020}.}  (the corresponding wave vectors being $k_\perp^w\sim1/\lambda_\perp^w$ and $k_\|^w\sim1/\lambda_\|^w$). 
The Alfv\'enic fluctuations populating the turbulent background are characterized by certain scale-dependent relations for their Els\"asser amplitude $\delta z_{\lambda_\perp^z}^\pm$, their wavelength anisotropy $\lambda_\|^{z,\pm}/\lambda_\perp^{z,\pm}$ (for which the corresponding wave vectors can be denoted as $k_\perp^{z,\pm}\sim1/\lambda_\perp^{z,\pm}$ and $k_\|^{z,\pm}\sim1/\lambda_\|^{z,\pm}$), and, if allowed, for the alignment angle between $\delta\boldsymbol{z}_{\lambda_\perp^z}^{+}$ and $\delta \boldsymbol{z}_{\lambda_\perp^z}^{-}$(i.e., $\sin\theta_{\lambda_\perp^{z,\pm}}^{\,z}$).
It is now instructive to derive the general relations first, leaving the explicit scaling belonging to different turbulence theories for later.
Hereafter we consider the case of balanced background turbulence, and thus drop the $\pm$ superscript everywhere for simplicity of notation.

While propagating, the AW packet  interacts nonlinearly only with counter-propagating Alfv\'enic fluctuations of the background. In terms of Els\"asser variables, the nonlinear interaction is   described by the term $(\delta\bb{z}\cdot\bb{\nabla})\,\delta\bb{w}$. 
The strength of this nonlinear interaction can then be determined by comparing the above nonlinear term with the term describing its linear propagation $(\bb{v}_{\rm A,0}\cdot\bb{\nabla})\delta\bb{w}$.
The interaction between the AW packet and the pre-existing fluctuations is described by the nonlinear parameter of the packet
\begin{equation}\label{eq:chi_w_general}
    \chi^w\sim\frac{\tau_{\rm A}^w}{\tau_{\rm nl}^w}\sim \frac{(\lambda_\|^w/v_{\rm A,0})}{(\lambda_\perp^w/\delta z_{\lambda_\perp^w})}\sim \left(\frac{\lambda_\|^w}{\lambda_\perp^w}\right) \left(\frac{\delta z_{\lambda_\perp^w}}{v_{\rm A,0}}\right)\,,
\end{equation}
where  local-in-scale interactions are assumed, so that $\delta z_{\lambda_\perp^z}$ is substituted with $\delta z_{\lambda_\perp^w}$ in the timescale associated with the nonlinear interaction between the AW packet and pre-existing turbulence:  $\tau_{\rm nl}^w\sim\lambda_\perp^w/\delta z_{\lambda_\perp^z}\sim\lambda_\perp^w/\delta z_{\lambda_\perp^w}$. A parameter $\chi^w\gtrsim1$ means strong nonlinear interactions, while $\chi^w<1$ denotes the weakly nonlinear regime. 
We note that the parameter in \eqref{eq:chi_w_general} is different from the nonlinear parameter that characterizes background turbulence (i.e., $\chi^z\sim \tau_{\rm A}^z/\tau_{\rm nl}^z\sim (\lambda_\|^z/\lambda_\perp^z)(\delta z_{\lambda_\perp^z}/v_{\rm A,0}))$,\footnote{Or, if scale-dependent (dynamic) alignment is taken into account, it is $\chi^z\sim \sin\theta_{\lambda_\perp^z}^{\,z} (\lambda_\|^z/\lambda_\perp^z)(\delta z_{\lambda_\perp^z}/v_{\rm A,0})$ (see Appendix~\ref{app:MHDturb_scalings}).} and we point out that, while background fluctuations can have $\chi^z\gtrsim1$ at scale $\lambda_\perp^z\sim\lambda_\perp^w$ (strong pre-existing turbulence), the condition $\chi^w\gtrsim1$ does not necessarily hold at these same scales.

It  should be noted that $\chi^w$ is not only proportional to the amplitude of background fluctuations at the scale $\lambda_\perp^w$ (i.e., $\delta z_{\lambda_\perp^w}/v_{\rm A,0}$), but it also depends   on the AW packet's propagation angle with respect to the mean magnetic field $\langle\boldsymbol{B}\rangle_{\lambda^w}$ at that scale: $\lambda_\|^w/\lambda_\perp^w\sim k_\perp^w/k_\|^w = \tan\Theta_{kB}^w$, where $\Theta_{kB}^w$ is the angle between $\boldsymbol{k}^w$ and $\langle\boldsymbol{B}\rangle_{\lambda^w\sim1/k^w}$.
As a result, if the amplitude $\delta z_0$ of background fluctuations at injection scale $\ell_0$ is such that $\delta z_0/v_{\rm A,0}\lesssim1$, then strong nonlinear interactions at scales $\lambda_\perp^w\ll\ell_0$ (where $\delta z_{\lambda_\perp^w}\ll\delta z_0$) require $\lambda_\|^w/\lambda_\perp^w\sim v_{\rm A,0}/\delta z_{\lambda_\perp^w}\gg1$. This regime is thus achieved only by quasi-perpendicular AW packets. In critically balanced pre-existing turbulence, for instance, fluctuations obey the relation $\delta z_{\lambda_\perp^z}/v_{\rm A,0}\sim\lambda_\|^z(\lambda_\perp^z)/\lambda_\perp^z$. Therefore, assuming local-in-scale interactions ($\lambda_\perp^z\sim\lambda_\perp^w$), the condition $\lambda_\|^w/\lambda_\perp^w\sim v_{\rm A,0}/\delta z_{\lambda_\perp^w}$ means that an AW packet undergoes strong nonlinear interactions (and thus severe turbulent damping) only if its wave vector matches the anisotropy of background turbulence associated with the perpendicular scale $\lambda_\perp^w$ (i.e., $\lambda_\|^w\approx \lambda_{\|}^z(\lambda_\perp^w))$. Therefore, the nonlinear interaction between a quasi-parallel AW (characterized by $\lambda_\|^w/\lambda_\perp^w\ll1$), and anisotropic pre-existing turbulence (characterized by $\lambda_\perp^z/\lambda_\|^z\ll1$) is always weak: $\chi^w\ll1$.

This can be shown explicitly by considering the quasi-parallel propagation limit, which is the case of interest for CR-generated AW packets. In this case, the propagation can only be as parallel as the external turbulence allows, meaning that  the propagation angle cannot be smaller than the amount $\delta b_{\lambda_\perp^w}/\langle B\rangle_{\lambda^w}\sim\delta z_{\lambda_\perp^w}/v_{\rm A,0}$ because of the field-line distortions induced by pre-existing turbulent fluctuations $\delta b_{\lambda_\perp^w}$ over the wavelength $\lambda_\perp^w$ of the packet. Therefore, the quasi-parallel (q$\|$) propagation limit is set by
\begin{equation}\label{eq:AW_q-prl-limit_general}
    \left.\frac{\lambda_\|^w}{\lambda_\perp^w}\right|_{\rm min} = \frac{\lambda_\|^{w,{\rm q\|}}}{\lambda_\perp^{w,{\rm q\|}}}\sim \frac{\delta z_{\lambda_\perp^{w,{\rm q\|}}}}{v_{\rm A,0}}\,.
\end{equation}
Hence, the associated nonlinear parameter in this limit is
\begin{equation}\label{eq:chi_w_q-prl-limit_general}
    \chi^{w,{\rm q\|}}\sim \left(\frac{\delta z_{\lambda_\perp^{w,{\rm q\|}}}}{v_{\rm A,0}}\right)^2\,.
\end{equation}

The strongly nonlinear regime can thus be achieved only at scales $\lambda$ where the AW packet interacts with pre-existing super-Alfv\'enic fluctuations ($\delta z_{\lambda}/v_{\rm A,0}>1$). In this regime, the concept of quasi-parallel propagation in \eqref{eq:AW_q-prl-limit_general} does not apply because at scales where $\delta b_{\lambda^w}/\langle B\rangle_{\lambda^w}>1$, the distinction between $\lambda_\perp^w$ and $\lambda_\|^w$ is lost and $\lambda_\perp^w\sim\lambda_\|^w\sim\lambda$ holds; hence, $\chi^w\sim \delta z_{\lambda}/v_{\rm A,0}\gtrsim1$. 
However, even for external turbulence that is injected with super-Aflv\'enic amplitude (i.e., $\delta z_0/v_{\rm A,0}\approx M_{\rm A,0}>1$ at scale $\ell_0$)  the fluctuation amplitude  decreases with decreasing scale. As a result, the nonlinear interaction between the AW and external fluctuations  becomes weak at  small-enough scales (i.e., at scales $\lambda_\perp^w<\ell_{\rm A}=\ell_0/M_{\rm A,0}^{\,3}\ll\ell_0$) for which the initially super-Alfv\'enic fluctuations become sub-Alfv\'enic, $\delta z_{\lambda_\perp^w}/v_{\rm A,0}<1$, and anisotropic (see Appendix~\ref{app:MHDturb_scalings}). Another way to visualize this is by rewriting \eqref{eq:chi_w_q-prl-limit_general} as
\begin{equation}\label{eq:chi_w_q-prl-limit_general_vs_chi_z}
   \chi^{w,{\rm q\|}}\sim  \left(\frac{\lambda_\perp^z}{\lambda_\|^z}\right)^2(\,\chi^z)^2\,,
\end{equation}
which is $\ll\chi^z$ for anisotropic background fluctuations, $\lambda_\perp^z\ll\lambda_\|^z$, and thus $\chi^{w,{\rm q\|}}\ll1$ even if pre-existing turbulence is critically balanced ($\chi^z\sim1$).\\
The difference between the intrinsic nonlinear parameter of background fluctuations ($\chi^{\,z})$ and the nonlinear parameters of a quasi-parallel Alfv\'en wave interacting with those fluctuations ($\chi^w$) for explicit MHD turbulent scaling in different regimes is reported the last two columns of Table~\ref{tab:MHDturb_scaling_nonlin-parameters}.

Having clarified the packet's nonlinear regimes that have to be considered in terms of background turbulence, we can now estimate the associated rate of turbulent damping. This means estimating the timescale over which an AW packet undergoes a cascade process due to its nonlinear interaction with pre-existing turbulent fluctuations. The cascade time of the packet is given by $\tau_{\rm casc}^w\sim(\tau_{\rm nl}^w)^2/\tau_{\rm A}^w\sim\tau_{\rm nl}^w/\chi^w$ for the weak regime ($\chi^w<1$), while it is $\tau_{\rm casc}^w\sim\tau_{\rm nl}^w$ in the strongly nonlinear case ($\chi^w\gtrsim1$, which we recall also implies that $\lambda_\|^w\sim\lambda_\perp^w\sim\lambda^w$; see discussion after equation \eqref{eq:chi_w_q-prl-limit_general}).

As a result, the turbulent damping rate is
\begin{equation}\label{eq:Turb-damp_w_generic}
    \Gamma_{\rm turb}^w
    \sim
    \frac{1}{\tau_{\rm casc}^w}
    \sim
    \left\{
    \begin{array}{lcr}
      \left(\frac{\lambda_\|^w}{\lambda_\perp^w}\right)\left(\frac{\lambda_\perp^w}{\ell_0}\right)^{-1}\left(\frac{\delta z_{\lambda_\perp^w}}{v_{\rm A,0}}\right)^2\frac{v_{\rm A,0}}{\ell_0} & &  {\rm if}\,\,\,\chi^w<1\\
      & & \\
      \left(\frac{\lambda^w}{\ell_0}\right)^{-1} \left(\frac{\delta z_{\lambda^w}}{v_{\rm A,0}}\right)\frac{v_{\rm A,0}}{\ell_0}  &  & {\rm if}\,\,\,\chi^w\gtrsim1
    \end{array}
    \right.
,\end{equation}
which reduces to 
\begin{equation}\label{eq:Turb-damp_w_q-prl-limit_generic}
    \Gamma_{\rm turb}^{w,{\rm q\|}}\,
    \sim
    \left(\frac{\lambda_\perp^{w,{\rm q\|}}}{\ell_0}\right)^{-1}\left(\frac{\delta z_{\lambda_\perp^{w,{\rm q\|}}}}{v_{\rm A,0}}\right)^3\frac{v_{\rm A,0}}{\ell_0}
\end{equation}
for quasi-parallel propagation and $\chi^{w,{\rm q\|}}<1$. We recall that $\chi^w$ should not be identified with the nonlinear parameter $\chi^z$ that describes the strength of background turbulence.
Therefore, when $\chi^w<1$ the turbulent damping rate of an AW packet is nonlinear with respect to the background-fluctuation amplitude and depends on the propagation angle of the wave, becoming a third-order quantity of the pre-existing turbulent amplitude in the quasi-parallel limit.

We conclude by highlighting that in \eqref{eq:Turb-damp_w_q-prl-limit_generic} there is a factor $\lambda_\perp^{-1}$ in front of the $\delta z_{\lambda_\perp}^3$ term. 
Therefore, a turbulent perpendicular scaling for $\delta z_{\lambda_\perp}\propto\lambda_\perp^\alpha$ with a spectral index $\alpha>1/3$ will produce a turbulent damping rate of quasi-parallel AW packets that decreases with decreasing scale. That would be the case of weak Alfv\'enic turbulence, as in \citet{GaltierJPP2000}, or the tearing-mediated regime, as in \citet{BoldyrevLoureiroAPJ2017} and \citet{MalletMNRAS2017}. On the other hand, fluctuations that scale with $\alpha<1/3$ result in a damping rate that increases with decreasing $\lambda_\perp^w$. 
This would be the case of critically balanced strong Alfv\'enic turbulence with scale-dependent alignment, as in \citet{BoldyrevPRL2006}.
Finally, for a Kolmogorov-like perpendicular scaling $\delta z_{\lambda_\perp}\propto\lambda_{\perp}^{1/3}$, as in critically balanced strong Alfv\'enic turbulence without dynamic alignment \citep{GoldreichSridharAPJ1995}, we can expect that the turbulent damping of quasi-parallel AW packets becomes scale-independent (i.e., $\Gamma_{\rm GS95}^{w,{\rm q\|}}\sim {\rm const}$).

\section{Turbulent damping with explicit MHD scalings}\label{sec:explicit_scalings}

The injection-scale Alfv\'enic Mach number is defined as the ratio $M_{\rm A,0} \approx \delta z_0/v_{\rm A,0}$, where $\delta z_0$ is the fluctuation amplitude at injection scale $\ell_0$, and determines the cascading regimes of background fluctuations. 
The Lunquist number at injection scale, $S_0=\ell_0\,v_{\rm A,0}/\eta$, is related to the system's resitivity $\eta$ and determines the dissipation scale of turbulent fluctuations;  we recall that here we have assumed $\nu=\eta$.
If we assume isotropic injection, the nonlinear parameter of background turbulence at scale $\ell_0$ indeed corresponds to the injection-scale Alfv\'enic Mach number, $\chi_0^z \approx M_{\rm A,0}$. If $M_{\rm A,0} < 1$, the turbulence is called  sub-Alfv\'enic: it starts as an anisotropic weak cascade that transitions into critically balanced strong turbulence at smaller scales (still anisotropic, but in a different fashion).  Trans-Alfv\'enic turbulence ($M_{\rm A,0}\approx 1$) consists of an anisotropic strong cascade of critically balanced fluctuations at all scales. When $M_{\rm A,0} > 1$ (large-amplitude injection), turbulence is called  super-Alfv\'enic and fluctuations initially undergo an isotropic (``hydrodynamic-like'') cascade until sub-Alfv\'enic amplitudes are attained at smaller scales, and turbulence becomes critically balanced and anisotropic. 
Then, if scale-dependent (dynamic) alignment of fluctuations is allowed in the critical-balance range, an additional transition to a different regime of anisotropic, strong turbulence can occur at even smaller scales due to magnetic reconnection (if the injection-scale Lundquist number $S_0$ is large enough; see Section~\ref{subsec:MHDscalings_alignment-only-strong}). In the following we only summarize the relevant scaling of background turbulent fluctuations in the different ranges. These scaling relations, along with the intrinsic nonlinear parameter $\chi^{\,z}$ of background fluctuations and the nonlinear parameter $\chi^{w}$ of a quasi-parallel AW propagating through such background turbulence, are also shown in Table~\ref{tab:MHDturb_scaling_nonlin-parameters} for convenience. A more detailed derivation of these ranges and of the associated scaling is provided in Appendix~\ref{app:MHDturb_scalings}.

The turbulent damping rates in this section were derived as follows. 
First, we employed the known perpendicular scaling of background turbulence, $\delta z_{\lambda_\perp}$, and locality of interactions ($\lambda_\perp^z\sim\lambda_\perp^{w,{\rm q\|}}$) in \eqref{eq:Turb-damp_w_q-prl-limit_generic} to obtain the turbulent damping rate as a function of the packet's perpendicular wavelength, $\Gamma_{\rm turb}^{w,{\rm q\|}}(\lambda_\perp^{w,{\rm q\|}})$. 
Then we used the quasi-parallel condition in \eqref{eq:AW_q-prl-limit_general} to retrieve the scaling of $\Gamma_{\rm turb}^{w,{\rm q\|}}$ with respect to the parallel wavelength $\lambda_\|^{w,{\rm q\|}}$.

The scaling of the turbulent damping rate for different background cascades and the corresponding range of scales where that is valid are given  in Table~\ref{tab:TurbDamping_scaling_comparison}, along with an explicit comparison with the rates available in the existing literature (i.e., from \citealt{FarmerGoldreichAPJ2004}, hereafter FG04, and from \citealt{LazarianAPJ2016}, hereafter L16). The behavior  of these damping rates and the comparison with previous estimates for two choices of sub-Alfv\'enic and super-Alfv\'enic injection ($M_{\rm A,0}=0.1$, $10$) and for $S_0=10^{14}$ are also shown in Figure~\ref{fig:TurbDamp_examples}, for convenience.
We note that previous results not only overestimate the damping rate by a factor that could be several orders of magnitude, but in most cases they also obtain a completely different result on how this damping rate depends upon the packet's parallel wavelength $\lambda_\|^w$. 

\begin{table*}[ht]
\caption[]{\bf Quasi-parallel Alfv\'en waves in MHD turbulence}
\centering
\begin{tabular}{|cccccc|}
\hline
  \multirow{2}{*}{\bf\em background}  &  \multicolumn{2}{c}{\multirow{3}{*}{\bf\em scale range of validity}}   &  \multirow{2}{*}{\bf\em scaling of background} & \multirow{2}{*}{\bf\em nonlinear parameter of} &  \multirow{2}{*}{\bf\em nonlinear parameter} \\  
   \multirow{2}{*}{\bf\em cascade} & & & \multirow{2}{*}{\bf\em turbulent fluctuations} & \multirow{2}{*}{\bf\em background turbulence}  & \multirow{2}{*}{\bf\em of quasi-parallel AWs} \\    
   \multirow{2}{*}{(acronym)}  &  \multirow{2}{*}{$\lambda_{\rm\perp,max}$}  & \multirow{2}{*}{$\lambda_{\rm\perp,min}$} & \multirow{2}{*}{$\delta z_{\lambda_\perp}$}   & \multirow{2}{*}{$\chi_{\lambda_\perp}^{\,z}$}  & \multirow{2}{*}{$\chi_{\lambda_\perp}^{w,{\rm q\|}}$} \\    
   & & & & & \\
\hline
\hline
  \multicolumn{6}{|c|}{\multirow{2}{*}{\bf without scale-dependent alignment}}\\
   & & & & & \\
\hline
  \multicolumn{6}{|c|}{\multirow{2}{*}{{\it sub-Alfv\'enic injection} ($M_{\rm A,0}<1$):}}\\
   & & & & & \\
   \multirow{2}{*}{(W0)} & \multirow{2}{*}{$\ell_0$} & \multirow{2}{*}{$M_{\rm A,0}^2\,\ell_0$} & \multirow{2}{*}{$v_{\rm A,0}\,M_{\rm A,0}\,(\lambda_\perp/\ell_0)^{1/2}$} & \multirow{2}{*}{$M_{\rm A,0}\,(\lambda_\perp/\ell_0)^{-1/2}$} & \multirow{2}{*}{$M_{\rm A,0}^2\,(\lambda_\perp/\ell_0)$} \\ 
   & & & & & \\
  \multirow{2}{*}{(GS95)} & \multirow{2}{*}{$M_{\rm A,0}^2\,\ell_0$} & \multirow{2}{*}{$M_{\rm A,0}^{-1}\,S_0^{-3/4}\,\ell_0$} & \multirow{2}{*}{$v_{\rm A,0}\,M_{\rm A,0}^{4/3}\,(\lambda_\perp/\ell_0)^{1/3}$} & \multirow{2}{*}{$\sim1$} & \multirow{2}{*}{$M_{\rm A,0}^{8/3}\,(\lambda_\perp/\ell_0)^{2/3}$} \\ 
   & & & & & \\
\hline
  \multicolumn{6}{|c|}{\multirow{2}{*}{{\it trans-Alfv\'enic injection} ($M_{\rm A,0}\approx1$):}}\\
   & & & & & \\
  \multirow{2}{*}{(GS95)} & \multirow{2}{*}{$\ell_0$} & \multirow{2}{*}{$S_0^{-3/4}\,\ell_0$} & \multirow{2}{*}{$v_{\rm A,0}\,(\lambda_\perp/\ell_0)^{1/3}$} & \multirow{2}{*}{$\sim1$} & \multirow{2}{*}{$(\lambda_\perp/\ell_0)^{2/3}$} \\ 
   & & & & & \\
\hline
  \multicolumn{6}{|c|}{\multirow{2}{*}{{\it super-Alfv\'enic injection} ($M_{\rm A,0}>1$):}}\\
   & & & & & \\
  \multirow{2}{*}{(K41)} & \multirow{2}{*}{$\ell_0$} & \multirow{2}{*}{$M_{\rm A,0}^{-3}\,\ell_0$} & \multirow{2}{*}{$v_{\rm A,0}\,M_{\rm A,0}\,(\lambda/\ell_0)^{1/3}$} & \multirow{2}{*}{$M_{\rm A,0}\,(\lambda/\ell_0)^{1/3}$} & \multirow{2}{*}{$M_{\rm A,0}\,(\lambda/\ell_0)^{1/3}$} \\ 
   & & & & & \\
  \multirow{2}{*}{(GS95)} & \multirow{2}{*}{$M_{\rm A,0}^{-3}\,\ell_0$} & \multirow{2}{*}{$(M_{\rm A,0}\,S_0)^{-3/4}\,\ell_0$} & \multirow{2}{*}{$v_{\rm A,0}\,M_{\rm A,0}\,(\lambda_\perp/\ell_0)^{1/3}$} & \multirow{2}{*}{$\sim1$} & \multirow{2}{*}{$M_{\rm A,0}^2\,(\lambda_\perp/\ell_0)^{2/3}$} \\ 
   & & & & & \\
\hline
\hline
  \multicolumn{6}{|c|}{\multirow{2}{*}{\bf with scale-dependent alignment}}\\
   & & & & & \\
\hline
  \multicolumn{6}{|c|}{\multirow{2}{*}{{\it sub-Alfv\'enic injection} ($M_{\rm A,0}<1$):}}\\
   & & & & & \\
   \multirow{2}{*}{(W0)} & \multirow{2}{*}{$\ell_0$} & \multirow{2}{*}{$M_{\rm A,0}^2\,\ell_0$} & \multirow{2}{*}{$v_{\rm A,0}\,M_{\rm A,0}\,(\lambda_\perp/\ell_0)^{1/2}$} & \multirow{2}{*}{$M_{\rm A,0}\,(\lambda_\perp/\ell_0)^{-1/2}$} & \multirow{2}{*}{$M_{\rm A,0}^2\,(\lambda_\perp/\ell_0)$} \\ 
   & & & & & \\
  \multirow{2}{*}{(B06)} & \multirow{2}{*}{$M_{\rm A,0}^2\,\ell_0$} & \multirow{2}{*}{$M_{\rm A,0}^{-2/7}\, S_0^{-4/7}\,\ell_0$} & \multirow{2}{*}{$v_{\rm A,0}\,M_{\rm A,0}^{3/2}\,(\lambda_\perp^z/\ell_0)^{1/4}$} & \multirow{2}{*}{$\sim1$} & \multirow{2}{*}{$M_{\rm A,0}^{3}\,(\lambda_\perp^z/\ell_0)^{1/2}$} \\ 
   & & & & & \\
  \multirow{2}{*}{(TMT)} & \multirow{2}{*}{$M_{\rm A,0}^{-2/7}\, S_0^{-4/7}\,\ell_0$} & \multirow{2}{*}{$M_{\rm A,0}^{-1}\,S_0^{-3/4}\,\ell_0$} & \multirow{2}{*}{$v_{\rm A,0}\,S_0^{1/5}\,M_{\rm A,0}^{8/5}\,(\lambda_\perp/\ell_0)^{3/5}$} & \multirow{2}{*}{$\sim1$} & \multirow{2}{*}{$S_0^{2/5}\,M_{\rm A,0}^{16/5}\,(\lambda_\perp/\ell_0)^{6/5}$} \\ 
   & & & & & \\
\hline
  \multicolumn{6}{|c|}{\multirow{2}{*}{{\it trans-Alfv\'enic injection} ($M_{\rm A,0}\approx1$):}}\\
   & & & & & \\
  \multirow{2}{*}{(B06)} & \multirow{2}{*}{$\ell_0$} & \multirow{2}{*}{$S_0^{-4/7}\,\ell_0$} & \multirow{2}{*}{$v_{\rm A,0}\,(\lambda_\perp/\ell_0)^{1/4}$} & \multirow{2}{*}{$\sim1$} & \multirow{2}{*}{$(\lambda_\perp/\ell_0)^{1/2}$} \\ 
   & & & & & \\
  \multirow{2}{*}{(TMT)} & \multirow{2}{*}{$S_0^{-4/7}\,\ell_0$} & \multirow{2}{*}{$S_0^{-3/4}\,\ell_0$} & \multirow{2}{*}{$v_{\rm A,0}\,S_0^{1/5}\,(\lambda_\perp/\ell_0)^{3/5}$} & \multirow{2}{*}{$\sim1$} & \multirow{2}{*}{$S_0^{2/5}\,(\lambda_\perp/\ell_0)^{6/5}$} \\ 
   & & & & & \\
\hline
  \multicolumn{6}{|c|}{\multirow{2}{*}{{\it super-Alfv\'enic injection} ($M_{\rm A,0}>1$):}}\\
   & & & & & \\
  \multirow{2}{*}{(K41)} & \multirow{2}{*}{$\ell_0$} & \multirow{2}{*}{$M_{\rm A,0}^{-3}\,\ell_0$} & \multirow{2}{*}{$v_{\rm A,0}\,M_{\rm A,0}\,(\lambda/\ell_0)^{1/3}$} & \multirow{2}{*}{$M_{\rm A,0}\,(\lambda/\ell_0)^{1/3}$} & \multirow{2}{*}{$M_{\rm A,0}\,(\lambda/\ell_0)^{1/3}$} \\ 
   & & & & & \\
  \multirow{2}{*}{(B06)} & \multirow{2}{*}{$M_{\rm A,0}^{-3}\,\ell_0$} & \multirow{2}{*}{$M_{\rm A,0}^{-9/7}\,S_0^{-4/7}\,\ell_0$} & \multirow{2}{*}{$v_{\rm A,0}\,M_{\rm A,0}^{3/4}\,(\lambda_\perp/\ell_0)^{1/4}$} & \multirow{2}{*}{$\sim1$} & \multirow{2}{*}{$M_{\rm A,0}^{3/2}\,(\lambda_\perp/\ell_0)^{1/2}$} \\ 
   & & & & & \\
  \multirow{2}{*}{(TMT)} & \multirow{2}{*}{$M_{\rm A,0}^{-9/7}\,S_0^{-4/7}\,\ell_0$} & \multirow{2}{*}{$(M_{\rm A,0}\,S_0)^{-3/4}\,\ell_0$} & \multirow{2}{*}{$v_{\rm A,0}\,S_0^{1/5}\,M_{\rm A,0}^{6/5}\,(\lambda_\perp^z/\ell_0)^{3/5}$} & \multirow{2}{*}{$\sim1$} & \multirow{2}{*}{$S_0^{2/5}\,M_{\rm A,0}^{12/5}\,(\lambda_\perp^z/\ell_0)^{6/5}$} \\ 
   & & & & & \\
\hline
\end{tabular}
\tablefoot{Summary of the relevant scaling relations for balanced MHD turbulence in the different regimes mentioned in Section~\ref{sec:explicit_scalings}  (see Appendix~\ref{app:MHDturb_scalings} for the derivation). The last two columns explicitly show the difference between the intrinsic nonlinear parameter $\chi^{\,z}$ of background fluctuations and the nonlinear parameter $\chi^{w,{\rm q\|}}$ of a quasi-parallel Alfv\'en wave interacting with those fluctuations. The locality of interactions $\lambda_\perp^w\sim\lambda_\perp^z\sim\lambda_\perp$ has been implied everywhere and that in all the regimes with dynamic alignment, a transition to a tearing-mediated range (TMT) has been assumed, i.e., that the injection-scale Lunquist number satisfies the inequality $S_0\gg M_{\rm A,0}^{-4}$ ($S_0\gg M_{\rm A,0}^3$) for sub-Alfv\'enic (super-Alfv\'enic) injection. See Table~\ref{tab:TurbDamping_scaling_comparison} for the consequences of not taking into account this difference between $\chi^{w,{\rm q\|}}$ and $\chi^{\,z}$ on the inferred turbulent damping rate of quasi-parallel Alfv\'en waves.}
\label{tab:MHDturb_scaling_nonlin-parameters}
\end{table*}

\subsection{Magnetohydrodynamic turbulence without scale-dependent alignment}\label{subsec:MHDscalings_no_alignment}

We first consider the  classic picture in which dynamic alignment of turbulent fluctuations does not occur. 
In this case, we have three possible regimes for background   turbulence: 
\begin{itemize}

    \item[] {\bf[W0]} A weak anisotropic cascade with fluctuation scaling $\delta z_{\lambda_\perp^z}^{\rm(W0)}/v_{\rm A,0} \sim M_{\rm A,0}\,(\lambda_\perp^z/\ell_0)^{1/2}$ that only generates smaller perpendicular scales (i.e., $\lambda_\|^z\sim\ell_0\approx{\rm const.}$) and transitions into a strong cascade at the critical balance (CB) scale $\lambda_{\perp,{\rm CB}}^z \sim M_{\rm A,0}^{\,2}\,\ell_0$. This cascade is realized in the range of scales $\lambda_{\perp,{\rm CB}}^z\lesssim\lambda_\perp^z\lesssim\ell_0$, and only for sub-Alfv\'enic injection ($M_{\rm A,0} < 1$).\\
    
    \item[] {\bf[K41]} A strong, isotropic (hydrodynamic-like) cascade characterized by the scaling $\delta z_{\lambda^z}^{\rm(K41)}/v_{\rm A,0} \sim M_{\rm A,0}\,(\lambda^z/\ell_0)^{1/3}$. These fluctuations attain sub-Alfv\'enic amplitudes, becoming anisotropic and critically balanced, at a scale $\ell_{\rm A} \sim M_{\rm A,0}^{\,-3}\,\ell_0$. This cascade is realized at scales $\ell_{\rm A}\lesssim\lambda^z\lesssim\ell_0$, and only for super-Alfv\'enic injection ($M_{\rm A,0} > 1$). \\
    
    \item[] {\bf[GS95]} A strong anisotropic cascade of critically balanced fluctuations with perpendicular scaling $\delta z_{\lambda_\perp^z}^{\rm(GS95)} \propto (\lambda_\perp^z)^{1/3}$.  
    This type of cascade is realized either for trans-Alfv\'enic injection ($M_{\rm A,0} \approx 1$), or when cascading fluctuations of the two  regimes above reach the scale $\lambda_{\perp,CB}^z$ and $\ell_{\rm A}$, respectively. For trans-/sub-Alfv\'enic injection ($M_{\rm A,0}\lesssim1$), the dependence on $M_{\rm A,0}$ of the scaling is $\delta z_{\lambda_\perp^z}^{\rm(GS95)}/v_{\rm A,0} \sim M_{\rm A,0}^{\,4/3}\,(\lambda_\perp^z/\ell_0)^{1/3}$, and the cascade achieves dissipation at a scale $\lambda_{\perp,{\rm min}}^{z\,{\rm(subA)}}/\ell_0\sim M_{\rm A,0}^{\,-1}\,S_0^{-3/4}$. For super-Alfv\'enic injection ($M_{\rm A,0}>1$), the fluctuation scaling with $M_{\rm A,0}$ is linear (i.e., $\delta z_{\lambda_\perp^z}^{\rm(GS95)}/v_{\rm A,0} \sim M_{\rm A,0}\,(\lambda_\perp^z/\ell_0)^{1/3}$), and the dissipation scale is given by $\lambda_{\perp,{\rm min}}^{z\,{\rm(supA)}}/\ell_0\sim (M_{\rm A,0}\,S_0)^{-3/4}$. Here $S_0=\ell_0\,v_{\rm A,0}/\eta$ is the Lundquist number at injection scale.

\end{itemize}

By using \eqref{eq:chi_w_q-prl-limit_general} we can verify that the nonlinear interaction between a quasi-parallel AW packets with wavelength $\lambda_\perp^{w,{\rm q\|}}$ and the pre-existing turbulence is weak, $\chi^{w,{\rm q\|}} \ll 1$, in the range of scales where the cascade of background fluctuations is either weak (W0) or critically balanced (GS95). This means $\chi^{w,{\rm q\|}}\ll1$ at scales $\lambda_\perp^{w,{\rm q\|}}<\ell_0$ and $\lambda_\perp^{w,{\rm q\|}}<\ell_{\rm A}$ respectively for sub-Alfv\'enic and super-Alfv\'enic injection (see Table~\ref{tab:MHDturb_scaling_nonlin-parameters} for the explicit scaling of $\chi^{w,{\rm q\|}}$ in these different regimes). Hence, the cascade time $\tau_{\rm casc}^{w,{\rm q\|}}\sim\tau_{\rm nl}^w/\chi^{w,{\rm q\|}}$ of the AW packets for these cases is not just the nonlinear time $\tau_{\rm nl}^w$, and the turbulent damping rate is given by \eqref{eq:Turb-damp_w_q-prl-limit_generic}. 
In \citet{FarmerGoldreichAPJ2004}, for instance, the nonlinear time $\tau_{\rm nl}^z$ instead of $\tau_{\rm casc}^{w,{\rm q\|}}$ was used to compute the turbulent damping rate. In a subsequent work by \citet{LazarianAPJ2016},  the nonlinear parameter of background turbulence $\chi^z$ was used instead of $\chi^{w,{\rm q\|}}\ll\chi^z$ to compute a cascade time $\tau_{\rm nl}^w/\chi^z$. This resulted in an estimated timescale for turbulent damping that was  notably shorter than the actual cascade time that should be used.
Taking properly into account the difference between $\chi^z$ and $\chi^{w,{\rm q\|}}$ thus changes significantly the effectiveness of turbulent damping in pre-existing turbulence with respect to all these previous estimates (see Table~\ref{tab:TurbDamping_scaling_comparison} for the generic case, or Figure~\ref{fig:TurbDamp_examples} for  two specific examples of sub- and super-Alfv\'enic injection regimes).

\subsubsection{Sub- and trans-Alfv\'enic turbulence ($M_{\rm A,0} \leq 1$) without dynamic alignment}\label{subsubsec:subA_no_alignment}

In sub-Alfv\'enic background turbulence without dynamic alignment, a quasi-parallel AW packet with  normalized  parallel wavelength $\hat{\lambda}_\|^w=\lambda_\|^{w,{\rm q\|}}/\ell_0$ is subjected to the following turbulent damping rate:
\begin{center}
    {\bf [$M_{\rm A,0}<1$, no dynamic alignment]}
\end{center}
\begin{equation}\label{eq:Turb-damp_subA_no_alignment}
    \Gamma_{\rm subA}^{w,{\rm q\|}}
    \sim
    \left\{
    \begin{array}{lcr}
      M_{\rm A,0}^{\,8/3}\left(\hat{\lambda}_\|^w\right)^{1/3}\frac{v_{\rm A,0}}{\ell_0} & \,\,\, & M_{\rm A,0}^{\,4} \lesssim \hat{\lambda}_\|^w\lesssim M_{\rm A,0}\\
      & & \\
      M_{\rm A,0}^{\,4}\,\frac{v_{\rm A,0}}{\ell_0} & \,\,\, & \hat{\lambda}_{\|,{\rm min}}^{w\,{\rm(subA)}} \lesssim \hat{\lambda}_\|^w\lesssim M_{\rm A,0}^{\,4} 
    \end{array}
    \right.
.\end{equation}
Here $\hat{\lambda}_{\|,{\rm min}}^{w\,{\rm(subA)}} \sim (M_{\rm A,0}\,\hat{\lambda}_{\perp,{\rm min}}^{z\,{\rm(subA)}})^{4/3} \sim S_0^{-1}$ is the minimum packet wavelength that is effectively subjected to turbulent damping, with $\hat{\lambda}_{\perp,{\rm min}}^{z\,{\rm(subA)}}\sim M_{\rm A,0}^{\,-1}\,S_0^{-3/4}$ being the normalized dissipation scale of the turbulence.
The ranges of $\hat{\lambda}_\|^w$ in \eqref{eq:Turb-damp_subA_no_alignment} were determined according  to the quasi-parallel condition in \eqref{eq:AW_q-prl-limit_general} and, assuming local interactions $\lambda_\perp^{w,{\rm q\|}}\sim \lambda_\perp^z$, employing the $\lambda_\perp^z$ range of validity for each turbulent regime (see Appendix~\ref{app:MHDturb_scalings}).\\

The trans-Alfv\'enic regime is trivially obtained from \eqref{eq:Turb-damp_subA_no_alignment} when the initial weak cascade does not occur:
\begin{center}
    {\bf [$M_{\rm A,0}\simeq1$, no dynamic alignment]}
\end{center}
\begin{equation}\label{eq:Turb-damp_transA_no_alignment}
    \quad\Gamma_{\rm transA}^{w,{\rm q\|}}
    \sim   
    \frac{v_{\rm A,0}}{\ell_0}
    \qquad\qquad\qquad\,\, 
    (\hat{\lambda}_{\perp,{\rm min}}^z)^{4/3} \lesssim \hat{\lambda}_\|^w\lesssim 1
.\end{equation}
Here $\hat{\lambda}_{\perp,{\rm min}}^z\sim S_0^{-3/4}$ is the (normalized) dissipation scale of GS95 turbulence in the trans-Alfv\'enic regime.

The damping rates in \eqref{eq:Turb-damp_subA_no_alignment} differ from the values previously derived in the literature (see Table~\ref{tab:TurbDamping_scaling_comparison}) because here the nonlinear parameter of the AW packet is properly taken into account (see Table~\ref{tab:MHDturb_scaling_nonlin-parameters}). We can verify that the turbulent damping rate in the W0 range of sub-Alfv\'enic turbulence (i.e., equation (46) in \citealt{LazarianAPJ2016}) can be recovered if the nonlinear parameter $\chi^z$ of background turbulence is employed instead of the nonlinear parameter $\chi^w$ of the AW packet. Analogously, the result in the GS95 range of sub-Alfv\'enic turbulence in Eq. (34) of the same paper is recovered by assuming strong interactions between the quasi-parallel AW packet and background fluctuations (i.e., identifying $\chi^w$ with $\chi^z\sim1$ at those scales). However, given the expression for $\chi^w$ in \eqref{eq:chi_w_general}, the assumption $\chi^w\sim1$ would require $\lambda_\|^w/\lambda_\perp^w\sim v_{\rm A,0}/\delta z_{\lambda_\perp^w}\gg1$, which is inconsistent with the quasi-parallel limit $\lambda_\|^w/\lambda_\perp^w\ll1$. The same argument applies when comparing the damping rate for the trans-Alfv\'enic case in \eqref{eq:Turb-damp_transA_no_alignment} with Eq. (9) in \citet{FarmerGoldreichAPJ2004}.

It is important to note that the results obtained here strongly change the effectiveness of the turbulent damping of CR-generated Alfv\'en-wave packets. 
Depending on the Alfv\'enic Mach number $M_{\rm A,0}$ and on the Lunquist number $S_0$, the damping rates in \eqref{eq:Turb-damp_subA_no_alignment} and \eqref{eq:Turb-damp_transA_no_alignment} can be several orders of magnitude lower than the  previously derived rates, those  usually employed in CR studies (see Table~\ref{tab:TurbDamping_scaling_comparison} and the left panel in Figure~\ref{fig:TurbDamp_examples}).
In particular, we can see that the damping rate of an AW packet interacting with pre-existing weak turbulence is at least a factor $M_{\rm A,0}$ lower than   previously estimated (i.e., at scale $\lambda_{\|,{\rm max}}^{w,{\rm q\|}}\sim M_{\rm A,0}\,\ell_0$, when this difference is at its minimum, then it increases even further due to the different dependence on $\lambda_\|^{w,{\rm q\|}}$). When the packet starts to interact with strong turbulence (i.e., for $\lambda_{\|,{\rm CB}}^{w,{\rm q\|}}\sim M_{\rm A,0}^4\,\ell_0$), the damping rate becomes at least a factor $M_{\rm A,0}^4\ll1$ lower than   has been derived in the literature (a difference that, again, increases even further with decreasing packet's parallel wavelength due to the radically different wavelength dependence of $\Gamma_{\rm GS95}^{w,{\rm q\|}}$ in \eqref{eq:Turb-damp_subA_no_alignment} and \eqref{eq:Turb-damp_transA_no_alignment} with respect to the results in \citealt{FarmerGoldreichAPJ2004} and in \citealt{LazarianAPJ2016}). This is also true   for trans-Alfv\'enic injection ($M_{\rm A,0}\simeq1$), in which case the damping rate in \eqref{eq:Turb-damp_transA_no_alignment} would be the same as that in the literature, only at scales $\lambda_{\|}^{w,{\rm q\|}}\sim\ell_0$, and then the two results would rapidly diverge with decreasing wavelength of the quasi-parallel AW packet at $\lambda_{\|}^{w,{\rm q\|}}<\ell_0$.
Finally, the maximum difference between the damping rate obtained here and those found in the literature is achieved at the minimum wavelength for which this damping mechanism is effective; at $\lambda_{\|}^{w,{\rm q\|}}\sim \lambda_{\|,{\rm min}}^{w}$ the actual damping rate is a factor $\sim S_0^{-1/2}M_{\rm A,0}^2\ll1$ lower than the results in \citet{FarmerGoldreichAPJ2004} and in \citet{LazarianAPJ2016}; in astrophysical systems this factor can represent many orders of magnitude since $S_0$ can be extremely large, for example larger than $10^{20}$~\citep{PriestForbesBOOK2007}. 
We also point out that in sub-Alfv\'enic turbulence the ordering $S_0\gg M_{\rm A,0}^{-4}$ is implied in order to have a significant GS95 range (see Appendix~\ref{app:MHDturb_scalings}).

\subsubsection{Super-Alfv\'enic turbulence ($M_{\rm A,0} > 1$) without dynamic alignment}\label{subsubsec:supA_no_alignment}

When the injection regime of background fluctuations is super-Alfv\'enic, an AW packet is instead subjected to a turbulent damping given by
\begin{center}
    {\bf [$M_{\rm A,0}>1$, no dynamic alignment]}
\end{center}
\begin{equation}\label{eq:Turb-damp_supA_no_alignment}
    \Gamma_{\rm supA}^{w,{\rm q\|}}
    \sim
    \left\{
    \begin{array}{lcr}
      M_{\rm A,0}\,\left(\hat{\lambda}^w\right)^{-2/3}\frac{v_{\rm A,0}}{\ell_0} &  & M_{\rm A,0}^{\,-3} \lesssim \hat{\lambda}^w\lesssim 1\\
      & & \\
      M_{\rm A,0}^{\,3}\,\frac{v_{\rm A,0}}{\ell_0} &  & \hat{\lambda}_{\|,{\rm min}}^{w\,{\rm(supA)}} \lesssim \hat{\lambda}_\|^w \lesssim M_{\rm A,0}^{\,-3} 
    \end{array}
    \right.
,\end{equation}
where the damping rate in the range $M_{\rm A,0}^{\,-3} \lesssim \hat{\lambda}^w\lesssim 1$ is obtained from the $\chi^w\gtrsim1$ part of \eqref{eq:Turb-damp_w_generic}, and the shortest wavelength affected by turbulent damping is $\hat{\lambda}_{\|,{\rm min}}^{w\,{\rm(supA)}} \sim M_{\rm A,0}\,(\hat{\lambda}_{\perp,{\rm min}}^{z\,{\rm(supA)}})^{4/3} \sim S_0^{-1}$. 
We also point out that the isotropic normalized wavelength $\hat{\lambda}^w=\lambda^w/\ell_0$ enters the damping rate in the range $\ell_{\rm A}\lesssim\lambda^w\lesssim\ell_0$ (i.e., where the AW packet interacts with hydrodynamic-like  pre-existing turbulence). 

In the range $M_{\rm A,0}^{\,-3} \lesssim \hat{\lambda}^w\lesssim 1$, the packet's nonlinear parameter is larger than unity (i.e., $\chi^w\approx\delta z_\lambda/v_{\rm A,0}$, and $\delta z_\lambda/v_{\rm A,0}>1$ at those scales; see Table~\ref{tab:MHDturb_scaling_nonlin-parameters}). Thus, the result obtained above for this range of scales agrees with the corresponding result provided in equation (55) of \citet{LazarianAPJ2016}. This is a consequence of the fact that the quasi-parallel condition \eqref{eq:AW_q-prl-limit_general} does not apply in the range $\ell_0\gtrsim\lambda\gtrsim\ell_{\rm A}\approx M_{\rm A,0}^{-3}\ell_0$ and the distinction between $\lambda_\|^w$ and $\lambda_\perp^w$ is lost. As a result, for local and isotropic interactions (meaning $\lambda^w\sim\lambda^z\sim\lambda$), there is no  difference between the expressions for $\chi^w$ and for $\chi^z$ (see Table~\ref{tab:MHDturb_scaling_nonlin-parameters}). On the other hand, at smaller scales ($\lambda^w \lesssim \ell_{\rm A}$), we recover a distinction between $\lambda_\|^w$ and $\lambda_\perp^w$ because background fluctuations become sub-Alfv\'enic and anisotropic (i.e., $\delta z_{\lambda_\perp}/v_{\rm A,0}<1$ and $\lambda_\perp^z\ll\lambda_\|^z$), and the quasi-parallel condition \eqref{eq:AW_q-prl-limit_general} does apply again, affecting $\chi^w$. Therefore, a quasi-parallel AW packet with $\lambda_\|^{w,{\rm q\|}}<\ell_{\rm A}$ experiences a weak nonlinear interaction with background turbulence (i.e., $\chi_{\lambda_\perp}^{w,{\rm q\|}}\approx(\delta z_{\lambda_\perp}/v_{\rm A,0})^2\ll1$), while background turbulence is critically balanced, $\chi_{\lambda_\perp}^{\,z}\sim1$ (cf. equation \eqref{eq:chi_w_q-prl-limit_general_vs_chi_z} in Section~\ref{sec:general_damping} and Table~\ref{tab:MHDturb_scaling_nonlin-parameters}). Hence, the two nonlinear parameters need not  be confused at scales below $\ell_{\rm A}$, and this is why the turbulent damping rate of quasi-parallel AWs that we obtain for this range of scales is again different from equation (52) of \citet{LazarianAPJ2016}.

As for the sub-Alfv\'enic case discussed earlier, we note that also for this $M_{\rm A,0}>1$ regime the result in \eqref{eq:Turb-damp_supA_no_alignment} implies a drastic change in the effectiveness of turbulent damping for CR-generated Alfv\'en-wave packets. 
While our result agrees with the turbulent damping rate found in the literature for the range of scales $\ell_{\rm A}\lesssim \lambda^w \lesssim \ell_0$, the corresponding rate at smaller scales, $\lambda_\|^{w,{\rm q\|}}<\ell_{\rm A}$, can be several orders of magnitude smaller than that usually employed in CR studies (see Table~\ref{tab:TurbDamping_scaling_comparison} and the right panel in Figure~\ref{fig:TurbDamp_examples}).
The damping rate in \eqref{eq:Turb-damp_supA_no_alignment} indeed rapidly diverges from the one given in \citet{LazarianAPJ2016} with decreasing wavelength of the quasi-parallel AW packet when $\lambda_{\|}^{w,{\rm q\|}}<\ell_{\rm A}$, reaching its maximum difference at $\lambda_{\|}^{w,{\rm q\|}}\sim \lambda_{\|,{\rm min}}^{w}$, where the actual damping rate in \eqref{eq:Turb-damp_supA_no_alignment} is a factor $\sim S_0^{-1/2}M_{\rm A,0}^{3/2}\ll1$ lower than the value provided in the literature (see the explicit comparison in Table~\ref{tab:TurbDamping_scaling_comparison}). We also point out that for super-Alfv\'enic injection, the ordering $S_0\gg M_{\rm A,0}^3$ is implied in order to have a significant GS95 range (see Appendix~\ref{app:MHDturb_scalings}). 

\subsection{Magnetohydrodynamic turbulence with dynamic alignment}\label{subsec:MHDscalings_alignment-only-strong}

The  classic picture presented above is now extended to the case in which counter-propagating Els\"asser fields $\delta\bb{z}_{\lambda_\perp}^+$ and $\delta\bb{z}_{\lambda_\perp}^-$ (or, in a similar way, velocity and magnetic-field fluctuations, $\delta\bb{u}_{\lambda_\perp}$ and $\delta\bb{b}_{\lambda_\perp}$) tend to align with each other in a scale-dependent fashion \citep{BoldyrevPRL2006}. This  dynamic alignment not only modifies the fluctuation scaling and anisotropy by inducing a weakening of the nonlinear interaction, but can also open the possibility of a reconnection-mediated regime at small scales \citep[still within the MHD range of scales, not in the kinetic regime; see, e.g.,][]{BoldyrevLoureiroAPJ2017,MalletMNRAS2017}.  
In this section we consider the case when such a scale-dependent alignment occur only in critically balanced turbulent fluctuations\footnote{Addressing the case in which alignment could occur also at weak nonlinearities \citep{CerriAPJ2022} may be still premature at this point, and it requires us to account for the alignment induced by background fluctuations on the AW packet itself. For the sake of simplicity, this case is  not  treated here, but  will be addressed separately in a following work.}, $\chi^z\sim1$.
In this case, in addition to the (W0) and (K41) regimes of the previous Section~\ref{subsec:MHDscalings_no_alignment}, one can have two additional regimes for background turbulence (see also Table~\ref{tab:MHDturb_scaling_nonlin-parameters}): 
\begin{itemize}
   
    \item[] {\bf[B06]} An anisotropic, strong cascade of critically balanced and dynamically aligned fluctuations that replaces the GS95 regime. In this case, the fluctuation alignment angle decreases with decreasing scale so that $\sin\theta_{\lambda_\perp}^{\,z}\propto (\lambda_\perp^z/\ell_0)^{1/4}$. For sub- and trans-Alfv\'enic injection ($M_{\rm A,0} \leq 1$), the perpendicular scaling of turbulent fluctuations at scales $\lambda_\perp^z\lesssim\lambda_{\perp,{\rm CB}}^z$, turns out to be $\delta z_{\lambda_\perp^z}^{\rm(B06)}/v_{\rm A,0} \sim M_{\rm A,0}^{\,3/2}\,(\lambda_\perp^z/\ell_0)^{1/4}$. 
    When $S_0\gg M_{\rm A,0}^{-4}$, this cascade can further turn into a reconnection-mediated regime below a transition scale $\lambda_{\perp,*}^{z\,{\rm(subA)}}/\ell_0\sim M_{\rm A,0}^{\,-2/7}\, S_0^{-4/7}$.
    For super-Alfv\'enic injection ($M_{\rm A,0} > 1$), turbulent fluctuations at scales $\lambda_\perp^z\lesssim\ell_{\rm A}$ follow instead a perpendicular scaling given by $\delta z_{\lambda_\perp^z}^{\rm(B06)}/v_{\rm A,0} \sim M_{\rm A,0}^{\,3/4}\,(\lambda_\perp^z/\ell_0)^{1/4}$. In this super-Alfv\'enic regime, a transition to reconnection-mediated turbulence may occur at a scale $\lambda_{\perp,*}^{z\,{\rm(supA)}}/\ell_0\sim M_{\rm A,0}^{\,-9/7}\, S_0^{-4/7}$ if $S_0\gg M_{\rm A,0}^3$.  
    If $S_0\lesssim M_{\rm A,0}^{-4}$ and $S_0\lesssim M_{\rm A,0}^{3}$ respectively in the sub-Alfv\'enic and super-Alfv\'enic regime, the dissipation scale for a given regime is larger than the corresponding transition scale and the (B06) cascade does not transition into the tearing-mediated regime. When this is case, the dissipation scale is achieved at $\lambda_{\perp,{\rm min}}^{z\,{\rm(subA)}}/\ell_0\sim (M_{\rm A,0}\,S_0)^{-2/3}$ in the  trans-/sub-Alfv\'enic regime, or at $\lambda_{\perp,{\rm min}}^{z\,{\rm(supA)}}/\ell_0\sim M_{\rm A,0}^{\,-1}\, S_0^{-2/3}$ for super-Alfv\'enic injection.
    \\
    
    \item[] {\bf[TMT]} A strong anisotropic cascade of critically balanced and dinamically (mis-)aligned fluctuations that are generated by magnetic-reconnection processes. In this case, fluctuations scale as $\delta z_{\lambda_\perp^z}^{\rm(TMT)} \propto S_0^{1/5}(\lambda_\perp^z)^{3/5}$ and are subjected to a scale-dependent mis-alignment given by $\sin\theta_{\lambda_\perp^{\,z}}^z\propto(\lambda_\perp^z/\ell_0)^{-4/5}$.  
    For sub- and trans-Alfv\'enic injection ($M_{\rm A,0}\leq1$), the perpendicular scaling of tearing-mediated turbulent fluctuations is given by $\delta z_{\lambda_\perp^z}^{\rm(TMT)}/v_{\rm A,0} \sim S_0^{1/5}\,M_{\rm A,0}^{\,8/5}\,(\lambda_\perp^z/\ell_0)^{3/5}$, while in the super-Alfv\'enic regime ($M_{\rm A,0}>1$) they scale as $\delta z_{\lambda_\perp^z}^{\rm(TMT)}/v_{\rm A,0} \sim S_0^{1/5}\,M_{\rm A,0}^{\,6/5}\,(\lambda_\perp^z/\ell_0)^{3/5}$. In this regime the dissipation scale is the same as for the GS95 cascade (i.e., $\lambda_{\perp,{\rm min}}^{z\,{\rm(subA)}}/\ell_0\sim M_{\rm A,0}^{\,-1}\,S_0^{-3/4}$ for trans-/sub-Alfv\'enic turbulence, or $\lambda_{\perp,{\rm min}}^{z\,{\rm(supA)}}/\ell_0\sim (M_{\rm A,0}\,S_0)^{-3/4}$ for super-Alfv\'enic injection).

\end{itemize}

We can verify that the nonlinear interaction between a quasi-parallel AW packet and the anisotropic turbulent fluctuations populating the background is also weak  for these cascades (i.e., $\chi_{\lambda_\perp}^{w,{\rm q\|}}<1$), except for the case of super-Alfv\'enic injection at scales $\lambda^w\gtrsim\ell_{\rm A}$, where instead $\chi_\lambda^w\sim\chi_\lambda^z>1$ holds (see Table~\ref{tab:MHDturb_scaling_nonlin-parameters}). 

We note that a tearing-mediated range emerges either when $S_0\gg M_{\rm A,0}^{-4}$ and $S_0\gg M_{\rm A,0}^3$ for sub-Alfv\'enic and super-Alfv\'enic turbulence injection, respectively (see Appendix~\ref{app:MHDturb_scalings}). Even admitting a wide range of values for the injection-scale Alfv\'enic Mach number $M_{\rm A,0}$, it seems reasonable to assume that these conditions would be met quite easily in many astrophysical systems. This is because the turbulent plasmas hosted by these environments are typically very weakly collisional, and thus characterized by large Lundquist numbers~\citep[see, e.g.,][and references therein]{PriestForbesBOOK2007,JiDaughtonPOP2011}. Nevertheless, for a TMT range to exist, 3D anisotropy of turbulent fluctuations is required. Hence, scale-dependent (dynamic) alignment is absolutely necessary. How and under what circumstances dynamic alignment occurs is still largely unexplored and matter of ongoing debate~\citep[see, e.g.,][and references therein]{SchekochihinJPP2022,CerriAPJ2022}.

\subsubsection{Sub- and trans-Alfv\'enic turbulence ($M_{\rm A,0} \leq 1$) with dynamic alignment}\label{subsubsec:subA_alignment-only-strong}

A quasi-parallel AW packet with normalized parallel wavelength $\hat{\lambda}_\|^w=\lambda_\|^{w,{\rm q\|}}/\ell_0$ injected in pre-existing sub-Alfv\'enic turbulence for which dynamic alignment of critically balanced fluctuations occurs is subjected to the following turbulent damping rate:
\begin{center}
    {\bf [$M_{\rm A,0}<1$, with dynamic alignment]}
\end{center}
\begin{equation}\label{eq:Turb-damp_subA_alignment-only-strong}
    \Gamma_{\rm subA}^{w,{\rm q\|}}
    \sim
    \left\{
    \begin{array}{lcr}
      M_{\rm A,0}^{\,8/3}\,\left(\hat{\lambda}_\|^w\right)^{1/3}\frac{v_{\rm A,0}}{\ell_0}\, & & M_{\rm A,0}^{\,4} \lesssim \hat{\lambda}_\|^w\lesssim M_{\rm A,0}\\
      & & \\
      M_{\rm A,0}^{\,24/5}\,\left(\hat{\lambda}_\|^w\right)^{-1/5}\,\frac{v_{\rm A,0}}{\ell_0}\, & & \hat{\lambda}_{\|,*}^{w\,{\rm(subA)}} \lesssim \hat{\lambda}_\|^w\lesssim M_{\rm A,0}^{\,4} \\
      & & \\
      M_{\rm A,0}^{\,4}\,\left(S_0\,\hat{\lambda}_\|^w\right)^{1/2}\,\frac{v_{\rm A,0}}{\ell_0}\, & & \hat{\lambda}_{\|,{\rm min}}^{w\,{\rm(subA)}}\lesssim \hat{\lambda}_\|^w\lesssim \hat{\lambda}_{\|,*}^{w\,{\rm(subA)}} 
    \end{array}
    \right.
.\end{equation}
Here $\hat{\lambda}_{\|,*}^{w\,{\rm(subA)}} \sim S_0^{1/5}(M_{\rm A,0}\hat{\lambda}_{\perp,*}^{z\,{\rm(subA)}})^{8/5} \sim S_0^{-5/7}\,M_{\rm A,0}^{\,8/7}$ is the wavelength below which the AW packet interacts with background fluctuations in the TMT regime, while $\hat{\lambda}_{\|,{\rm min}}^{w\,{\rm(subA)}} \sim S_0^{1/5}(M_{\rm A,0}\,\hat{\lambda}_{\perp,{\rm min}}^{z\,{\rm(subA)}})^{8/5}\sim S_0^{-1}$ is the shortest wavelength at which the turbulent damping is effective.

The trans-Alfv\'enic regime is obtained from the above case (i.e., when there is no W0 range):
\begin{center}
    {\bf [$M_{\rm A,0}\simeq1$, with dynamic alignment]}
\end{center}
\begin{equation}\label{eq:Turb-damp_transA_alignment-only-strong}
    \Gamma_{\rm transA}^{w,{\rm q\|}}
    \sim
    \left\{
    \begin{array}{lcr}
      \left(\hat{\lambda}_\|^w\right)^{-1/5}\,\frac{v_{\rm A,0}}{\ell_0}\, & \quad & S_0^{-5/7} \lesssim \hat{\lambda}_\|^w\lesssim 1 \\
      & &\\
      \left(S_0\,\hat{\lambda}_\|^w\right)^{1/2}\,\frac{v_{\rm A,0}}{\ell_0}\, & \quad & S_0^{-1} \lesssim \hat{\lambda}_\|^w\lesssim S_0^{-5/7} 
    \end{array}
    \right.
.\end{equation}
Here we have explicitly written the transition and dissipation scales:  $\hat{\lambda}_{\|,*}^{w\,{\rm(transA)}} \sim S_0^{1/5}(\hat{\lambda}_{\perp,*}^{z\,{\rm(transA)}})^{8/5} \sim S_0^{-5/7}$ and $\hat{\lambda}_{\|,{\rm min}}^{w\,{\rm(transA)}} \sim S_0^{1/5}(\hat{\lambda}_{\perp,{\rm min}}^{z\,{\rm(transA)}})^{8/5}\sim S_0^{-1}$, respectively.

One can see that when it comes to the interaction of the AW packet with anisotropic background fluctuations, including dynamic alignment in the picture changes the behavior of the turbulent damping rate significantly with respect to the  classic scenario (cf. Table~\ref{tab:TurbDamping_scaling_comparison}).
In the range of scales for which the packet interacts with critically balanced turbulence (i.e., $\lambda_{\|}^{w,{\rm q\|}}\lesssim\lambda_{\|,{\rm CB}}^{w,{\rm q\|}}$), the damping rate due to this nonlinear interaction is always higher than the corresponding rate obtained without dynamic alignment (cf. Eqs. \eqref{eq:Turb-damp_subA_no_alignment} and \eqref{eq:Turb-damp_transA_no_alignment} and Table~\ref{tab:TurbDamping_scaling_comparison}; see also the left panel of Figure~\ref{fig:TurbDamp_examples} for an immediate visual example).
This can be understood by considering that dynamic alignment means a shallower perpendicular spectrum of background fluctuations ($-3/2$ instead of $-5/3$), and thus at any scale $\lambda_\perp^z<\lambda_{\perp,{\rm CB}}^z$ there is more turbulent power to nonlinearly damp the AW packet. 

In general, if a CR-driven Alfv\'en-wave is injected in a background of sub-Alfv\'enic turbulence with dynamic alignment, now the damping rate interestingly exhibits two breaks that separate the three distinct regimes available in this scenario (contrary to the single break that would be present without dynamic alignment). This is a consequence of the new tearing-mediated regime that is only possible when a scale-dependent alignment takes place, and   is well summarized in Tables~\ref{tab:MHDturb_scaling_nonlin-parameters} and \ref{tab:TurbDamping_scaling_comparison} (see also Appendix~\ref{app:MHDturb_scalings}). The first break is the same as in turbulence without dynamic alignment, and it occurs for wavelengths interacting with background fluctuations at the transition scale between weak and strong turbulence ($\lambda_{\|}^{w,{\rm q\|}}\sim\lambda_{\|,{\rm CB}}^{w,{\rm q\|}}\sim M_{\rm A,0}^4\,\ell_0$). The second break instead emerges when the wavelength corresponds to a scale for which the AW packet starts to interact with tearing-mediated turbulence ($\lambda_{\|}^{w,{\rm q\|}}\sim \lambda_{\|,*}^{w\,{\rm(subA)}}\sim S_0^{-5/7}\,M_{\rm A,0}^{\,8/7}\,\ell_0$). 
In astrophysical situations for which this damping mechanism is the main process that determines the efficiency of CR confinement, these breaks could leave a signature at the corresponding energies in the propagated spectrum of these cosmic particles (see Section~\ref{sec:conclusions} for a brief discussion about the values of $M_{\rm A,0}$ and $S_0$ for which these breaks in the damping rate could be responsible for the features that are observed in the propagated CR spectrum).

\begin{table*}[!ht]
\caption[]{\bf Turbulent damping of quasi-parallel Alfv\'en waves with parallel wavelength $\lambda_\|^w$}
\centering
\begin{tabular}{|cccccc|}
\hline
  \multirow{2}{*}{\bf\em background}  &  \multicolumn{2}{c}{\multirow{3}{*}{\bf\em scale range of interaction}}   &  \multicolumn{3}{c|}{\multirow{3}{*}{{\bf{\em scaling of the turbulent damping rate} (}$\Gamma_{\rm turb}^{\,w}${\bf):}}} \\  
   \multirow{2}{*}{\bf\em cascade} & & & & & \\    
   \multirow{2}{*}{(acronym)}  &  \multirow{2}{*}{$\lambda_{\rm\|,max}^{w}$}  & \multirow{2}{*}{$\lambda_{\rm\|,min}^{w}$} & \multirow{2}{*}{``FG04''}   & \multirow{2}{*}{``L16''}  & \multirow{2}{*}{\em this work} \\    
   & & & & & \\
\hline
\hline
  \multicolumn{6}{|c|}{\multirow{2}{*}{\bf without scale-dependent alignment}}\\
   & & & & & \\
\hline
  \multicolumn{6}{|c|}{\multirow{2}{*}{{\it sub-Alfv\'enic injection} ($M_{\rm A,0}<1$):}}\\
   & & & & & \\
   \multirow{2}{*}{(W0)} & \multirow{2}{*}{$M_{\rm A,0}\,\ell_0$} & \multirow{2}{*}{$M_{\rm A,0}^4\,\ell_0$} & \multirow{2}{*}{---} & \multirow{2}{*}{$\omega_{\rm A,0}\,M_{\rm A,0}^{8/3}\,(\lambda_\|^w/\ell_0)^{-2/3}$} & \multirow{2}{*}{$\omega_{\rm A,0}\,M_{\rm A,0}^{8/3}\,(\lambda_\|^w/\ell_0)^{1/3}$} \\ 
   & & & & & \\
  \multirow{2}{*}{(GS95)} & \multirow{2}{*}{$M_{\rm A,0}^4\,\ell_0$} & \multirow{2}{*}{$S_0^{-1}\,\ell_0$} & \multirow{2}{*}{---} & \multirow{2}{*}{$\omega_{\rm A,0}\,M_{\rm A,0}^2\,(\lambda_\|^w/\ell_0)^{-1/2}$} & \multirow{2}{*}{$\omega_{\rm A,0}\,M_{\rm A,0}^{4}$} \\ 
   & & & & & \\
\hline
  \multicolumn{6}{|c|}{\multirow{2}{*}{{\it trans-Alfv\'enic injection} ($M_{\rm A,0}\approx1$):}}\\
   & & & & & \\
  \multirow{2}{*}{(GS95)} & \multirow{2}{*}{$\ell_0$} & \multirow{2}{*}{$S_0^{-1}\,\ell_0$} & \multirow{2}{*}{$\omega_{\rm A,0}\,(\lambda_\|^w/\ell_0)^{-1/2}$} & \multirow{2}{*}{---} & \multirow{2}{*}{$\omega_{\rm A,0}$} \\ 
   & & & & & \\
\hline
  \multicolumn{6}{|c|}{\multirow{2}{*}{{\it super-Alfv\'enic injection} ($M_{\rm A,0}>1$):}}\\
   & & & & & \\
  \multirow{2}{*}{(K41)} & \multirow{2}{*}{$\ell_0$} & \multirow{2}{*}{$M_{\rm A,0}^{-3}\,\ell_0$} & \multirow{2}{*}{---} & \multirow{2}{*}{$\omega_{\rm A,0}\,M_{\rm A,0}\,(\lambda_\|^w/\ell_0)^{-2/3}$} & \multirow{2}{*}{$\omega_{\rm A,0}\,M_{\rm A,0}\,(\lambda_\|^w/\ell_0)^{-2/3}$} \\ 
   & & & & & \\
  \multirow{2}{*}{(GS95)} & \multirow{2}{*}{$M_{\rm A,0}^{-3}\,\ell_0$} & \multirow{2}{*}{$S_0^{-1}\,\ell_0$} & \multirow{2}{*}{---} & \multirow{2}{*}{$\omega_{\rm A,0}\,M_{\rm A,0}^{3/2}\,(\lambda_\|^w/\ell_0)^{-1/2}$} & \multirow{2}{*}{$\omega_{\rm A,0}\,M_{\rm A,0}^{3}$} \\ 
   & & & & & \\
\hline
\hline
  \multicolumn{6}{|c|}{\multirow{2}{*}{\bf with scale-dependent alignment}}\\
   & & & & & \\
\hline
  \multicolumn{6}{|c|}{\multirow{2}{*}{{\it sub-Alfv\'enic injection} ($M_{\rm A,0}<1$):}}\\
   & & & & & \\
   \multirow{2}{*}{(W0)} & \multirow{2}{*}{$M_{\rm A,0}\,\ell_0$} & \multirow{2}{*}{$M_{\rm A,0}^4\,\ell_0$} & \multirow{2}{*}{---} & \multirow{2}{*}{$\omega_{\rm A,0}\,M_{\rm A,0}^{8/3}\,(\lambda_\|^w/\ell_0)^{-2/3}$} & \multirow{2}{*}{$\omega_{\rm A,0}\,M_{\rm A,0}^{8/3}\,(\lambda_\|^w/\ell_0)^{1/3}$} \\ 
   & & & & & \\
  \multirow{2}{*}{(B06)} & \multirow{2}{*}{$M_{\rm A,0}^4\,\ell_0$} & \multirow{2}{*}{$M_{\rm A,0}^{8/7}\, S_0^{-5/7}\,\ell_0$} & \multirow{2}{*}{---} & \multirow{2}{*}{---} & \multirow{2}{*}{$\omega_{\rm A,0}\,M_{\rm A,0}^{24/5}\,(\lambda_\|^w/\ell_0)^{-1/5}$} \\ 
   & & & & & \\
  \multirow{2}{*}{(TMT)} & \multirow{2}{*}{$M_{\rm A,0}^{8/7}\, S_0^{-5/7}\,\ell_0$} & \multirow{2}{*}{$S_0^{-1}\,\ell_0$} & \multirow{2}{*}{---} & \multirow{2}{*}{---} & \multirow{2}{*}{$\omega_{\rm A,0}\,M_{\rm A,0}^{4}\,S_0^{1/2}\,(\lambda_\|^w/\ell_0)^{1/2}$} \\ 
   & & & & & \\
\hline
  \multicolumn{6}{|c|}{\multirow{2}{*}{{\it trans-Alfv\'enic injection} ($M_{\rm A,0}\approx1$):}}\\
   & & & & & \\
  \multirow{2}{*}{(B06)} & \multirow{2}{*}{$\ell_0$} & \multirow{2}{*}{$S_0^{-5/7}\,\ell_0$} & \multirow{2}{*}{---} & \multirow{2}{*}{---} & \multirow{2}{*}{$\omega_{\rm A,0}\,(\lambda_\|^w/\ell_0)^{-1/5}$} \\ 
   & & & & & \\
  \multirow{2}{*}{(TMT)} & \multirow{2}{*}{$S_0^{-5/7}\,\ell_0$} & \multirow{2}{*}{$S_0^{-1}\,\ell_0$} & \multirow{2}{*}{---} & \multirow{2}{*}{---} & \multirow{2}{*}{$\omega_{\rm A,0}\,S_0^{1/2}\,(\lambda_\|^w/\ell_0)^{1/2}$} \\ 
   & & & & & \\
\hline
  \multicolumn{6}{|c|}{\multirow{2}{*}{{\it super-Alfv\'enic injection} ($M_{\rm A,0}>1$):}}\\
   & & & & & \\
  \multirow{2}{*}{(K41)} & \multirow{2}{*}{$\ell_0$} & \multirow{2}{*}{$M_{\rm A,0}^{-3}\,\ell_0$} & \multirow{2}{*}{---} & \multirow{2}{*}{$\omega_{\rm A,0}\,M_{\rm A,0}\,(\lambda_\|^w/\ell_0)^{-2/3}$} & \multirow{2}{*}{$\omega_{\rm A,0}\,M_{\rm A,0}\,(\lambda_\|^w/\ell_0)^{-2/3}$} \\ 
   & & & & & \\
  \multirow{2}{*}{(B06)} & \multirow{2}{*}{$M_{\rm A,0}^{-3}\,\ell_0$} & \multirow{2}{*}{$M_{\rm A,0}^{-6/7}\,S_0^{-5/7}\,\ell_0$} & \multirow{2}{*}{---} & \multirow{2}{*}{---} & \multirow{2}{*}{$\omega_{\rm A,0}\,M_{\rm A,0}^{12/5}\,(\lambda_\|^w/\ell_0)^{-1/5}$} \\ 
   & & & & & \\
  \multirow{2}{*}{(TMT)} & \multirow{2}{*}{$M_{\rm A,0}^{-6/7}\,S_0^{-5/7}\,\ell_0$} & \multirow{2}{*}{$S_0^{-1}\,\ell_0$} & \multirow{2}{*}{---} & \multirow{2}{*}{---} & \multirow{2}{*}{$\omega_{\rm A,0}\,M_{\rm A,0}^3\,S_0^{1/2}\,(\lambda_\|^w/\ell_0)^{1/2}$} \\ 
   & & & & & \\
\hline
\end{tabular}
\tablefoot{Summary of the scaling relations for the turbulent damping rate $\Gamma_{\rm turb}^{\,w}$ of a quasi-parallel Alfv\'en wave in background MHD turbulence for the different regimes mentioned in Section~\ref{sec:explicit_scalings}. The damping rates obtained in this work are compared with tose available in the existing literature, namely in \citet{FarmerGoldreichAPJ2004} (FG04) and in \citet{LazarianAPJ2016} (L16). The notation $\omega_{\rm A,0}=v_{\rm A,0}/\ell_0$ has been used. In all the regimes with dynamic alignment, a transition to a tearing-mediated range (TMT) has been assumed, i.e., that $S_0\gg M_{\rm A,0}^{-4}$ ($S_0\gg M_{\rm A,0}^3$) holds for sub-Alfv\'enic (super-Alfv\'enic) injection.}
\label{tab:TurbDamping_scaling_comparison}
\end{table*}

\subsubsection{Super-Alfv\'enic turbulence ($M_{\rm A,0} > 1$) with dynamic alignment}\label{subsubsec:supA_alignment-only-strong}

When background fluctuations are injected with $M_{\rm A,0} > 1$ and dynamic alignment of critically balanced turbulent fluctuations takes place, an AW packet undergoes  turbulent damping with the following rate:
\begin{center}
    {\bf [$M_{\rm A,0}>1$, with dynamic alignment]}
\end{center}
\begin{equation}\label{eq:Turb-damp_supA_alignment-only-strong}
    \Gamma_{\rm supA}^{w,{\rm q\|}}
    \sim
    \left\{
    \begin{array}{lcr}
      M_{\rm A,0}\,\left(\hat{\lambda}^w\right)^{-2/3}\frac{v_{\rm A,0}}{\ell_0}\, &  & M_{\rm A,0}^{\,-3} \lesssim \hat{\lambda}^w\lesssim 1\\
      & & \\
      M_{\rm A,0}^{\,12/5}\,\left(\hat{\lambda}_\|^w\right)^{-1/5}\,\frac{v_{\rm A,0}}{\ell_0}\, & & \hat{\lambda}_{\|,*}^{w\,{\rm(supA)}} \lesssim \hat{\lambda}_\|^w\lesssim M_{\rm A,0}^{\,-3} \\
      & & \\
      M_{\rm A,0}^{\,3}\, \left(S_0\,\hat{\lambda}_\|^w\right)^{1/2}\,\frac{v_{\rm A,0}}{\ell_0}\, & & \hat{\lambda}_{\|,{\rm min}}^{w\,{\rm(supA)}} \lesssim \hat{\lambda}_\|^w\lesssim \hat{\lambda}_{\|,*}^{w\,{\rm(supA)}} 
    \end{array}
    \right.
.\end{equation}
Here $\hat{\lambda}_{\|,*}^{w\,{\rm(supA)}} \sim S_0^{1/5} M_{\rm A,0}^{6/5} (\hat{\lambda}_{\perp,*}^{z\,{\rm(supA)}})^{8/5}\sim S_0^{-5/7}\,M_{\rm A,0}^{\,-6/7}$, and the shortest wavelength for turbulent damping to be effective is $\hat{\lambda}_{\|,{\rm min}}^{w\,{\rm(supA)}} \sim S_0^{1/5}\,M_{\rm A,0}^{\,6/5}(\hat{\lambda}_{\perp,{\rm min}}^{z\,{\rm(supA)}})^{8/5}\sim S_0^{-1}$.

   \begin{figure*}[!ht]
   \includegraphics[width=0.5\textwidth]{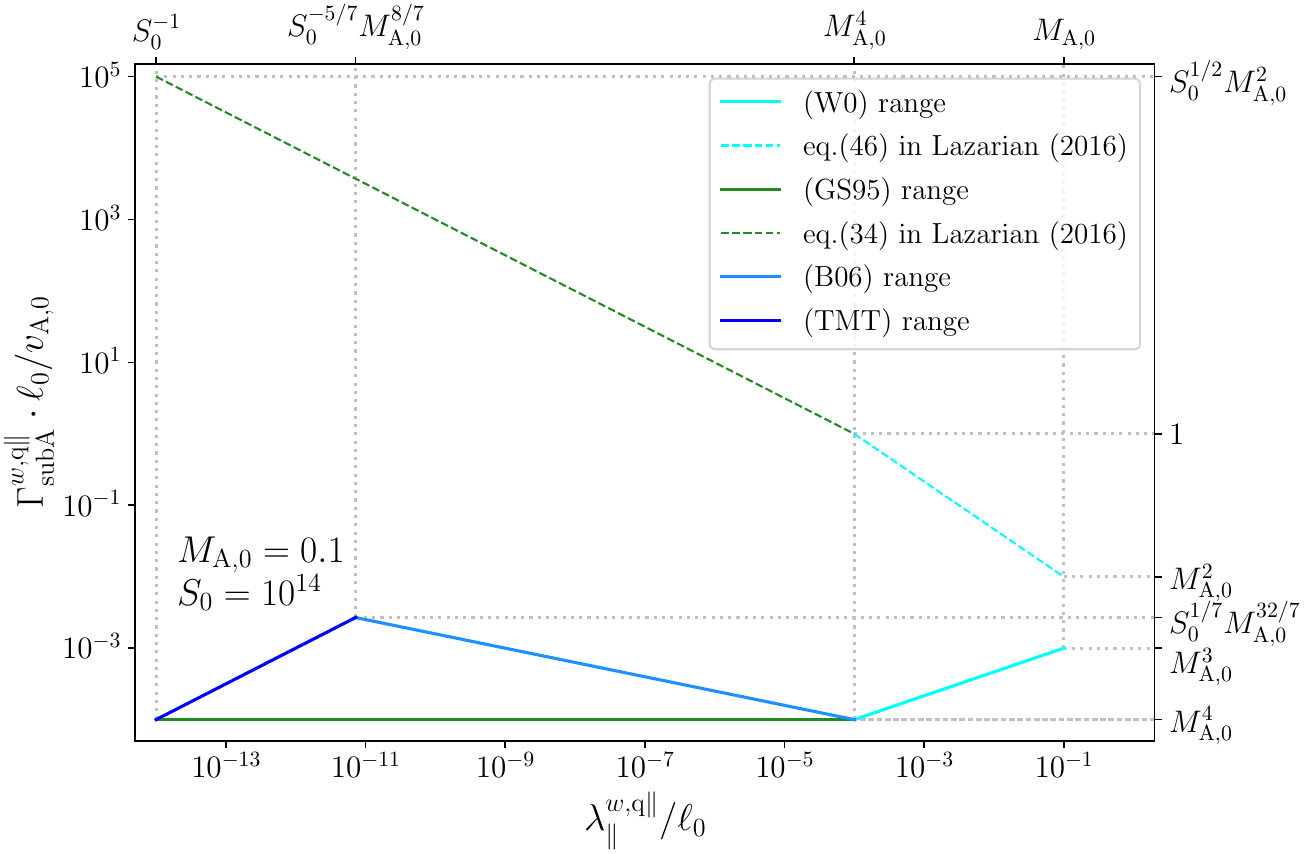}
   \includegraphics[width=0.5\textwidth]{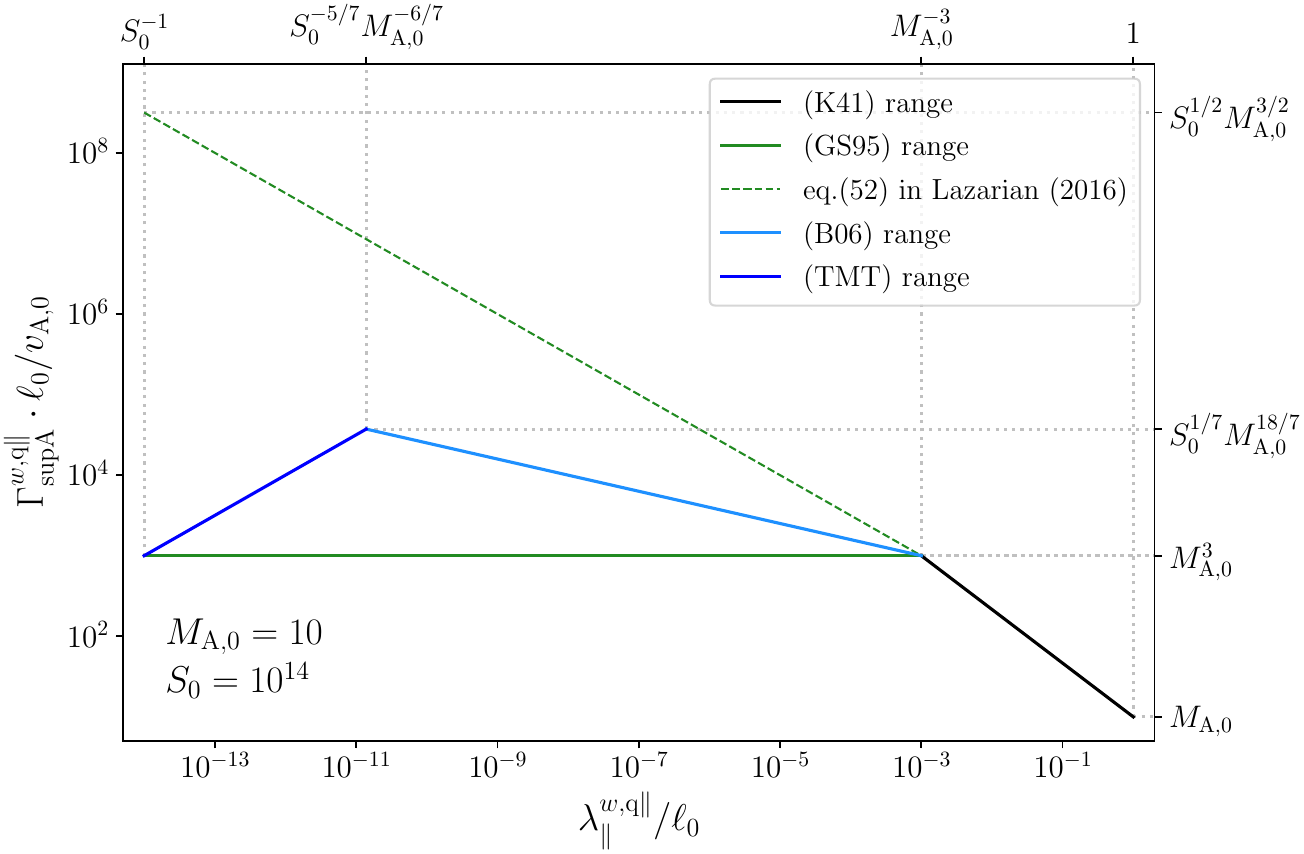}
   \caption{Normalized turbulent damping rate $\frac{\ell_0}{v_{\rm A,0}}\,\Gamma_{\rm turb}^{w,{\rm q\|}}$ for quasi-parallel AW packets with normalized parallel wavelength $\lambda_\|^{w,{\rm q\|}}/\ell_0$, in a background plasma with Lunquist number $S_0=10^{14}$ and different turbulent regimes (see Appendix~\ref{app:MHDturb_scalings}). Solid lines represent damping rates derived in this work (equations \eqref{eq:Turb-damp_subA_no_alignment}, \eqref{eq:Turb-damp_supA_no_alignment}, \eqref{eq:Turb-damp_subA_alignment-only-strong}, and \eqref{eq:Turb-damp_supA_alignment-only-strong}), while dashed lines report the damping rates in \citet{LazarianAPJ2016} for reference. General expressions for transition scales and damping-rate values are reported on the right and upper axes. Left: Damping rates in sub-Alfv\'enic turbulence ($M_{\rm A,0}=0.1$). Right: Damping rates in super-Alfv\'enic turbulence ($M_{\rm A,0}=10$).}
   \label{fig:TurbDamp_examples}
   \end{figure*}
Again, we note that the isotropic normalized wavelength $\hat{\lambda}^w=\lambda^w/\ell_0$ enters the damping rate in the range $\ell_{\rm A}\lesssim\lambda^w\lesssim\ell_0$ (i.e., where the AW packet interacts with hydrodynamic-like pre-existing turbulence). In this range of scales, the result is unchanged with respect to the turbulent damping rate obtained without dynamic alignment (Section~\ref{subsubsec:supA_no_alignment}). 
At smaller scales $\lambda_\|^{w,{\rm q\|}} \lesssim \ell_{\rm A}$, the quasi-parallel condition in \eqref{eq:AW_q-prl-limit_general} applies again and the damping rate depends explicitly on the normalized parallel wavelength $\hat{\lambda}_\|^w=\lambda_\|^{w,{\rm q\|}}/\ell_0$. In this regime, the turbulent damping rate differs significantly from the one obtained without dynamic alignment. 
When a scale-dependent alignment of fluctuations is taken into account, the turbulent damping is always much more effective than in the case obtained without dynamic alignment. This leads to a damping rate that can be higher  by orders of magnitude with respect to that in \eqref{eq:Turb-damp_supA_no_alignment}, depending on the Alfv\'enic Mach number $M_{\rm A,0}$ and on the Lunquist number $S_0$ at injection scales (see Table~\ref{tab:TurbDamping_scaling_comparison} for a comparison of the scaling and the right panel of Figure~\ref{fig:TurbDamp_examples} for and explicit graphic example).
Finally, analogously to the case with $M_{\rm A,0}<1$, the damping rate also exhibits two breaks  in a background of super-Alfv\'enic turbulence, if dynamic alignment can occur. The first break emerges for wavelengths comparable to the transition scale between hydrodynamic-like and critically balanced turbulence ($\lambda^{w}\sim\ell_{\rm A}\sim M_{\rm A,0}^{-3}\,\ell_0$). A second break occurs at wavelengths corresponding to the scale marking the transition between a dynamically aligned cascade and the tearing-mediated range ($\lambda_{\|}^{w,{\rm q\|}}\sim\lambda_{\|,*}^{w\,{\rm(supA)}}\sim S_0^{-5/7}\,M_{\rm A,0}^{\,-6/7}\,\ell_0$).
For which values of $M_{\rm A,0}$ and $S_0$ these breaks in the damping rate could be responsible for the features that are observed in the propagated CR spectrum is briefly discussed in Section~\ref{sec:conclusions}.

\section{Feedback on background fluctuations}\label{sec:AW_feedback}

When deriving the scaling of $\Gamma_{\rm turb}^{w,{\rm q\|}}$ in the previous section, we   neglected the feedback that the injected Alfv\'en-wave packets could have on pre-existing turbulent fluctuations. 
In general, background fluctuations could also be affected by their nonlinear interaction with these AW packets, which we call   feedback. 
It is therefore instructive to understand when this effect has to be taken into account. The relevance of this feedback can be estimated by comparing two timescales. The first is the  intrinsic cascade time of background fluctuations, which is the timescale of the cascade process induced by pre-existing (counter-propagating) fluctuations on themselves, $\tau_{\rm casc}^{(z|z)}\sim\tau_{\rm nl}^{(z|z)}/\chi^{(z|z)}$ (or just $\tau_{\rm casc}^{(z|z)}\sim\tau_{\rm nl}^{(z|z)}$, if $\chi^{(z|z)}\gtrsim1$). The second is  CR-induced cascade time, which is  the timescale of the cascade that would be induced by the injected AW packets on the background fluctuations, $\tau_{\rm casc}^{(z|w)}\sim\tau_{\rm nl}^{(z|w)}/\chi^{(z|w)}$ (or just $\tau_{\rm casc}^{(z|w)}\sim\tau_{\rm nl}^{(z|w)}$, if $\chi^{(z|w)}\gtrsim1$). 

Hereafter, the simpler superscript ``$z$'' is  used instead of ``$(z|z)$'' for the sake of homogeneity of notation with the previous sections. 
Moreover, for the sake of clarity in the qualitative discussion that  follows, the effect of dynamic alignment is not   taken into account in this section.
The nonlinear parameter $\chi^{(z|w)}$ describing the interaction of background fluctuations with CR-driven AW packets is $\chi_{\lambda_\perp}^{(z|w)}\sim(\lambda_{\|,\lambda_\perp}^z/\lambda_\perp^z)(\delta w_{\lambda_\perp}/v_{\rm A,0})\sim (\delta w_{\lambda_\perp}/\delta z_{\lambda_\perp})\,\chi_{\lambda_\perp}^{\,z}$, while the timescale associated with this nonlinear interaction is $\tau_{\rm nl,\lambda_\perp}^{(z|w)}\sim \lambda_\perp/\delta w_{\lambda_\perp}\sim (\delta z_{\lambda_\perp}/\delta w_{\lambda_\perp})\,\tau_{\rm nl,\lambda_\perp}^{\,z}$.
The ratio of the two cascade timescales is thus given by 
\begin{equation}\label{eq:AW_feedback_generic}
    \hspace{-0.1cm}\frac{\tau_{\rm casc}^{(z|w)}}{\tau_{\rm casc}^{z}}
    \sim
    \left\{
    \begin{array}{ll}
      \left(\delta z_{\lambda_\perp}/\delta w_{\lambda_\perp}\right)^2 &  \text{if} \left\{\begin{array}{c}
        \chi_{\lambda_\perp}^z<1\,\text{and}\,\,\chi_{\lambda_\perp}^{(z|w)}\lesssim1 \\
        \text{or} \\
        \chi_{\lambda_\perp}^z\sim1\,\text{and}\,\,\chi^{(z|w)}_{\lambda_\perp}<1
      \end{array}\right.\\
      & \\
      \left(\delta z_{\lambda_\perp}/\delta w_{\lambda_\perp}\right)\,\chi_{\lambda_\perp}^z  &  \text{if}\,\,\,  \chi_{\lambda_\perp}^z<1\,\text{and}\,\,\chi_{\lambda_\perp}^{(z|w)}>1\\
      & \\
      \left(\delta z_{\lambda_\perp}/\delta w_{\lambda_\perp}\right) &  \text{if}\,\,\, \chi_{\lambda_\perp}^z\sim1\,\text{and}\,\,\chi_{\lambda_\perp}^{(z|w)}\gtrsim1\\
      & \\
      \left(\delta z_{\lambda}/\delta w_{\lambda}\right)(\delta w_{\lambda}/v_{\rm A,0})^{-1}  &  \text{if}\,\,\, \chi_{\lambda}^z>1\,\text{and}\,\,\chi_{\lambda}^{(z|w)}<1\\
      & \\
      \left(\delta z_{\lambda}/\delta w_{\lambda}\right) &  \text{if}\,\,\, \chi_{\lambda}^z>1\,\text{and}\,\,\chi_{\lambda}^{(z|w)}\gtrsim1
    \end{array}
    \right.
.\end{equation}
Feedback effects are   taken into account if the nonlinear cascade process that would be induced by the injected Alfv\'en-wave packets becomes faster than the intrinsic cascading process of background fluctuations (i.e., at scales where $\tau_{\rm casc}^{(z|w)}/\tau_{\rm casc}^{z}\lesssim1$).
From \eqref{eq:AW_feedback_generic} we can see that  cosmic-ray feedback (CRF) is relevant at scales for which self-generated waves achieve non-negligible amplitudes with respect to the background fluctuations. Clearly, the weaker the damping, the larger the amplitude that self-generated fluctuations can attain. Hence, feedback on background turbulence becomes more important as the most-relevant damping mechanism becomes weaker; and this is a general statement that does not depend on which damping process determines the saturation level of CR-generated waves. The scales at which CRF has to be taken into account thus requires that we compare the scale-dependent amplitudes of both the CR-driven AW packets ($\delta w_\lambda$) and the pre-existing background fluctuations ($\delta z_\lambda$).
A precise estimate of these scales requires a detailed knowledge of how the CRSI saturation level in pre-existing turbulence depends upon plasma parameters and background conditions, which is not yet achieved. At this stage, we provide only a general, qualitative discussion.

We  consider cases in which the CRSI saturates at a level $(\delta w/v_{\rm A,0})^2\sim(\delta B^{\rm(CRSI)}/B_0)^2\ll1$. Then, from \eqref{eq:AW_feedback_generic} we can see that when the CR-driven AW packets interact with super-Alfv\'enic turbulence ($M_{\rm A,0}>1$), and at scales where the cascade is hydrodynamic-like ($\lambda^w\gtrsim\ell_{\rm A}$), the ratio $\tau_{\rm casc}^{(z|w)}/\tau_{\rm casc}^{z}$ is typically much higher than unity. Therefore, when the CR-driven instability produces fluctuations at a level $\delta B^{\rm(CRSI)}/B_0\ll1$ (which also depends   on the presence of a mean field $B_0$), CRF is likely negligible at all scales belonging to the K41 regime.
The situation is different for trans- and sub-Alfv\'enic injection ($M_{\rm A,0}\lesssim1$), or for super-Alfv\'enic injection at scales below which the hydrodynamic-like cascade transitions to the critically balanced regime ($\lambda_\|^w <\ell_{\rm A}$). In these cases, the CR feedback on pre-existing turbulence is negligible only at scales for which the packet amplitudes are smaller than the background fluctuation level. 
As a consequence, cosmic-ray feedback should be taken into account at scales $\lambda_\perp\lesssim \lambda_\perp^{\rm CRF}$, where the perpendicular CR-feedback scale $\lambda_\perp^{\rm CRF}$ is defined as the scale at which $\delta w_{\lambda_\perp^{\rm CRF}}\sim\delta z_{\lambda_\perp^{\rm CRF}}$ holds. If $\delta B^{\rm(CRSI)}/B_0$ is sufficiently low, such a scale may be smaller than the turbulent dissipation scale, and thus the feedback could be neglected for all purposes of CR transport. However, the growth and saturation level of the CRSI depends upon the CR-to-thermal density ratio $n_{\rm CR}/n_{\rm th}$. In the Galactic halo this ratio is negligibly low, and the above reasoning likely applies. This may not be the case near CR sources, where such a density ratio is not as low and CRs can further evacuate the thermal gas~\citep{SchroerAPJL2021,SchroerMNRAS2022}. We thus expect this feedback to be relevant in these environments. This issue will be addressed in more detail and quantitatively in the accompanying Paper II.

Finally, we note  that the discussion above regarding the scale at which CRF could become relevant was done in terms of perpendicular scales $\lambda_\perp$ (see equation \eqref{eq:AW_feedback_generic}). This means that we denoted with $\lambda_\perp^{\rm CRF}$ the perpendicular scale at which CR-generated waves affect pre-existing turbulence. 
However, regardless of the damping mechanism that saturates the amplitude of the fluctuations, the CRSI growth rate is such that quasi-parallel Alfv\'en waves $\lambda_\|^w\ll\lambda_\perp^w$ are mainly produced. 
It is thus convenient to relate the scale $\lambda_\perp^{\rm CRF}$ at which the CR-driven fluctuation amplitude becomes comparable to the amplitude of background fluctuations to the injected parallel wavelength $\lambda_\|^{w,{\rm q\|}}$. This can be done by using the quasi-parallel condition \eqref{eq:AW_q-prl-limit_general}: 
\begin{equation}\label{eq:CRFscale_parallel}
    \lambda_\|^{\rm CRF}
    \,\sim\,
    \left(\frac{\delta z_{\lambda_\perp^{w,{\rm q\|}}}}{v_{\rm A,0}}\right)\,
    \lambda_\perp^{\rm CRF}
    \,\ll\,
    \lambda_\perp^{\rm CRF}\,.
\end{equation}
Hence,   quasi-parallel Alfv\'en waves $\delta w_{\lambda_\|}$ driven by CRs at scales $\lambda_\|^{w,{\rm q\|}}\lesssim\lambda_\|^{\rm CRF}$ can actually affect pre-existing turbulent fluctuations $\delta z_{\lambda_\perp}$ on much larger scales $\lambda_\perp^{\rm CRF}\gtrsim\lambda_\perp\gg\lambda_\|^{w,{\rm q\|}}$.
We recall that we assume that nonlinear interactions are local in perpendicular scale, so in what follows we do not need to distinguish between the two scales $\lambda_\perp^w$ and $\lambda_\perp^z$ (i.e., $\lambda_\perp^w\sim\lambda_\perp^z\sim\lambda_\perp$).

In the following, we attempt to provide two phenomenological models for the CR-modified cascade of background fluctuations. However, these are just plausible models at this stage. Focused numerical investigations will be necessary in order to verify if (and under what circumstances) they can be realized.

\subsection{A phenomenological model for  CR-modified scaling of pre-existing turbulent fluctuations}

A simple phenomenological model for the cascade modified by the CR-generated AW packets can be constructed as follows. 

Here we  assume that, at scales where $\chi_{\lambda_\perp}^z\lesssim1$, pre-existing fluctuations and their anisotropies follow the scaling $\delta z_{\lambda_\perp}^{(0)}\propto \lambda_\perp^{\,\alpha_\perp^{z}}$ and $\lambda_{\|,\lambda_\perp}\propto \lambda_\perp^{\,\delta^{z}}$, respectively, with $\alpha_\perp^{z}>0$. For critically balanced fluctuations ($\chi^z\sim1$) the anisotropy is such that $\delta^z=1-\alpha_\perp^z$, while in weak turbulence ($\chi^z<1$) it is $\delta^z=0$. 
Such scaling for the fluctuations corresponds to a perpendicular power spectrum $E_{\delta z}^{(0)}(k_\perp)\propto k_\perp^{\,-\xi_\perp^{z}}$, with $\xi_\perp^{z}=1+2\,\alpha_\perp^{z}$. At scales where $\chi_{\lambda_\perp}^z>1$, the scaling is isotropic (i.e., $\delta^{z}=1$), and so $\delta z_{\lambda}^{(0)}\propto \lambda^{\,\alpha^{z}}$ and $E_{\delta z}^{(0)}(k)\propto k^{\,-\xi^{z}}$, with $\xi^{z}=1+2\,\alpha^{z}$.
These are the  unperturbed properties of background fluctuations (i.e., without CR feedback), and are thus denoted by the superscript ``(0)''.
Then we assume a scaling $\delta w_{\lambda_\|}\propto(\lambda_\|^w)^{\,-\alpha_\|^{w}}$ for the CR-driven (quasi-parallel) fluctuations, corresponding to a parallel power spectrum $E_{\rm CRSI}(k_\|^w)\propto (k_\|^w)^{\,\xi_\|^{w}}$, where $\xi_\|^{w}=2\,\alpha_\|^{w}-1$.
At scales $\lambda_\perp$ where background fluctuations are sub-Alfv\'enic and anisotropic (i.e., such that $\chi_{\lambda_\perp}^z\lesssim1$), the quasi-parallel condition \eqref{eq:AW_q-prl-limit_general} holds for CR-driven waves. Hence, the corresponding perpendicular scaling for self-generated fluctuations is typically steeper than its parallel counterpart, and is related to the perpendicular scaling of pre-existing fluctuations, namely\footnote{The quasi-parallel condition,  $k_\|^w\sim(\delta z_{k_\perp}/v_{\rm A,0})^{-1}k_\perp$, and condition on the total energy,  $(k_\|^w)^{-1}(\delta w_{k_\|^w})^2 {\rm d}k_\|^w \sim k_\perp^{\,-1}(\delta w_{k_\perp})^2 {\rm d}k_\perp$, have been employed to derive the perpendicular scaling of the fluctuations, $\delta w_{k_\perp}$.} $\delta w_{\lambda_\perp}\propto \lambda_\perp^{\,-\alpha_\perp^w}$ with $\alpha_\perp^w=\alpha_\|^w\,(1+\alpha_\perp^z)$. This corresponds to a perpendicular spectrum $E_{\rm CRSI}(k_\perp)\propto (k_\perp)^{\,\xi_\perp^{w}}$, with $\xi_\perp^{w}=2\,\alpha_\|^{w}\,(1+\alpha_\perp^z)-1=\xi_\|^w\,+\,(\xi_\|^w+1)(\xi_\perp^z-1)/2$. On the other hand, at scales where $\chi_\lambda^z>1$, the distinction between $\alpha_\perp^w$ and $\alpha_\|^w$ is lost. In this case, we just assume a scaling $\delta w_{\lambda}\propto \lambda^{\,-\alpha^w}$ and a (isotropic) power spectrum $E_{\rm CRSI}(k)\propto (k)^{\,\xi^{w}}$ with $\xi^{w}=2\,\alpha^{w}-1$.

Before proceeding further, it is worth mentioning that here we are considering a generic  case where self-generated turbulence can be described by a power-law spectrum, without making any assumption on its spectral index nor on the damping mechanism that sets the saturation of the instability. The only condition that we  require is that the hierarchy of possible interactions between CR-generated waves and pre-existing turbulent fluctuations is self consistent (i.e., that background turbulence is affected by the self-generated waves before turbulence can affect the waves). In this regard, we can verify that when $\tau_{\rm casc}^{(z|w)}\ll\tau_{\rm casc}^{z}$ holds, then the condition $\tau_{\rm casc}^{(z|w)}\ll\tau_{\rm casc}^{w}\sim(\Gamma_{\rm turb}^{w,{\rm q\|}})^{-1}$ is automatically satisfied:\footnote{To show this, we multiply by $\Gamma_{\rm turb}^{w,{\rm q\|}}$ both sides of the condition $\tau_{\rm casc}^{(z|w)}\ll\tau_{\rm casc}^{z}$ and obtain the equivalent condition $\Gamma_{\rm turb}^{w,{\rm q\|}}\tau_{\rm casc}^{(z|w)}\ll\Gamma_{\rm turb}^{w,{\rm q\|}}\tau_{\rm casc}^{z}$. Then, using \eqref{eq:Turb-damp_w_generic}-\eqref{eq:Turb-damp_w_q-prl-limit_generic}, we can show that $\Gamma_{\rm turb}^{w,{\rm q\|}}\tau_{\rm casc}^{z}\lesssim1$ holds for any value of $\chi^w$ and $\chi^z$, which further implies the condition $\Gamma_{\rm turb}^{w,{\rm q\|}}\tau_{\rm casc}^{(z|w)}\ll1$.} this condition allows us to neglect mutual feedback between background fluctuations and CR-driven Alfv\'en waves (i.e., to consider only the modification to the scaling of pre-existing turbulence induced by a stationary spectrum of saturated self-generated fluctuations). 

We now want to know how the scaling of $\delta z^{(0)}$ is modified by the presence of $\delta w$, knowing that we are in a regime in which such feedback is faster that the intrinsic cascade time of $\delta z^{(0)}$ ({i.e.}, $\tau_{\rm casc}^{(z|w)}\ll\tau_{\rm casc}^{z}$). 
 Within these assumptions, we can derive the scaling of CR-modified turbulence by replacing the cascade timescale,
\begin{equation}\label{eq:CRfeedback_timescale_replacement}
    \tau_{\rm casc}^z
    \,\,\to\,\,
    \tau_{\rm casc}^{(z|w)}\,,
\end{equation}
by rescaling the unperturbed background fluctuations into a  first-order modified fluctuation denoted by the superscript ``(1)'',
\begin{equation}\label{eq:CRfeedback_fluct_rescaling}
    \delta z_{\lambda_\perp}^{(0)}
    \,\,\to\,\,
    \delta z_{\lambda_\perp}^{(1)}=\zeta_{\lambda_\perp}\delta z_{\lambda_\perp}^{(0)}\,,
\end{equation}
and by requiring that the cascade rate is still scale-inependent (i.e., $(\delta z_{\lambda_\perp}^{(1)})^2/\tau_{\rm casc}^{(z|w)}\sim\varepsilon\sim\text{const}$). This procedure readily provides the scaling factor
\begin{equation}\label{eq:CRfeedback_fluct_rescaling_zeta}
    \zeta_{\lambda_\perp}\sim\left(\frac{\tau_{\rm casc}^{(z|w)}}{\tau_{\rm casc}^z}\right)^{1/2}\,,
\end{equation}
which can be estimated using \eqref{eq:AW_feedback_generic} for the various $\chi^z$ and $\chi^w$ regimes and, as a first approximation, by employing the unperturbed scaling $\delta z_{\lambda_\perp}^{(0)}$; the rescaling factor computed in this way is  denoted as $\zeta_{\lambda_\perp}^{(0)}$.\footnote{We can perform    an expansion of the rescaling factor based on iteratively modified timescales $\tau_{\rm casc,\lambda_\perp}^{z^{(n)}}$, and rewrite it as a series $\zeta_{\lambda_\perp}=(\tau_{\rm casc,\lambda_\perp}^{(z|w)})^{1/2}\left(\sum_n 1/\tau_{\rm casc,\lambda_\perp}^{z^{(n)}}\right)^{1/2}=\zeta_{\lambda_\perp}^{(0)}\left(1+\sum_{n>0} \tau_{\rm casc,\lambda_\perp}^{z^{(0)}}/\tau_{\rm casc,\lambda_\perp}^{z^{(n)}}\right)^{1/2}$; we recall that we  assume that the CR-driven fluctuations $\delta w_{\lambda_\perp}$ are unaffected by background fluctuations. The ratio of the $0$th to the $n$th timescale is $\tau_{\rm casc,\lambda_\perp}^{z^{(0)}}/\tau_{\rm casc,\lambda_\perp}^{z^{(n)}}\propto\delta z_{\lambda_\perp}^{(n)}/\delta z_{\lambda_\perp}^{(0)}$ (or even $\propto(\delta z_{\lambda_\perp}^{(n)}/\delta z_{\lambda_\perp}^{(0)})^2$). Then, if the CR-induced modification of the pre-existing spectrum is a steepening, the cascade timescale would significantly increase with increasing $n$, i.e., $\tau_{\rm casc,\lambda_\perp}^{z^{(n)}}\gg\tau_{\rm casc,\lambda_\perp}^{z^{(n-1)}}\gg\dots\gg\tau_{\rm casc,\lambda_\perp}^{z^{(1)}}\gg\tau_{\rm casc,\lambda_\perp}^{z^{(0)}}$. So, if the series converges and its contribution is negligible (which shall be verified): $\zeta_{\lambda_\perp}\approx\zeta_{\lambda_\perp}^{(0)}$.}
The resulting CR-modified perpendicular power spectrum of background fluctuations is then given by 
$E_{\delta z}^{(1)}(k_\perp)\sim (\zeta_{\lambda_\perp}^{(0)})^2 E_{\delta z}^{(0)}(k_\perp)$ (i.e., $E_{\delta z}^{(1)}(k_\perp)\propto k_\perp^{\,-\,(\,\xi_\perp^z\,+\,\Delta\xi_\perp^{\rm CRF}\,)}$), where the CR-induced modification of the spectral index is
\begin{equation}\label{eq:CRfeedback_spectral_steepening}
    \Delta\xi_\perp^{\rm CRF}
    \sim
    \left\{
    \begin{array}{ll}
      \frac{1}{2}(\xi_\perp^z+1)(\xi_\|^w+3)-2 &  \text{if} \left\{\begin{array}{c}
        \chi_{\lambda_\perp}^z<1\,\text{and}\,\,\chi_{\lambda_\perp}^{(z|w)}\lesssim1 \\
        \text{or} \\
        \chi_{\lambda_\perp}^z\sim1\,\text{and}\,\,\chi^{(z|w)}_{\lambda_\perp}<1
      \end{array}\right.\\
      & \\
      \frac{1}{4}(\xi_\perp^z+1)(\xi_\|^w+5)-3  &  \text{if}\,\,\,  \chi_{\lambda_\perp}^z<1\,\text{and}\,\,\chi_{\lambda_\perp}^{(z|w)}>1\\
      & \\
      \frac{1}{4}(\xi_\perp^z+1)(\xi_\|^w+3)-1 &  \text{if}\,\,\, \chi_{\lambda_\perp}^z\sim1\,\text{and}\,\,\chi_{\lambda_\perp}^{(z|w)}\gtrsim1\\
      & \\
      \frac{1}{2}(\xi^z+2\xi^w+1)  &  \text{if}\,\,\, \chi_{\lambda}^z>1\,\text{and}\,\,\chi_{\lambda}^{(z|w)}<1\\
      & \\
      \frac{1}{2}(\xi^z+\xi^w) &  \text{if}\,\,\, \chi_{\lambda}^z>1\,\text{and}\,\,\chi_{\lambda}^{(z|w)}\gtrsim1
    \end{array}
    \right.
.\end{equation}
\noindent
We recall that this result is only valid at scales where $\zeta_{\lambda_\perp}<1$, and only if $\Gamma_{\rm turb}^{w,{\rm q\|}}\tau_{\rm casc}^{(z|w)}\ll1$ (i.e., if the CR-driven Alfv\'en-wave packets are unaffected by pre-existing fluctuations).
It is interesting to note that the anisotropy of the pre-existing cascade would be unaffected if $\chi^{(z|w)}<1$ (leaving $\ell_{\|,\lambda_\perp}\approx{\rm const.}$ if $\chi^z<1$, or $\ell_{\|,\lambda_\perp}\propto\lambda_\perp^{\delta^z}$ if $\chi^z\sim1$), or if $\chi^z>1$ (leaving $\ell_{\|,\lambda_\perp}\propto\lambda_\perp$). On the other hand, if $\chi^{(z|w)}\sim1$, a critical balance between the Alfv\'en time $\tau_{\rm A}^z$ and the nonlinear time $\tau_{\rm nl}^{(z|w)}$ would be established, leading to a modified anisotropy $\ell_{\|,\lambda_\perp}\propto\lambda_\perp^{1+\alpha_\perp^w}=\lambda_\perp^{\delta^z+\delta^{\rm CRF}}$ with $\delta^{\rm CRF}=\alpha_\|^w+\alpha_\perp^z(\alpha_\|^w-1)$. This means that the anisotropy of pre-existing fluctuations would be reduced if $|\alpha_\|^w|<\alpha_\perp^z/(1+\alpha_\perp^z)$ (e.g., for GS95 turbulence, this means $|\alpha_\|^w|<1/4$, corresponding to a CR-driven spectrum $\propto k_\|^{-1/2}$ or steeper), and it would be instead increased otherwise.

As mentioned before, the exact scaling and amplitude of the self-generated turbulent spectrum depends   on the properties of the CR distribution that drives the instability and on the different damping mechanisms that contribute to the instability saturation~\citep[see, e.g.,][and references therein]{MarcowithPOP2021}. However, as an example, we consider the results obtained with $1$D-$3$V kinetic simulations in \citet{HolcombSpitkovskyAPJ2019}, where the CR-driven fluctuations at saturation developed a scaling roughly consistent with $\delta B\propto k_\|^{-1/2}$ (i.e., $\alpha_\|^w\approx-1/2$ and thus $\xi_\|^w\approx-2$). Assuming a GS95 cascade of the background fluctuations (i.e., $\chi^z\sim 1$ and $\xi_\perp^z=5/3$) and strong nonlinearities induced by the CR-driven waves on these fluctuations (i.e., $\chi_{\lambda}^{(z|w)}\gtrsim1$), we obtain $\Delta\xi_\perp^{\rm CRF}\approx 1/3$. This means that the spectrum of background fluctuations below $\lambda_\perp^{\rm CRF}$ would be steepened from $k_\perp^{-5/3}$ to $k_\perp^{-2}$ due to the CR feedback (and also further suppressing the turbulent damping by lowering the amplitude of background fluctuations at those scales; cf. equation \eqref{eq:Turb-damp_w_q-prl-limit_generic}). The anisotropy of background turbulence would be also significantly increased, from $k_\|\propto k_\perp^{2/3}$ to $k_\|\propto k_\perp^{-1/3}$ (thus further reducing the effectiveness of CR scattering on pre-existing fluctuations). Another example can be set by assuming an Iroshnikov-Kraichnan spectrum for self-generated turbulence ($\propto k^{-3/2}$). This type of spectrum has  often been invoked to explain the observed $\gamma$-ray emission and local CR data~\citep[e.g.,][and references therein]{GaggeroAPJL2015}. In this case, still assuming a GS95-type of background turbulence and $\chi_{\lambda}^{(z|w)}\gtrsim1$, we now obtain  that the background spectrum is unchanged ($\Delta\xi_\perp^{\rm CRF}\approx 0$), but the anisotropy is enhanced by the CR feedback ($k_\|\approx\text{const}$).

\subsection{Overcritical interaction ($\chi^{(z|w)}>1$) and alternative CR-modified scaling of pre-existing fluctuations}

When nonlinear interactions between the CR-driven Alfv\'en-wave packets and background turbulence are  overcritical (i.e., $\chi^{(z|w)}>1$), it is reasonable to consider that pre-existing scaling is not just perturbatively modified. Therefore, we present an alternative model in which the intrinsic cascade time of pre-existing fluctuations $\tau_{\rm casc}^z$ is completely replaced by the nonlinear timescale $\tau_{\rm nl}^{(z|w)}$, without further re-scaling of $\delta z^{(0)}$ as in \eqref{eq:CRfeedback_fluct_rescaling_zeta}. This allows us to directly derive the CR-modified scaling of background fluctuations $\delta z^{\rm CRF}$ by requiring 
$(\delta z^{\rm CRF})^2/\tau_{\rm nl}^{(z|w)}\sim\varepsilon={\rm const}$.

If $\chi^z\lesssim1$, pre-existing scaling is anisotropic, and the condition above yields the perpendicular scaling 
\begin{equation}\label{eq:CRfeedback_overcritical_alternative}
    \frac{\delta w_{\lambda_\perp}(\delta z_{\lambda_\perp}^{\rm CRF})^2}{\lambda_\perp}
    \sim \varepsilon = {\rm const.}
    \,\,\,\Rightarrow\,\,\,
    \delta z_{\lambda_\perp}^{\rm CRF}
    \propto \lambda_\perp^{(1+\alpha_\perp^w)/2}\,.
\end{equation}
This corresponds to a modified perpendicular spectrum $E_{\delta z}^{\rm CRF}(k_\perp)\propto k_\perp^{-(\xi_\perp^z+\Delta\xi_\perp^{\rm CRF})}$ with
\begin{equation}\label{eq:CRfeedback_overcritical_spectral-modification}
    \Delta\xi_\perp^{\rm CRF}=\frac{\xi_\|^w+\xi_\perp^z(\xi_\|^w-3)+9}{4}\,,
\end{equation}
where the link to the original scaling of pre-existing fluctuations is a consequence of the quasi-parallel condition \eqref{eq:AW_q-prl-limit_general}.

If $\chi^z>1$, CR-modified fluctuations would follow the isotropic scaling $\delta z_{\lambda}^{\rm CRF}\propto \lambda^{(1+\alpha^w)/2}$, corresponding to a spectrum
\begin{equation}\label{eq:CRfeedback_overcritical_spectral-modification-2}
E_{\delta z}^{\rm CRF}(k)\propto k^{-(\xi^w+5)/2}\,,
\end{equation}
which does not depend on the original scaling of pre-existing fluctuations due to the loss of  quasi-parallel concept. 

\section{Discussion and conclusions}\label{sec:conclusions}

The turbulent damping of an Alfv\'en-wave (AW) packet excited by cosmic rays (CRs) in pre-existing incompressible magnetohydrodynamic (MHD) turbulence was re-examined by carefully taking into account the role of the  nonlinearity parameter $\chi^w$ that quantifies the strength of the nonlinear interaction between the packet and background fluctuations.
In particular, the difference between $\chi^w$ and the nonlinear parameter $\chi^z$ that instead describes the regime of background turbulence (i.e., the intrinsic strength of nonlinear interactions between pre-existing fluctuations) was   elucidated.
The derivation of turbulent damping rates in a classic MHD turbulence scenario (i.e., without the so-called  dynamic alignment) was thus   revised, taking into account the difference between $\chi^w$ and $\chi^z$, and new scaling relations for the damping rates were   obtained.
Furthermore, by considering the most recent theories of MHD turbulence that account for a scale-dependent (dynamic) alignment of fluctuations and the possibility of a reconnection-mediated regime, completely new damping rates were also obtained for the first time.
Finally, the role of cosmic-ray feedback (CRF) on pre-existing turbulence is also examined and a simple criterion for CRF effects is derived. Two very simple phenomenological models of CR-modified scaling of background fluctuations were also obtained.
In particular, this feedback can steepen the spectrum of background turbulence and further enhance its spectral anisotropy ($k_\|\ll k_\perp$). By reducing the amplitude of pre-existing fluctuations at the CRF scales, the former effect would have the consequence of further reducing the turbulent damping rate at those scales. At the same time, the increased anisotropy of background turbulence would also reduce the effectiveness of CR resonant scattering on pre-existing fluctuations at the CRF scales. These two CR-feedback effects may thus clear the stage for self-generated turbulence to dominate the CR transport and reinforce the self-confinement picture.
Taking into account the feedback of CR-generated fluctuations on pre-existing turbulence may be relevant in astrophysical environments where the density of cosmic-ray $n_{\rm CR}$ is non-negligible with respect to the density of the background thermal plasma $n_{\rm th}$ (e.g., near CR sources).
The issue of CRF effects, as well as the role of other damping mechanisms, will be addressed in more detail in the following Paper II.

The main features of the new turbulent damping rates obtained in this work can be summarized as follows:
\begin{itemize}

    \item The nonlinear interaction between a quasi-parallel (${\rm q\|}$) AW packet and pre-existing anisotropic turbulence is always weak (equations \eqref{eq:chi_w_q-prl-limit_general}-\eqref{eq:chi_w_q-prl-limit_general_vs_chi_z}). As a result, the turbulent damping rate of the packets depends on the background-fluctuation amplitude to the third power (equation \eqref{eq:Turb-damp_w_q-prl-limit_generic}), and thus is strongly suppressed with respect to what previously estimated. This is true at any wavelength when the AW packet interacts with sub- and trans-Alfv\'enic turbulence, and also for those packets whose wavelengths interact  with fluctuations at scales where they become critically balanced in the case of super-Alfv\'enic injection.\\
    
    \item How the turbulent damping rate $\Gamma_{\rm turb}^{w,{\rm q\|}}$ depends on (i) the AW-packet's parallel wavelength $\lambda_\|$ (and thus on the CR gyro-radius from which it is excited) and on (ii) the injection-scale Alfv\'enic Mach number $M_{\rm A,0}$, in the  classic MHD turbulence scenario is significantly different from what is presented in the existing literature (equations \eqref{eq:Turb-damp_subA_no_alignment}--\eqref{eq:Turb-damp_supA_no_alignment}).
    The damping rate agrees with the literature only when the AW packet interacts with isotropic  K41 turbulence, namely $\Gamma_{\rm K41}^{w}\propto M_{\rm A,0}\,\lambda^{-2/3}$. On the contrary, when the packet interacts with weak turbulence  W0, the damping rate scales as $\Gamma_{\rm W0}^{w,{\rm q\|}}\propto M_{\rm A,0}^{8/3}\, \lambda_\|^{1/3}$ (instead of $\propto M_{\rm A,0}^{8/3}\,\lambda_\|^{-2/3}$ as previously obtained), while when it interacts with critically balanced turbulence GS95, turbulent damping does not depend on the wavelength (i.e., $\Gamma_{\rm GS95}^{w,{\rm q\|}}\sim{\rm const.)}$ and it is $\propto M_{\rm A,0}^4$ for $M_{\rm A,0}<1$ or $\propto M_{\rm A,0}^3$ if $M_{\rm A,0}>1$, instead of $\propto M_{\rm A,0}^2\,\lambda_\|^{-1/2}$ or $\propto M_{\rm A,0}\,\lambda_\|^{-1/2}$, respectively, as reported in the existing literature.\\
    
    \item Including dynamic alignment of pre-existing fluctuations in the picture, and thus also allowing   for the possibility of a reconnection-mediated range, introduces novel regimes and breaks in the turbulent damping rate (equations \eqref{eq:Turb-damp_subA_alignment-only-strong}--\eqref{eq:Turb-damp_supA_alignment-only-strong}). When a quasi-parallel AW packet interacts with critically balanced and dynamically aligning anisotropic turbulence  B06, it is subjected to a damping rate $\Gamma_{\rm B06}^{w,{\rm q\|}}\propto M_{\rm A,0}^{24/5}\,\lambda_\|^{-1/5}$ if $M_{\rm A,0}<1$ or $\Gamma_{\rm B06}^{w,{\rm q\|}}\propto M_{\rm A,0}^{12/5}\,\lambda_\|^{-1/5}$ if $M_{\rm A,0}>1$. Alfv\'en-wave packets that interact with tearing-mediated turbulence (TMT) are instead subjected to a damping rate that is now also sensitive   to the injection-scale Lundquist number $S_0$, and it scales as $\Gamma_{\rm TMT}^{w,{\rm q\|}}\propto M_{\rm A,0}^{4}(S_0\,\lambda_\|)^{1/2}$ if $M_{\rm A,0}<1$ or $\Gamma_{\rm TMT}^{w,{\rm q\|}}\propto M_{\rm A,0}^{3}(S_0\,\lambda_\|)^{1/2}$ if $M_{\rm A,0}>1$.\\
    
    \item Accounting for dynamic alignment (and TMT) introduces two breaks in the turbulent damping rate, instead of the single break that is present in the  classic picture. For sub-Alfv\'enic turbulence, the first break corresponds to the transition scale between weak and strong turbulence (i.e., $\lambda_{\|}\sim M_{\rm A,0}^4\,\ell_0$ in terms of parallel wavelength of the AW packet), while it corresponds to the transition scale between isotropic and anisotropic turbulence for super-Alfv\'enic injection (i.e., $\lambda\sim M_{\rm A,0}^{-3}\,\ell_0$). This is the same type of break found in  classic MHD turbulence. A second break, on the other hand, emerges due to the transition to tearing-mediated turbulence, which is only possible if dynamic alignment occurs (i.e., at a packet's parallel wavelength $\lambda_{\|}\sim S_0^{-5/7}\,M_{\rm A,0}^{\,8/7}\,\ell_0$ in sub-Alfv\'enic turbulence, or at $\lambda_{\|}\sim S_0^{-5/7}\,M_{\rm A,0}^{\,-6/7}\,\ell_0$ for super-Alfv\'enic injection). We recall that $M_{\rm A,0}$ and $S_0$ are respectively the Alfv\'enic Mach number of turbulent fluctuations and Lunquist number of the background plasma at injection scale $\ell_0$. 
    Since CR  self-confinement relies on a balance between the growth of these CR-driven Alfv\'en waves and their damping, it is reasonable to imagine that in astrophysical situations where turbulent damping is the most relevant damping mechanism, these breaks would also emerge   in the propagated CR spectrum. We note that this is a simple damping-rate effect, and does not consider CR feedback on background fluctuations. It is thus interesting  that, assuming a Galactic magnetic field $B\sim 1$-$3~\mu$G and an injection scale of background turbulence $\ell_0\sim30$-$100$~pc, the above breaks in the damping rate could be translated to CR energies $E_{\rm CR}$ (assuming $\lambda_\|\sim r_{\rm L}$, where $r_{\rm L}$ is the Larmor radius of the cosmic ray). A first break at $E_{\rm CR,1}\sim 10$~TeV would indeed emerge if the injection-scale Alfv\'enic Mach number $M_{\rm A,0}$ of pre-existing turbulence is in the range $0.07\lesssim M_{\rm A,0}\lesssim0.14$ (i.e., sub-Alfv\'enic injection with $M_{\rm A,0}$ of order $\sim0.1$) or in the range $15\lesssim M_{\rm A,0}\lesssim30$ (i.e., super-Alfv\'enic injection with $M_{\rm A,0}$ of order $\sim10$). Additionally, for both $M_{\rm A,0}$ regimes determined above, a second break at CR energies $E_{\rm CR,2}\sim 300$~GeV would be consistently recovered if the Lundquist number of the background plasma is of order $S_0\sim10^5$--$10^7$. Clearly, this represents only an interesting feature in a very simplified scenario, and in general many other mechanisms that can affect CR transport may need to be taken into account~\citep[e.g.,][and references therein]{FornieriMNRAS2021,LazarianXuAPJ2021,ChernyshovAPJ2022,KempskiQuataertMNRAS2022,KempskiMNRAS2023,LemoineJPP2023,PezziBlasiMNRAS2024}.
\end{itemize}

In conclusion, it is worth noting once more that the turbulent damping rates obtained in this work differ dramatically from those found in the literature, even for  classic MHD turbulence due to the confusion between $\chi^w$ and $\chi^z$. All the existing CR studies that assume the turbulent damping rate as a fundamental ingredient in their calculations have thus employed an incorrect version of this damping rate. 
Hence, a number of previous works on CR self-confinement may need to be revised in view of the results presented here.
\\

\begin{acknowledgements}
The author is grateful to the referee for several comments that helped to improve the manuscript.
The author warmly acknowledges the kind hospitality of the Gran Sasso Science Institute (GSSI) in L'Aquila, and, in particular, Pasquale Blasi and Ottavio Fornieri, who introduced him to the paper by Farmer \& Goldreich on turbulent damping during his first visit in November 2022. He also greatly acknowledges Alexandre Marcowith, Thierry Passot, and Steve Shore for helpful comments and feedback on the manuscript. The author is grateful to Stas Boldyrev, Ben Chandran, Matt Kunz, Alex Lazarian, Bill Matthaeus, Thierry Passot, Alex Schekochihin, and Pierre-Louis Sulem for useful conversations (and different opinions) about MHD turbulence theory over the years.
This work has been partially supported by the French government, through the UCA$^\text{JEDI}$ Investments in the Future project managed by the National Research Agency (ANR) with the reference number ANR-15-IDEX-01. 
The author is also supported by the ANR grant  MiCRO'' with the reference number ANR-23-CE31-0016.
\end{acknowledgements}



\begin{appendix}

\section{Scaling of turbulent fluctuations at magnetohydrodynamic  scales}\label{app:MHDturb_scalings}

In this appendix we briefly review the turbulent scaling of Alfv\'enic fluctuations at  fluid (MHD) scales \citep[see, e.g.,][for a more detailed review on the topic]{SchekochihinJPP2022}.
First, we present the standard scenario without scale-dependent alignment of turbulent fluctuations. 
In the second part, an alternative scenario in which such  dynamic alignment is taking place in the critically balanced regime is presented.

In the following, isotropic injection is assumed and $\ell_0$  denotes the injection scale (i.e., the injection properties is not   affected by the presence of a mean magnetic-field direction $\bb{B}_0$ at that scale). 
Balanced injection is also assumed (i.e., that the same amount of energy is injected in the Els\"asser fields at $\ell_0$, $|\delta\bb{z}_0^{+}|^2\approx|\delta\bb{z}_0^{-}|^2$ = $\delta z_0^2$). We thus define an injection-scale Alfv\'enic Mach number $M_{\rm A,0}=\delta z_0/v_{\rm A,0}\approx\delta b_0/B_0$, where $v_{\rm A,0}=B_0/\sqrt{4\pi\rho_0}$ is the Alfv\'en speed associated with the background plasma (having mass density $\rho_0$ and being embedded in a mean field $\bb{B}_0$). For isotropic injection, the nonlinear parameter at injection scales $\chi_0=(k_{\perp,{\rm inj}}\,\delta z_0)/(k_{\|,{\rm inj}}\,v_{\rm A,0})$ (see \eqref{eq:MHD_Elsasser_chi_pm}) identifies with the injection-scale Alfv\'enic Mach number (i.e., $\chi_0=M_{\rm A,0}$).
Analogously, we define the injection-scale Lundquist number $S_0=\ell_0\,v_{\rm A,0}/\eta$, where $\eta$ is the resistivity of the background plasma. The Lunquist and the Alfv\'enic Mach numbers can be combined to provide the  injection-scale  magnetic Reynolds number $Rm_0 = \ell_0\,\delta z_0/\eta = M_{\rm A,0}\,S_0$.

An example of the resulting turbulent spectra and fluctuations' anisotropy for different injection regimes and type of cascades are summarized in Figures~\ref{fig:app:Spectra_examples} and \ref{fig:app:Anisotropy_examples}.

\subsection{Magnetohydrodynamic turbulence without dynamic alignment}\label{app:subsec:MHDturb_NoAlignment}

In this section we present what can be called   the  classic cascade of Alfv\'enic fluctuations (i.e., the standard scenario in which a scale-dependent alignment of turbulent fluctuations is not taken into account). 
In this case, depending on the large-scale regime of injection, turbulence can start as either fluid-like \citep{Kolmogorov1941} or wave-like \citep{NgBhattacharjeePOP1997,GaltierJPP2000}, until the point at which the cascade transitions into a critically balanced state \citep{GoldreichSridharAPJ1995} and eventually reaches dissipation.

\paragraph{\bf Sub- and trans-Alfv\'enic injection ($M_{\rm A,0}\leq1$).}
When the injection conditions are isotropic and sub-Alfv\'enic (i.e., such that $\chi_0=M_{\rm A,0}<1$), then fluctuations initially cascade in a weakly nonlinear regime. During that weak cascade only smaller perpendicular scales $\lambda_\perp<\ell_0$ are generated, while $\lambda_\|\sim\ell_0\sim{\rm const}$.\footnote{This result is formally obtained through wave-turbulence theory~\citep{NgBhattacharjeePOP1997,GaltierJPP2000}. In weak MHD turbulence, the main contribution to the cascade is the three-wave interaction, where an Alfv\'en wave with frequency $\omega_1^\pm$ and wave-vector $\bb{k}_1^\pm$ nonlinearly interacts with a counter-propagating Alfv\'en wave having frequency $\omega_2^\mp$ and wave-vector $\bb{k}_2^\mp$ in order to generate a third wave with $\omega_3$ and $\bb{k}_3$. The resonance conditions for this process essentially correspond to  momentum and  energy conservation laws: $\bb{k}_1^\pm + \bb{k}_2^\mp=\bb{k}_3$ and $\omega_1^\pm+\omega_2^\mp=\omega_3$. Since for Alfv\'en waves these conditions on parallel wave-vectors become $k_{1,\|}^\pm - k_{2,\|}^\mp = \pm k_{3,\|}$  and $k_{1,\|}^\pm + k_{2,\|}^\mp = k_{3,\|}$, the only nontrivial solution requires that either $k_{2,\|}^\mp=0$ and $k_{3,\|}=k_{1,\|}^\pm$, or $k_{1,\|}^\pm = 0$ and $k_{3,\|}=k_{2,\|}^\mp$. This means that the parallel wave-vector does not change during the three-wave interaction and only smaller perpendicular scales with $\bb{k}_{3,\perp}=\bb{k}_{1,\perp}+\bb{k}_{2,\perp}$ are generated by the weak cascade.} The cascade timescale in such weak regime is
\begin{equation}\label{app:eq:MHDturb_subA_NoAlignment_TauCasc}
    \tau_{{\rm casc},\lambda_\perp}^{\rm(subA)}
    \sim 
    \frac{\tau_{{\rm nl},\lambda_\perp}^{\rm(subA)}}{\chi_{\lambda_\perp}^{\rm(subA)}}
    \sim
    \frac{v_{\rm A,0}}{\ell_0}\left(\frac{\lambda_\perp}{\delta z_{\lambda_\perp}^{\rm(subA)}}\right)^2\,,
\end{equation}
from which the fluctuation scaling in the inertial range are obtained by requiring a constant energy cascading rate $\varepsilon$ through scales
\begin{equation}\label{app:eq:MHDturb_subA_NoAlignment_FluctScal_W0}
    \frac{(\delta z_{\lambda_\perp}^{\rm(subA)})^2}{\tau_{{\rm casc},\lambda_\perp}^{\rm(subA)}}
    \sim \varepsilon = {\rm const.}
    \,\,\,\Rightarrow\,\,\,
    \frac{\delta z_{\lambda_\perp}^{\rm(subA)}}{v_{\rm A,0}}
    \sim M_{\rm A,0}\left(\frac{\lambda_\perp}{\ell_0}\right)^{1/2}\,,
\end{equation}
where we have used the fact that the cascading rate is the same as the injection rate (i.e., $\varepsilon\sim\varepsilon_0\sim\delta z_0^2/\tau_{{\rm casc},0}^{\rm(subA)}\sim M_{\rm A,0}^{\,4}\,v_{\rm A,0}^3/\ell_0$). The fluctuation power spectrum is obtained as $\mathcal{E}_{\delta z}\sim (\delta z_{k_\perp})^2/k_\perp$, and thus the one associated with the weak cascade is $\mathcal{E}_{\delta z}^{\rm(subA)}(k_\perp)\propto k_\perp^{-2}$; here and in the following we explicitly employ the more familiar wave-vector notation $k_\perp\sim\lambda_\perp^{-1}$ for the spectrum.\\
The weak cascade would reach a dissipation scale $\lambda_{\perp,{\rm diss}}^{\rm(subA)}$ if the nonlinear timescale $\tau_{{\rm nl},\lambda_\perp}\sim \lambda_\perp/\delta z_{\lambda_\perp}$ becomes comparable to the characteristic dissipation time $\tau_{{\rm diss},\lambda_\perp}\sim\lambda_\perp^2/\eta$: 
\begin{equation}\label{app:eq:MHDturb_subA_NoAlignment_DissScale_W0}
    \frac{\lambda_{\perp,{\rm diss}}^{\rm(subA)}}{\delta z_{\lambda_{\perp,{\rm diss}}}^{\rm(subA)}}
    \sim
    \frac{(\lambda_{\perp,{\rm diss}}^{\rm(subA)})^2}{\eta}
    \,\,\,\Rightarrow\,\,\,
    \frac{\lambda_{\perp,{\rm diss}}^{\rm(subA)}}{\ell_0}
    \sim
    (M_{\rm A,0}\,S_0)^{-2/3}\,.
\end{equation}
However, the scaling in \eqref{app:eq:MHDturb_subA_NoAlignment_FluctScal_W0} implies that the nonlinear parameter $\chi_{\lambda_\perp}$ increases with decreasing scales, 
\begin{equation}\label{app:eq:MHDturb_subA_NoAlignment_ChiScal}
    \chi_{\lambda_\perp}^{\rm(subA)}
    \sim \frac{\ell_0/v_{\rm A,0}}{\lambda_\perp^{\rm(subA)}/\delta z_{\lambda_\perp}^{\rm(subA)}}
    \sim
    M_{\rm A,0}\left(\frac{\lambda_\perp}{\ell_0}\right)^{-1/2}\,,
\end{equation}
and will thus achieve critical balance (CB) at a perpendicular scale
\begin{equation}\label{app:eq:MHDturb_subA_NoAlignment_CBscale}
    \chi_{\lambda_{\rm \perp,CB}}^{\rm(subA)}
    \sim 1
    \quad\Rightarrow\quad
    \frac{\lambda_{\perp,{\rm CB}}}{\ell_0}
    \sim
    M_{\rm A,0}^{\,2}\,.
\end{equation}
A transition to strong turbulence occurs only if $\lambda_{\perp,{\rm CB}}\gg\lambda_{\perp,{\rm diss}}^{\rm(subA)}$, and comparing \eqref{app:eq:MHDturb_subA_NoAlignment_DissScale_W0} and \eqref{app:eq:MHDturb_subA_NoAlignment_CBscale}, this means only if $S_0\gg M_{\rm A,0}^{\,-4}$. 
At scales below $\lambda_{\perp,{\rm CB}}$ turbulence stays critically balanced and the cascade timescale identifies with the nonlinear time:  
\begin{equation}\label{app:eq:MHDturb_subA_NoAlignment_NLtimeScaling}
    \tau_{{\rm casc},\lambda_\perp<\lambda_{\perp,{\rm CB}}}^{\rm(subA)}
    \sim 
    \tau_{{\rm nl},\lambda_\perp}^{\rm(subA)}
    \sim
    \frac{\lambda_\perp}{\delta z_{\lambda_\perp}^{\rm(subA)}}
.\end{equation}
As a result, the scaling of turbulent fluctuations at $\lambda_\perp<\lambda_{\perp,{\rm CB}}$ is such that
\begin{equation}\label{app:eq:MHDturb_subA_NoAlignment_FluctScal_CB}
    \frac{(\delta z_{\lambda_\perp<\lambda_{\perp,{\rm CB}}}^{\rm(subA)})^2}{\tau_{{\rm nl},\lambda_\perp<\lambda_{\perp,{\rm CB}}}^{\rm(subA)}}
    \sim \varepsilon = {\rm const.}
    \,\,\,\Rightarrow\,\,\,
    \frac{\delta z_{\lambda_\perp<\lambda_{\perp,{\rm CB}}}^{\rm(subA)}}{v_{\rm A,0}}
    \sim M_{\rm A,0}^{\,4/3}\left(\frac{\lambda_\perp}{\ell_0}\right)^{1/3}\,,
\end{equation}
while the critical-balance condition $\tau_{\rm A,\lambda_\perp}\sim\tau_{{\rm nl},\lambda_\perp}$ sets the scale-dependent anisotropy of turbulent fluctuations,
\begin{equation}\label{app:eq:MHDturb_NoAlignment_FluctScal_subA_CB_aniso}
    \frac{\lambda_{\|,\lambda_\perp<\lambda_{\perp,{\rm CB}}}}{v_{\rm A,0}}
    \sim \frac{\lambda_\perp}{\delta z_{\lambda_\perp<\lambda_{\perp,{\rm CB}}}^{\rm(subA)}}
    \,\,\Rightarrow\,\,\,
    \frac{\lambda_{\|,\lambda_\perp<\lambda_{\perp,{\rm CB}}}}{\ell_0}
    \sim M_{\rm A,0}^{\,-4/3}\left(\frac{\lambda_\perp}{\ell_0}\right)^{2/3}
.\end{equation}
Equation \eqref{app:eq:MHDturb_NoAlignment_FluctScal_subA_CB_aniso} implies that below $\lambda_{\perp,{\rm CB}}$ the critically balanced cascade starts to generate  smaller parallel scales. 
From \eqref{app:eq:MHDturb_subA_NoAlignment_FluctScal_CB} we can obtain a reduced (one-dimensional) perpendicular spectrum $\mathcal{E}_{\delta z}^{\rm(subA)}(k_\perp\lambda_{\perp,{\rm CB}}>1)\propto k_\perp^{\,-5/3}$ and a reduced parallel spectrum $\mathcal{E}_{\delta z}^{\rm(subA)}(k_\|)\propto k_\|^{\,-2}$.\footnote{There are different ways to obtain the reduced parallel spectrum, given the anisotropy in \eqref{app:eq:MHDturb_NoAlignment_FluctScal_subA_CB_aniso}. One option is to invert the anisotropy relation to obtain the scaling of $\delta z_{k_\|}\propto k_\|^{\,-1/2}$, and then use the critical-balance condition to employ $\tau_{{\rm A},k_\|}\sim (k_\|\,v_{\rm A,0})^{-1}$ instead of $\tau_{\rm nl}$ in the condition $(\delta z_{k_\|})^2/\tau_{{\rm A},k_\|}\sim\varepsilon$. Another way is to use the condition that the total energy must be obtained by integrating both one-dimensional spectra independently, i.e., $\int {\rm d}k_\|\,\mathcal{E}(k_\|)=E_{\rm tot}=\int {\rm d}k_\perp\,\mathcal{E}(k_\perp)$, and using the anisotropy relation to rewrite $\mathcal{E}(k_\perp)$ and ${\rm d}k_\perp$.}\\
The above cascade eventually reaches dissipation at a scale $\lambda_{\perp,{\rm diss}}^{\rm(subA)}$ for which the nonlinear and dissipation timescales become comparable,  $\tau_{{\rm nl},\lambda_\perp}^{\rm(subA)} \sim \tau_{{\rm diss},\lambda_\perp}$:\footnote{Here we implicitly assume that the condition $\lambda_{\perp,{\rm diss}}^{\rm(subA)}\ll\lambda_{\perp,{\rm CB}}$ holds, also when the dissipation scale is computed using \eqref{app:eq:MHDturb_subA_NoAlignment_DissScale_GS95}, which is  automatically fulfilled as long as $S_0\gg M_{\rm A,0}^{\,-4}$.}
\begin{equation}\label{app:eq:MHDturb_subA_NoAlignment_DissScale_GS95}
    \frac{\lambda_{\perp,{\rm diss}}^{\rm(subA)}}{\delta z_{\lambda_{\perp,{\rm diss}}}^{\rm(subA)}}
    \sim
    \frac{(\lambda_{\perp,{\rm diss}}^{\rm(subA)})^2}{\eta}
    \,\,\,\Rightarrow\,\,\,
    \frac{\lambda_{\perp,{\rm diss}}^{\rm(subA)}}{\ell_0}
    \sim
    M_{\rm A,0}^{-1}\,S_0^{-3/4}.
\end{equation}

\paragraph{\bf Super-Alfv\'enic injection ($M_{\rm A,0}>1$).} When fluctuations are injected (isotropically) with $\chi_0=M_{\rm A,0}>1$, the resulting turbulence starts as a strong  hydrodynamic-like cascade; in other words,  turbulence is isotropic and nearly insensitive to the presence of a background magnetic field for as long as $\delta b/B_0>1$ holds (i.e., until the presence of a mean field starts to play a role in the cascade, at smaller scales). In the hydrodynamic-like range, fluctuations cascade with the nonlinear characteristic timescale,
\begin{equation}\label{app:eq:MHDturb_supA_NoAlignment_TauCasc}
    \tau_{{\rm casc},\lambda}^{\rm(supA)}
    \sim 
    \tau_{{\rm nl},\lambda}^{\rm(supA)}
    \sim
    \frac{\lambda}{\delta z_{\lambda}^{\rm(supA)}}\,,
\end{equation}
where $\lambda$ is the isotropic wavelength of the fluctuations. The scaling for the fluctuating Els\"asser variable immediately follow from the constancy of the energy cascade rate $\varepsilon$,
\begin{equation}\label{app:eq:MHDturb_supA_NoAlignment_FluctScal_hyrdo}
    \frac{(\delta z_{\lambda}^{\rm(supA)})^2}{\tau_{{\rm nl},\lambda}^{\rm(supA)}}
    \sim \varepsilon = {\rm const.}
    \,\,\,\Rightarrow\,\,\,
    \frac{\delta z_\lambda^{\rm(supA)}}{v_{\rm A,0}}
    \sim M_{\rm A,0}\,\left(\frac{\lambda}{\ell_0}\right)^{1/3}\,,
\end{equation}
which corresponds to a Kolmogorov-like, isotropic fluctuation power spectrum $\mathcal{E}_{\delta z}^{\rm(supA)}(k)\propto k^{-5/3}$.
This cascading regime goes on until it reaches dissipation: $\lambda_{\rm diss}^{\rm(supA)}\sim(M_{\rm A,0}\,S_0)^{-3/4}\ell_0$. 
However, there is another important scale usually referred to as the  Alfv\'en scale $\ell_{\rm A}$ for which $\delta z_\lambda^{\rm(supA)}\sim v_{\rm A,0}$, given by
\begin{equation}\label{app:eq:MHDturb_supA_NoAlignment_AlfvenScale}
    \frac{\ell_{\rm A}}{\ell_0}
    \sim 
    M_{\rm A,0}^{\,-3}\,,
\end{equation}
which is attained well before dissipation (i.e., $\ell_{\rm A}\gg\lambda_{\rm diss}^{\rm(supA)}$) only if $S_0\gg M_{\rm A,0}^{\,3}$, and
below which the cascade becomes critically balanced ($\chi_{\rm\ell_A}^{\rm(supA)}\sim1$) and thus anisotropic. Turbulent fluctuations at scales $\lambda_\perp<\ell_{\rm A}$ thus follow the (GS95) scaling 
\begin{equation}\label{app:eq:MHDturb_supA_NoAlignment_FluctScal_CB}
    \frac{\delta z_{\lambda_\perp<\ell_{\rm A}}^{\rm(supA)}}{v_{\rm A,0}}
    \sim 
    \left(\frac{\lambda_\perp}{\ell_{\rm A}}\right)^{1/3}
    \sim 
    M_{\rm A,0}\,\left(\frac{\lambda_\perp}{\ell_0}\right)^{1/3}\,,
\end{equation}
with a fluctuation wavelength anisotropy 
that now follows the relation 
$\lambda_{\|,\lambda_\perp<\ell_{\rm A}}/\ell_0\sim M_{\rm A,0}^{\,-1}\,(\lambda_\perp/\ell_0)^{2/3}$. This corresponds to reduced perpendicular and parallel power spectra at $k_\perp\ell_{\rm A}\gtrsim 1$, which  are $\propto k_\perp^{\,-5/3}$ and $\propto k_\|^{\,-2}$, respectively.\\
The dissipation scale $\lambda_{\perp,{\rm diss}}^{\rm(supA)}$ in the super-Alfv\'enic regime is given again by matching the scale-dependent nonlinear timescale $\tau_{{\rm nl},\lambda_\perp}^{\rm(supA)}$ and the dissipation timescale $\tau_{{\rm diss},\lambda_\perp}$ for this type of cascade: 
\begin{equation}\label{app:eq:MHDturb_supA_NoAlignment_DissScale}
    \frac{\lambda_{\perp,{\rm diss}}^{\rm(supA)}}{\delta z_{\lambda_{\perp,{\rm diss}}}^{\rm(supA)}}
    \sim
    \frac{(\lambda_{\perp,{\rm diss}}^{\rm(supA)})^2}{\eta}
    \,\,\,\Rightarrow\,\,\,
    \frac{\lambda_{\perp,{\rm diss}}^{\rm(supA)}}{\ell_0}
    \sim
    (M_{\rm A,0}\,S_0)^{-3/4}\,.
\end{equation}
The condition $S_0\gg M_{\rm A,0}^{\,3}$ ensures that the above scale is well below the Alfv\'en scale:  $\lambda_{\perp,{\rm diss}}^{\rm(supA)}\ll\ell_{\rm A}$ (cf. \eqref{app:eq:MHDturb_supA_NoAlignment_AlfvenScale} and \eqref{app:eq:MHDturb_supA_NoAlignment_DissScale}).

\subsection{Magnetohydrodynamic turbulence with scale-dependent alignment}\label{app:subsec:MHDturb_DynamicAlignment}

In this section we present an alternative model to the  classic picture of the Alfv\'enic cascade presented in Section~\ref{app:subsec:MHDturb_NoAlignment}.
In this case, after a hydrodynamic-like or wave-like range, the cascade transitions into a critically balanced state in which fluctuations undergo a  dynamic (i.e., scale-dependent) alignment process~\citep{BoldyrevPRL2006}. 
Such a dynamically aligned, critically balanced cascade can further transition into a  tearing-mediated regime at MHD scales \citep{BoldyrevLoureiroAPJ2017,MalletMNRAS2017} before reaching the actual dissipation scales.

\paragraph{\bf Sub- and trans-Alfv\'enic injection ($M_{\rm A,0}\leq1$) with dynamic alignment.}
For isotropic and sub-Alfv\'enic injeciton (i.e., such that $\chi_0=M_{\rm A,0}<1$), fluctuations initially develop a weak cascade following the same scaling as in \eqref{app:eq:MHDturb_subA_NoAlignment_FluctScal_W0}. Dynamic alignment enters the picture only as soon as critical balance is reached (i.e., at scales $\lambda_\perp\leq\lambda_{\perp,{\rm CB}}\sim M_{\rm A,0}^{\,2}\,\ell_0$). 
The idea behind this effect is that Els\"asser fields tend to align in order to reduce the strength of nonlinearities.\footnote{Another effect of dynamic alignment is that fluctuations exhibit three-dimensional anisotropy. If we call $\lambda$ the length-scale of these 3D anisotropic turbulent eddies in the direction perpendicular to both  mean-field $\langle\bb{B}\rangle_\lambda$ and magnetic-field   $\delta\bb{B}_{\perp,\lambda}$ fluctuations at this scale ($\delta\bb{B}_{\perp,\lambda}$ being perpendicular to $\langle\bb{B}\rangle_\lambda$), then $\ell_\lambda$ and $\xi_\lambda$ denote the length-scales along $\langle\bb{B}\rangle_\lambda$ and $\delta\bb{B}_{\perp,\lambda}$, respectively \citep{BoldyrevPRL2006}. 
In the following, $k_\perp \sim \lambda_\perp^{\,-1}$ refers to the shortest length-scale $\lambda$, and we neglect the distinction between the two transverse directions $k_\lambda\sim\lambda^{\,-1}$ and $k_\xi\sim\xi^{\,-1}$;  an angular average of fluctuation properties in a wave-vector plane transverse to $\langle\bb{B}\rangle_\lambda$ would be dominated by the scaling with $k_\lambda$.} This process produces a scale-dependent angle between $\delta\bb{z}_{\lambda_\perp}^{+}$ and $\delta\bb{z}_{\lambda_\perp}^{-}$ that scales as
\begin{equation}\label{app:eq:MHDturb_subA_DynamicAlignment_ThetaScaling}
    \sin\theta_{\lambda_\perp<\,\lambda_{\perp,{\rm CB}}}^{\rm(subA)}
    \sim 
    M_{\rm A,0}^{\,-1/2}\,\left(\frac{\lambda_\perp}{\ell_0}\right)^{1/4}\,,
\end{equation}
which in turn appears explicitly in the nonlinear time scaling
\begin{equation}\label{app:eq:MHDturb_subA_DynamicAlignment_NLtimeScaling}
    \tau_{{\rm nl},\lambda_\perp<\,\lambda_{\perp,{\rm CB}}}^{\rm(subA)}
    \sim
    \frac{\lambda_\perp}{\delta z_{\lambda_\perp<\,\lambda_{\perp,{\rm CB}}}^{\rm(subA)}\sin\theta_{\lambda_\perp<\,\lambda_{\perp,{\rm CB}}}^{\rm(subA)}}
    \sim
    M_{\rm A,0}^{\,1/2}\, \frac{\lambda_\perp^{3/4}\,\ell_0^{1/4}}{\delta z_{\lambda_\perp<\,\lambda_{\perp,{\rm CB}}}^{\rm(subA)}}
,\end{equation}
which decreases more slowly than the corresponding timescale when dynamic alignment is not present (cf. \eqref{app:eq:MHDturb_subA_NoAlignment_NLtimeScaling}).
As a result, in the presence of a scale-dependent alignment, turbulent fluctuations at $\lambda_\perp<\lambda_{\perp,{\rm CB}}$ scale as
\begin{equation}\label{app:eq:MHDturb_subA_DynamicAlignment_FluctScal_B06}
    \frac{\delta z_{\lambda_\perp<\,\lambda_{\perp,{\rm CB}}}^{\rm(subA)}}{v_{\rm A,0}}
    \sim M_{\rm A,0}^{\,3/2}\left(\frac{\lambda_\perp}{\ell_0}\right)^{1/4}
\end{equation}
while the critical-balance condition $\tau_{\rm A,\lambda_\perp}\sim\tau_{{\rm nl},\lambda_\perp<\lambda_{\perp,{\rm CB}}}$ sets the scale-dependent anisotropy of dinamically aligned turbulent fluctuations:
\begin{equation}\label{app:eq:MHDturb_subA_DynamicAlignment_AnisoScal_B06}
    \frac{\lambda_{\|,\lambda_\perp<\,\lambda_{\perp,{\rm CB}}}}{\ell_0}
    \sim M_{\rm A,0}^{\,-1}\,\left(\frac{\lambda_\perp}{\ell_0}\right)^{1/2}\,.
\end{equation}
Equation \eqref{app:eq:MHDturb_subA_DynamicAlignment_AnisoScal_B06} implies that below $\lambda_{\perp,{\rm CB}}$ a dynamically aligned, critically balanced cascade exhibits a stronger anisotropy than the corresponding cascade without a scale-dependent alignment.\footnote{This scaling involves the parallel length-scale $\ell_\lambda$ and the shortest perpendicular length-scale $\lambda$. However, fluctuations are 3D anisotropic. Since $\xi_\lambda\propto\lambda^{\,3/4}$ \citep{BoldyrevPRL2006}, the anisotropy scales as  $\ell_\xi\propto \xi^{\,2/3}$, when considering the longest perpendicular length-scale $\xi$.} 
Using \eqref{app:eq:MHDturb_subA_DynamicAlignment_FluctScal_B06} we obtain the reduced perpendicular spectrum $\mathcal{E}_{\delta z}^{\rm(subA)}(k_\perp>k_{\perp,{\rm CB}})\propto k_\perp^{\,-3/2}$, which is slightly shallower that the $-5/3$ obtained without dynamic alignment; on the other hand, the parallel spectrum is still $\mathcal{E}_{\delta z}^{\rm(subA)}(k_\|)\propto k_\|^{\,-2}$.\\
At this point, if the Lundquist number is  not large enough (i.e., such that $S_0\lesssim M_{\rm A,0}^{\,-4}$; see below), this dynamically aligned, critically balanced cascade  reaches the dissipation scale $\lambda_{\perp,{\rm diss}}^{\rm(subA)}$ 
when $\tau_{{\rm nl},\lambda_\perp}\sim\tau_{{\rm diss},\lambda_\perp}$. 
Using \eqref{app:eq:MHDturb_subA_DynamicAlignment_NLtimeScaling}, this means
\begin{equation}\label{app:eq:MHDturb_subA_DynamicAlignment_DissScale-B06}
    \frac{\lambda_{\perp,{\rm diss}}^{\rm(subA)}}{\ell_0}
    \sim
    (M_{\rm A,0}\,S_0)^{-2/3}\,.
\end{equation}
However, in most cases of interest, the Lundquist number $S_0$ is large enough that this critically balanced cascade of dynamically aligning fluctuations transitions to a  tearing-mediated cascade. Such a transition occurs at a perpendicular scale $\lambda_{\perp,*}$ for which the timescale associated with the (linear) growth rate of the tearing instability, $\gamma_{\lambda_\perp}^{\rm t}\sim S_0^{\,-1/2}(\lambda_\perp/\ell_0)^{-3/2}(\delta z_{\lambda_\perp}/v_{\rm A,0})^{1/2}(v_{\rm A,0}/\ell_0)$, becomes comparable to the eddy turnover time at that scale $\tau_{\rm nl,\lambda_\perp}^{\rm(subA)}\sim \lambda_\perp/\delta z_{\lambda_\perp}^{\rm(subA)}$:  $\gamma_{\lambda_{\perp,*}}^{\rm t}\tau_{\rm nl,\lambda_{\perp,*}}^{\rm(subA)}\sim 1$, yielding
\begin{equation}\label{app:eq:MHDturb_subA_DynamicAlignment_TMTscale}
    \frac{\lambda_{\perp,*}^{\rm(subA)}}{\ell_0}
    \sim
    M_{\rm A,0}^{\,-2/7}\,S_0^{\,-4/7}\,.
\end{equation}
Comparing \eqref{app:eq:MHDturb_subA_DynamicAlignment_TMTscale} and \eqref{app:eq:MHDturb_subA_DynamicAlignment_DissScale-B06}, we find  that a tearing-mediated range emerges only if $S_0\gg M_{\rm A,0}^{\,-4}$, so that $\lambda_{\perp,*}^{\rm(subA)}\gg\lambda_{\perp,{\rm diss}}^{\rm(subA)}$.
In this regime, the generation of turbulent fluctuations at scales $\lambda_\perp\lesssim\lambda_{\perp,*}$ is due to the disruption of the (dynamically aligning\footnote{A tearing-mediated regime fundamentally relies on the fact that turbulent fluctuations develop anisotropy in the plane perpendicular to a mean field ({i.e.}, $\lambda\ll\xi_\lambda$). Hence, tearing-mediated turbulence only exists if fluctuations  align in a scale-dependent fashion.}) turbulent eddies by magnetic reconnection; hence, the scale $\lambda_{\perp,*}$ is usually referred to as the  disruption scale. 
Thus, the tearing instability timescale $\tau_{\lambda_\perp}^{\rm t}\sim 1/\gamma_{\lambda_\perp}^{\rm t}$ is the  cascade time in this range of scales,\footnote{One can verify {a posteriori} that in this regime fluctuation scaling indeed preserves the condition $\tau_{\rm nl,\lambda_\perp}\sim 1/\gamma_{\lambda_\perp}^{\rm t}$ at all scales below $\lambda_{\perp,*}$.} and assuming a constant energy flux through scales, $(\delta z_{\lambda_\perp<\,\lambda_{\perp,*}}^{\rm(subA)})^2/\tau_{\lambda_\perp}^{\rm t}\sim\varepsilon={\rm const.}$, provides us with the fluctuation scaling in the tearing-mediated regime
\begin{equation}\label{app:eq:MHDturb_subA_DynamicAlignment_FluctScal_TMT}
    \frac{\delta z_{\lambda_\perp<\,\lambda_{\perp,*}}^{\rm(subA)}}{v_{\rm A,0}}
    \sim S_0^{\,1/5}\,M_{\rm A,0}^{\,8/5}\,\left(\frac{\lambda_\perp}{\ell_0}\right)^{3/5}\,,
\end{equation}
corresponding to a reduced spectrum $\mathcal{E}_{\delta z}^{\rm(subA)}(k_\perp>k_{\perp,*})\propto k_\perp^{\,-11/5}$.\\ 
Due to the nonlinear stage of the tearing instability, turbulent fluctuations in the reconnection-mediated range tend to misalign in a scale-dependent fashion, following the scaling\footnote{This is obtained as the ratio of the resistive inner scale $\delta$ to the longitudinal scale $\zeta$ of the current layer~\citep{BoldyrevLoureiroAPJ2017}.}
\begin{equation}\label{app:eq:MHDturb_subA_DynamicAlignment_ThetaScaling_TMT}
    \sin\theta_{\lambda_\perp<\lambda_{\perp,*}}^{\rm(subA)}
    \sim 
    S_0^{\,-3/5}\,M_{\rm A,0}^{\,-4/5}\,\left(\frac{\lambda_\perp}{\ell_0}\right)^{-4/5}\,,
\end{equation}
while the fluctuation anisotropy in this range is obtained from the CB-like condition $\gamma_{\lambda_\perp}^{\rm t}\tau_{\rm A,\lambda_\perp}\sim1$: 
\begin{equation}\label{app:eq:MHDturb_subA_DynamicAlignment_AnisoScal_TMT}
    \frac{\lambda_{\|,\lambda_\perp<\lambda_{\perp,*}}}{\ell_0}
    \sim S_0^{\,2/5}\,M_{\rm A,0}^{\,-4/5}\,\left(\frac{\lambda_\perp}{\ell_0}\right)^{6/5}\,.
\end{equation}
The tearing-mediated cascade eventually dissipates at a scale where the characteristic dissipation time becomes comparable with the tearing timescale,  
\begin{equation}\label{app:eq:MHDturb_subA_DynamicAlignment_DissScale_TMT}
    \gamma_{\lambda_\perp}^{\rm t}\tau_{\rm diss,\lambda_\perp}\sim1
    \,\,\,\Rightarrow\,\,\,
    \frac{\lambda_{\perp,{\rm diss}}^{\rm(subA)}}{\ell_0}
    \sim
    M_{\rm A,0}^{-1}\, S_0^{-3/4}\,,
\end{equation}
which, interestingly enough, is exactly the same dissipation scale \eqref{app:eq:MHDturb_subA_NoAlignment_DissScale_GS95} that was found for the GS95 cascade. 

\paragraph{\bf Super-Alfv\'enic injection ($M_{\rm A,0}>1$) with dynamic alignment.} 
In this regime the cascade develops in a hydrodynamic-like fashion until the Alfv\'en scale $\ell_{\rm A}\sim M_{\rm A,0}^{\,-3}\,\ell_0$ (i.e., without being affected by dynamic alignment). Thus, fluctuations follow the scaling in \eqref{app:eq:MHDturb_supA_NoAlignment_FluctScal_hyrdo} down to $\ell_{\rm A}$, and only below this scale does the cascade become critically balanced and dynamic alignment plays a role. At $\lambda_\perp<\ell_{\rm A}$, the fluctuation alignment angle scales as
\begin{equation}\label{app:eq:MHDturb_supA_DynamicAlignment_ThetaScaling}
    \sin\theta_{\lambda_\perp<\,\ell_{\rm A}}^{\rm(supA)}
    \sim 
    M_{\rm A,0}^{\,3/4}\,\left(\frac{\lambda_\perp}{\ell_0}\right)^{1/4}\,,
\end{equation}
and the nonlinear time at such scales is thus given by
\begin{equation}\label{app:eq:MHDturb_supA_DynamicAlignment_NLtimeScaling}
    \tau_{{\rm nl},\lambda_\perp<\,\ell_{\rm A}}^{\rm(supA)}
    \sim
    \frac{\lambda_\perp}{\delta z_{\lambda_\perp<\,\ell_{\rm A}}^{\rm(supA)}\sin\theta_{\lambda_\perp<\,\ell_{\rm A}}^{\rm(supA)}}
    \sim
    M_{\rm A,0}^{\,-3/4}\, \frac{\lambda_\perp^{3/4}\,\ell_0^{1/4}}{\delta z_{\lambda_\perp<\,\ell_{\rm A}}^{\rm(supA)}}\,.
\end{equation}
As a result, in the presence of a scale-dependent alignment, turbulent fluctuations at $\lambda_\perp<\ell_{\rm A}$ scale as
\begin{equation}\label{app:eq:MHDturb_supA_DynamicAlignment_FluctScal_B06}
    \frac{(\delta z_{\lambda_\perp<\,\ell_{\rm A}}^{\rm(supA)})^2}{\tau_{{\rm nl},\lambda_\perp<\,\ell_{\rm A}}^{\rm(supA)}}
    \sim \varepsilon = {\rm const.}
    \,\,\,\Rightarrow\,\,\,
    \frac{\delta z_{\lambda_\perp<\,\ell_{\rm A}}^{\rm(supA)}}{v_{\rm A,0}}
    \sim M_{\rm A,0}^{\,3/4}\,\left(\frac{\lambda_\perp}{\ell_0}\right)^{1/4}
,\end{equation}
corresponding to a $\propto k_\perp^{\,-3/2}$ spectrum for $k_\perp\ell_{\rm A}>1$. Fluctuation scale-dependent anisotropy is obtained via the critical-balance condition $\tau_{\rm A,\lambda_\perp<\,\ell_{\rm A}}\sim\tau_{{\rm nl},\lambda_\perp<\,\ell_{\rm A}}$:
\begin{equation}\label{app:eq:MHDturb_supA_DynamicAlignment_AnisoScal_B06}
    \frac{\lambda_{\|,\lambda_\perp<\,\ell_{\rm A}}}{\ell_0}
    \sim M_{\rm A,0}^{\,-3/2}\,\left(\frac{\lambda_\perp}{\ell_0}\right)^{1/2}\,.
\end{equation}
The above cascade of critically balanced, dynamically aligned fluctuations can either reach actual dissipation at a scale $\lambda_{\rm \perp,diss}^{\rm(supA)}/\ell_0\sim M_{\rm A,0}^{\,-1}\,S_0^{\,-2/3}$ or, if $S_0 \gg M_{\rm A,0}^{\,3}$ holds, will instead transition to the tearing-mediated regime at a (disruption) scale\footnote{We recall that the transition scale in \eqref{app:eq:MHDturb_supA_DynamicAlignment_TMTscale} is obtained using the condition $\gamma_{\lambda_{\perp,*}}^{\rm t}\tau_{\rm nl,\lambda_{\perp,*}}^{\rm(supA)}\sim 1$, where the growth rate of the tearing instability is given by $\gamma_{\lambda_\perp}^{\rm t}\sim S_0^{\,-1/2}(\lambda_\perp/\ell_0)^{-3/2}(\delta z_{\lambda_\perp}/v_{\rm A,0})^{1/2}(v_{\rm A,0}/\ell_0)$.}
\begin{equation}\label{app:eq:MHDturb_supA_DynamicAlignment_TMTscale}
    \frac{\lambda_{\perp,*}^{\rm(supA)}}{\ell_0}
    \sim
    M_{\rm A,0}^{\,-9/7}\,S_0^{\,-4/7}\,.
\end{equation}
At scales $\lambda_\perp<\lambda_{\perp,*}^{\rm(supA)}$, fluctuations   then follow the scaling 
\begin{equation}\label{app:eq:MHDturb_supA_DynamicAlignment_FluctScal_TMT}
    \frac{\delta z_{\lambda_\perp<\,\lambda_{\perp,*}}^{\rm(supA)}}{v_{\rm A,0}}
    \sim S_0^{\,1/5}\,M_{\rm A,0}^{\,6/5}\,\left(\frac{\lambda_\perp}{\ell_0}\right)^{3/5}\,,
\end{equation}
corresponding to a $\propto k_\perp^{\,-11/5}$ spectrum at $k_\perp\lambda_{\perp,*}^{\rm(supA)}>1$. In this range, fluctuations develop a scale-dependent (mis-)alignment angle 
\begin{equation}\label{app:eq:MHDturb_supA_DynamicAlignment_ThetaScaling_TMT}
    \sin\theta_{\lambda_\perp<\,\lambda_{\perp,*}}^{\rm(supA)}
    \sim 
    S_0^{\,-3/5}\,M_{\rm A,0}^{\,-3/5}\,\left(\frac{\lambda_\perp}{\ell_0}\right)^{-4/5}\,,
\end{equation}
and an aniosotropy given by
\begin{equation}\label{app:eq:MHDturb_supA_DynamicAlignment_AnisoScal_TMT}
    \frac{\ell_{\|,\lambda_\perp<\,\lambda_{\perp,*}}}{\ell_0}
    \sim S_0^{\,2/5}\,M_{\rm A,0}^{\,-3/5}\,\left(\frac{\lambda_\perp}{\ell_0}\right)^{6/5}\,.
\end{equation}
Finally, this tearing-mediated regime reaches dissipation at
\begin{equation}\label{app:eq:MHDturb_supA_DynamicAlignment_DissScale_TMT}
    \frac{\lambda_{\perp,{\rm diss}}^{\rm(supA)}}{\ell_0}
    \sim \frac{\eta^{3/4}}{\varepsilon^{1/4}\ell_0}
    \sim
    (M_{\rm A,0}\,S_0)^{-3/4}\,.
\end{equation}

\subsection{Summary of the scaling relations for incompressible magnetohydrodynamic turbulence}

In this section, we gather the scaling of turbulent fluctuations and of their anisotropy that are derived in Sections~\ref{app:subsec:MHDturb_NoAlignment} and \ref{app:subsec:MHDturb_DynamicAlignment}.

\subsubsection{Scaling without dynamic alignment}

If we put all the relations of Section~\ref{app:subsec:MHDturb_NoAlignment} back together, then the scaling for the normalized fluctuation amplitudes $\delta\hat{z}=\delta z/v_{\rm A,0}$ at MHD scales are as shown below. We recall that $\hat{\lambda}_\perp=\lambda_\perp/\ell_0$ is the normalized perpendicular wavelength.
\paragraph{\bf $M_{\rm A,0}\leq1$ regime (no dynamic alignment, $S_0\gg M_{\rm A,0}^{\,-4}$):}
\begin{equation}\label{app:eq:MHDturb_subA_NoAlignment_FluctScal_full}
    \delta\hat{z}_{\hat{\lambda}_\perp}^{\rm(subA)}
    \sim
    \left\{
    \begin{array}{lcrc}
      M_{\rm A,0}\,\hat{\lambda}_\perp^{\,1/2}\, & \qquad & \hat{\lambda}_{\perp,{\rm CB}} < \hat{\lambda}_\perp\leq 1 & \,\,{\rm[W0]}\\
      & & &\\
      M_{\rm A,0}^{\,4/3}\,\hat{\lambda}_\perp^{\,1/3}\, & \qquad &  \hat{\lambda}_{\perp,{\rm diss}}^{\rm(subA)} < \hat{\lambda}_\perp \leq \hat{\lambda}_{\perp,{\rm CB}} & \,\,{\rm[GS95]} 
    \end{array}
    \right.
,\end{equation}
where $\hat{\lambda}_{\perp,{\rm CB}}\sim M_{\rm A,0}^{\,2}$ and $\hat{\lambda}_{\perp,{\rm diss}}^{\rm(subA)}\sim M_{\rm A,0}^{\,-1}\,S_0^{\,-3/4}$, while the fluctuation anisotropy is given by  
\begin{equation}\label{app:eq:MHDturb_subA_NoAlignment_FluctAniso_full}
    \hat{\lambda}_{\|,\hat{\lambda}_\perp}^{\rm(subA)}
    \sim
    \left\{
    \begin{array}{lcrc}
      {\rm const.}\, & \qquad & \hat{\lambda}_{\perp,{\rm CB}} < \hat{\lambda}_\perp\leq 1 & \,\,{\rm[W0]}\\
      & & &\\
      M_{\rm A,0}^{\,-4/3}\,\hat{\lambda}_\perp^{\,2/3}\, & \qquad &  \hat{\lambda}_{\perp,{\rm diss}}^{\rm(subA)} < \hat{\lambda}_\perp \leq \hat{\lambda}_{\perp,{\rm CB}} & \,\,{\rm[GS95]} 
    \end{array}
    \right.
,\end{equation}
where $\hat{\lambda}_{\|,\hat{\lambda}_\perp}=\lambda_{\|,\hat{\lambda}_\perp}/\ell_0$ is the normalized parallel wavelength of turbulent fluctuations.

\paragraph{\bf $M_{\rm A,0}>1$ regime (no dynamic alignment, $S_0\gg M_{\rm A,0}^{\,3}$):}
\begin{equation}\label{app:eq:MHDturb_supA_NoAlignment_FluctScal_full}
    \delta\hat{z}_{\hat{\lambda}_\perp}^{\rm(supA)}
    \sim
    \left\{
    \begin{array}{lcrc}
      M_{\rm A,0}\,\hat{\lambda}^{\,1/3}\, & \qquad & \hat{\ell}_{\rm A} < \hat{\lambda}\leq 1 & \,\,{\rm[K41]}\\
      & & &\\
      M_{\rm A,0}\,\hat{\lambda}_\perp^{\,1/3}\, & \qquad &  \hat{\lambda}_{\perp,{\rm diss}}^{\rm(supA)} < \hat{\lambda}_\perp \leq \hat{\ell}_{\rm A} & \,\,{\rm[GS95]}
    \end{array}
    \right.
,\end{equation}
where $\hat{\ell}_{\rm A}\sim M_{\rm A,0}^{\,-3}$ and $\hat{\lambda}_{\perp,{\rm diss}}^{\rm(supA)}\sim (M_{\rm A,0}\,S_0)^{\,-3/4}$, while the fluctuations exhibit an anisotropy  
\begin{equation}\label{app:eq:MHDturb_supA_NoAlignment_FluctAniso_full}
    \hat{\lambda}_{\|,\hat{\lambda}_\perp}^{\rm(supA)}
    \sim
    \left\{
    \begin{array}{lcrc}
      \hat{\lambda}_\perp\sim\hat{\lambda}\, & \qquad & \hat{\ell}_{\rm A} < \hat{\lambda}\leq 1 & \,\,{\rm[K41]}\\
      & & &\\
      M_{\rm A,0}^{\,-1}\,\hat{\lambda}_\perp^{\,2/3}\, & \qquad &  \hat{\lambda}_{\perp,{\rm diss}}^{\rm(supA)} < \hat{\lambda}_\perp \leq \hat{\ell}_{\rm A} & \,\,{\rm[GS95]} 
    \end{array}
    \right.
\end{equation}
with $\hat{\lambda}_{\|,\hat{\lambda}_\perp}=\lambda_{\|,\hat{\lambda}_\perp}/\ell_0$.
\\
See Figures \ref{fig:app:Spectra_examples} and \ref{fig:app:Anisotropy_examples} for the resulting spectra and fluctuation anisotropy versus the perpendicular wavenumber $k_\perp$.

\subsubsection{Scaling with dynamic alignment}

We summarize here all the scaling of Section~\ref{app:subsec:MHDturb_DynamicAlignment} for the (normalized) fluctuation amplitudes $\delta\hat{z}=\delta z/v_{\rm A,0}$ with respect to the (normalized) perpendicular wavelength $\hat{\lambda}_\perp=\lambda_\perp/\ell_0$.

\paragraph{\bf $M_{\rm A,0}\leq1$ regime (with dynamic alignment, $S_0\gg M_{\rm A,0}^{\,-4}$):}
\begin{equation}\label{app:eq:MHDturb_subA_DynamicAlignment_FluctScal_full}
    \hspace{-0.05cm}\delta\hat{z}_{\hat{\lambda}_\perp}^{\rm(subA)}
    \hspace{-0.05cm}\sim\hspace{-0.05cm}
    \left\{
    \begin{array}{lcrc}
      M_{\rm A,0}\,\hat{\lambda}_\perp^{\,1/2} & \,\,\, & \hat{\lambda}_{\perp,{\rm CB}} < \hat{\lambda}_\perp\leq 1 & {\rm[W0]}\\
      & & &\\
      M_{\rm A,0}^{\,3/2}\,\hat{\lambda}_\perp^{\,1/4} & \,\,\, &  \hat{\lambda}_{\perp,*}^{\rm(subA)} < \hat{\lambda}_\perp \leq \hat{\lambda}_{\perp,{\rm CB}} & {\rm[B06]}\\ 
      & & &\\
      S_0^{\,1/5}\,M_{\rm A,0}^{\,8/5}\,\hat{\lambda}_\perp^{\,3/5} & \,\,\, &  \hat{\lambda}_{\perp,{\rm diss}}^{\rm(subA)} < \hat{\lambda}_\perp \leq \hat{\lambda}_{\perp,*}^{\rm(subA)} & {\rm[TMT]} 
    \end{array}
    \right.
,\end{equation}
with $\hat{\lambda}_{\perp,{\rm CB}}\sim M_{\rm A,0}^{\,2}$, $\hat{\lambda}_{\perp,*}^{\rm(subA)}\sim M_{\rm A,0}^{\,-2/7}\,S_0^{\,-4/7}$ and $\hat{\lambda}_{\perp,{\rm diss}}^{\rm(subA)}\sim M_{\rm A,0}^{\,-1}\,S_0^{\,-3/4}$, while the fluctuation anisotropy is given by  
\begin{equation}\label{app:eq:MHDturb_subA_DynamicAlignment_FluctAniso_full}
    \hspace{-0.05cm}\hat{\lambda}_{\|,\hat{\lambda}_\perp}^{\rm(subA)}
    \hspace{-0.05cm}\sim\hspace{-0.05cm}
    \left\{
    \begin{array}{lcrc}
      {\rm const.} &  \,\,\, & \hat{\lambda}_{\perp,{\rm CB}} < \hat{\lambda}_\perp\leq 1 & {\rm[W0]}\\
      & & &\\
      M_{\rm A,0}^{\,-1}\,\hat{\lambda}_\perp^{\,1/2} &  \,\,\, &  \hat{\lambda}_{\perp,*}^{\rm(subA)} < \hat{\lambda}_\perp \leq \hat{\lambda}_{\perp,{\rm CB}} & {\rm[B06]}\\
      & & &\\
      S_0^{\,2/5}\,M_{\rm A,0}^{\,-4/5}\,\hat{\lambda}_\perp^{\,6/5} & \,\,\, &  \hat{\lambda}_{\perp,{\rm diss}}^{\rm(subA)} < \hat{\lambda}_\perp \leq \hat{\lambda}_{\perp,*}^{\rm(subA)} & {\rm[TMT]} 
    \end{array}
    \right.
,\end{equation}
where $\hat{\lambda}_{\|,\hat{\lambda}_\perp}=\lambda_{\|,\hat{\lambda}_\perp}/\ell_0$ is the normalized parallel wavelength of turbulent fluctuations.

\begin{figure*}[!ht]
   \includegraphics[width=0.5\textwidth]{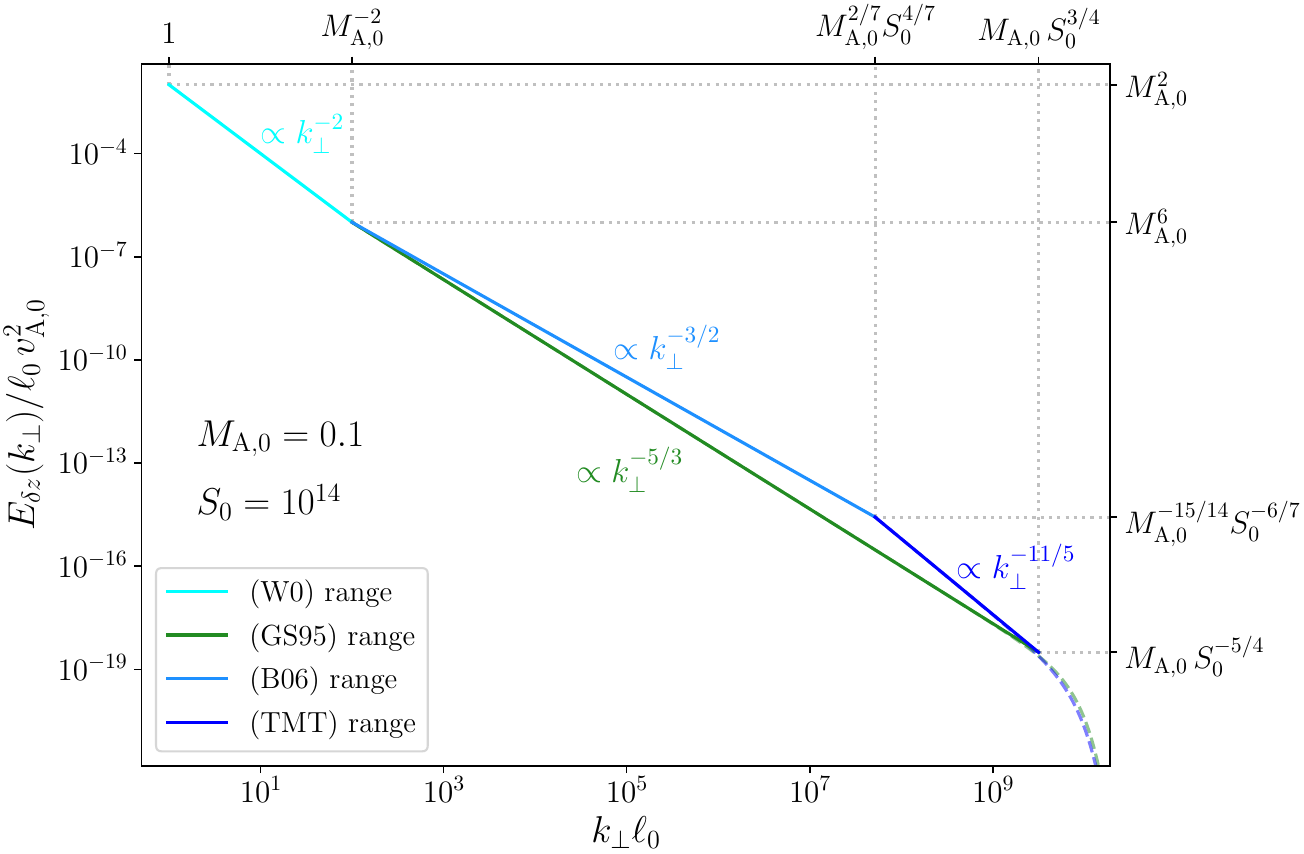}
   \includegraphics[width=0.5\textwidth]{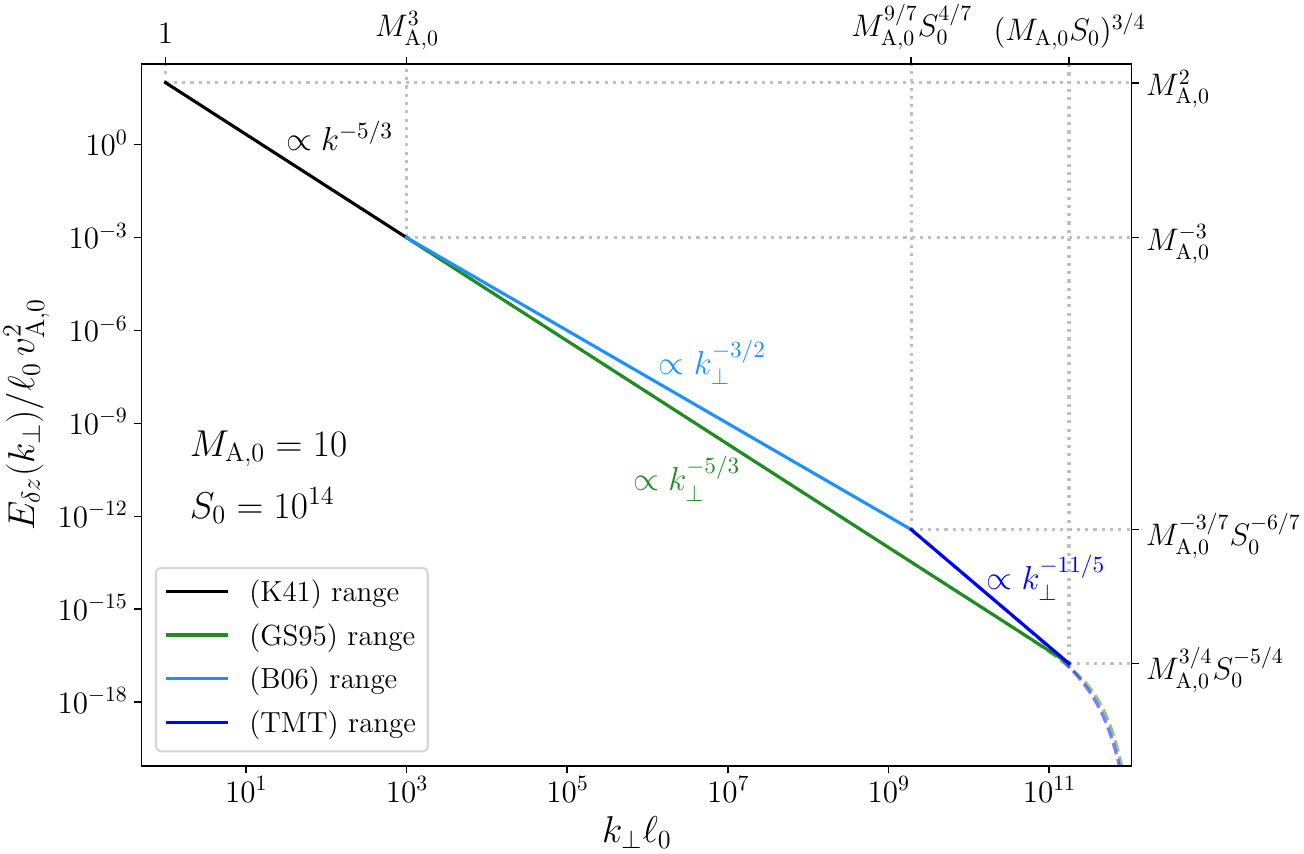}
   \caption{Normalized reduced spectrum, $E_{\delta z}(k_\perp)/\ell_0 v_{\rm A,0}^2$, vs. fluctuation perpendicular wave-vector, $k_\perp\ell_0$, in a plasma with Lunquist number $S_0=10^{14}$ and sub-Alfv\'enic ($M_{\rm A,0}=0.1$, left) or super-Aflv\'enic ($M_{\rm A,0}=10$, right) injection regimes. The different colors represent different cascading regimes (see legend), and general expressions for transition scales and fluctuation power level are reported on the right and upper axis. The solid lines show ideal scaling from \eqref{app:eq:MHDturb_subA_NoAlignment_FluctScal_full}, \eqref{app:eq:MHDturb_supA_NoAlignment_FluctScal_full}, \eqref{app:eq:MHDturb_subA_DynamicAlignment_FluctScal_full}, and \eqref{app:eq:MHDturb_supA_DynamicAlignment_FluctScal_full} for the nominal range $\ell_0^{-1}\lesssim k_\perp\lesssim \lambda_{\perp,{\rm diss}}^{-1}$, while the dashed lines represent their extension in the dissipation range with a damping factor $\sim\exp(-k_\perp^2\lambda_{\perp,{\rm diss}}^2)$.}
   \label{fig:app:Spectra_examples}
   \end{figure*}

\begin{figure*}[!ht]
   \includegraphics[width=0.5\textwidth]{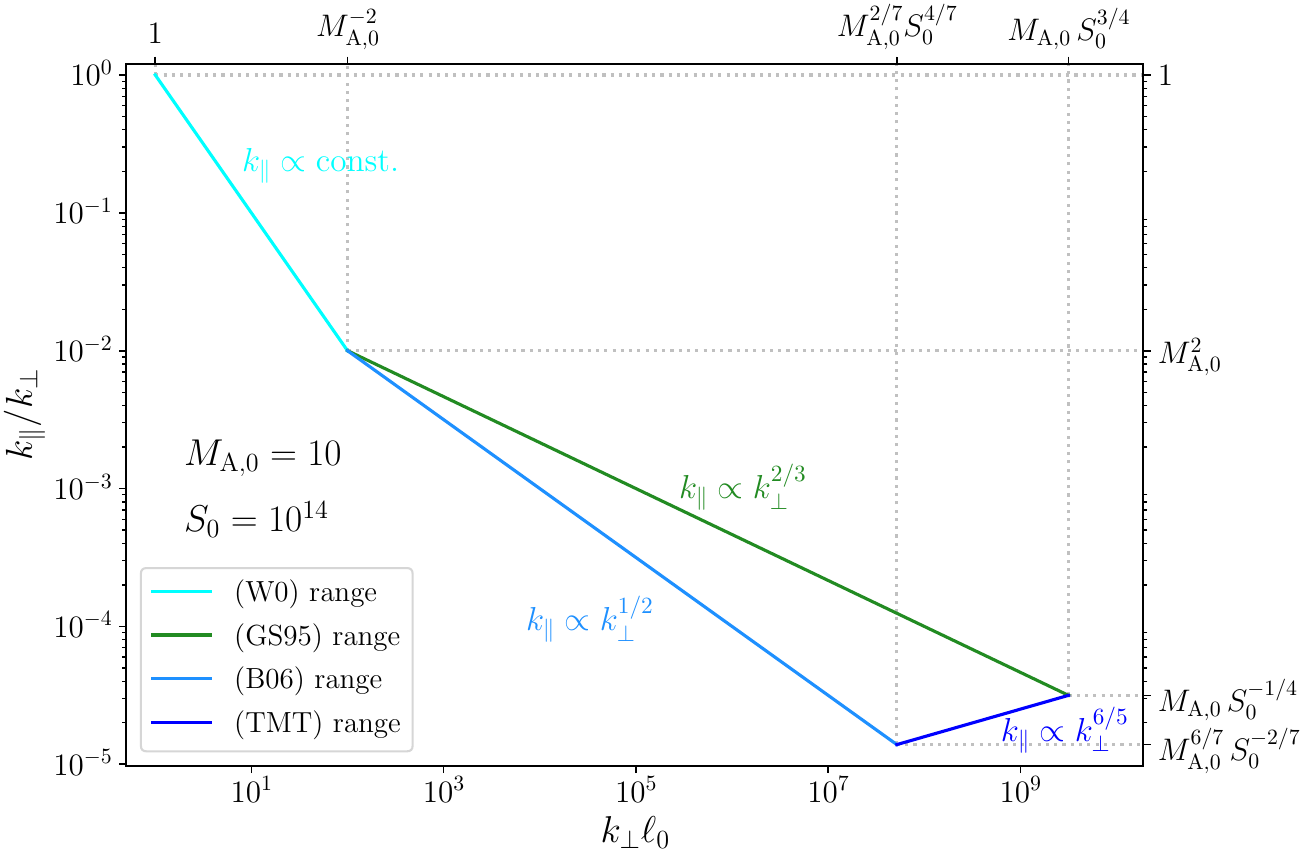}
   \includegraphics[width=0.5\textwidth]{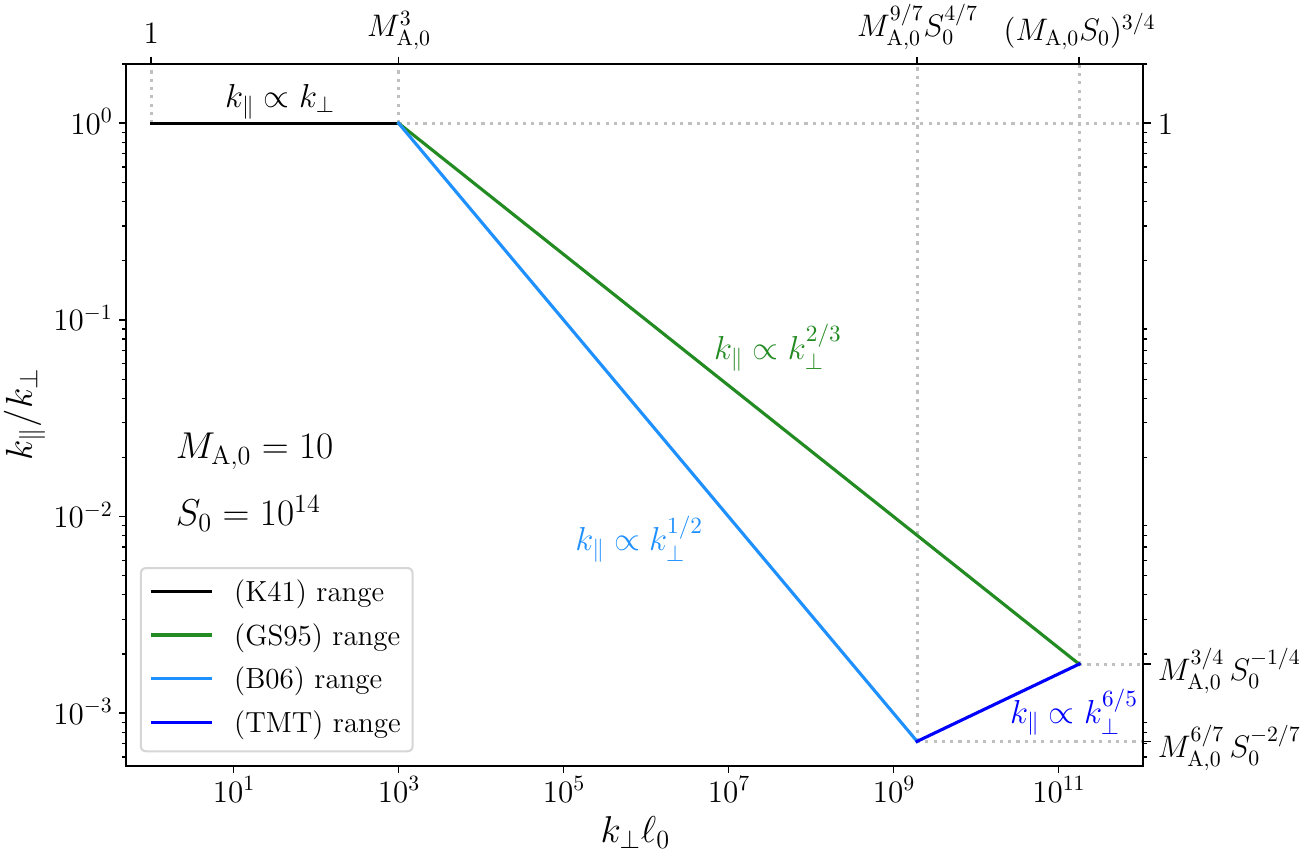}
   \caption{Wave-vector anisotropy of fluctuations, $k_\|/k_\perp$, vs. normalized fluctuation perpendicular wave-vector, $k_\perp\ell_0$, for the cascades shown in Figure~\ref{fig:app:Spectra_examples} in the nominal range $\ell_0^{-1}\lesssim k_\perp\lesssim \lambda_{\perp,{\rm diss}}^{-1}$ (cf. equations \eqref{app:eq:MHDturb_subA_NoAlignment_FluctAniso_full}, \eqref{app:eq:MHDturb_supA_NoAlignment_FluctAniso_full}, \eqref{app:eq:MHDturb_subA_DynamicAlignment_FluctAniso_full}, and \eqref{app:eq:MHDturb_supA_DynamicAlignment_FluctAniso_full}).}
   \label{fig:app:Anisotropy_examples}
   \end{figure*}

\paragraph{\bf $M_{\rm A,0}>1$ regime (with dynamic alignment, $S_0\gg M_{\rm A,0}^{\,3}$):}
\begin{equation}\label{app:eq:MHDturb_supA_DynamicAlignment_FluctScal_full}
    \hspace{-0.05cm}\delta\hat{z}_{\hat{\lambda}_\perp}^{\rm(supA)}
    \hspace{-0.05cm}\sim\hspace{-0.05cm}
    \left\{
    \begin{array}{lcrc}
      M_{\rm A,0}\,\hat{\lambda}^{\,1/3} & \,\,\, & \hat{\ell}_{\rm A} < \hat{\lambda}\leq 1 & {\rm[K41]}\\
      & & &\\
      M_{\rm A,0}^{\,3/4}\,\hat{\lambda}_\perp^{\,1/4} & \,\,\, &  \hat{\lambda}_{\perp,*}^{\rm(supA)} < \hat{\lambda}_\perp \leq \hat{\ell}_{\rm A} & {\rm[B06]}\\ 
      & & &\\
      S_0^{\,1/5}\,M_{\rm A,0}^{\,6/5}\,\hat{\lambda}_\perp^{\,3/5} & \,\,\, &  \hat{\lambda}_{\perp,{\rm diss}}^{\rm(supA)} < \hat{\lambda}_\perp \leq \hat{\lambda}_{\perp,*}^{\rm(supA)} & {\rm[TMT]} 
    \end{array}
    \right.
,\end{equation}
where $\hat{\ell}_{\rm A}\sim M_{\rm A,0}^{\,-3}$, $\hat{\lambda}_{\perp,*}^{\rm(supA)}\sim M_{\rm A,0}^{\,-9/7}\,S_0^{\,-4/7}$ and $\hat{\lambda}_{\perp,{\rm diss}}^{\rm(supA)}\sim (M_{\rm A,0}\,S_0)^{\,-3/4}$, while the fluctuation anisotropy is given by  
\begin{equation}\label{app:eq:MHDturb_supA_DynamicAlignment_FluctAniso_full}
    \hspace{-0.05cm}\hat{\lambda}_{\|,\hat{\lambda}_\perp}^{\rm(supA)}
    \hspace{-0.05cm}\sim\hspace{-0.05cm}
    \left\{
    \begin{array}{lcrc}
      \hat{\lambda}_\perp\sim\hat{\lambda} &  \,\,\, & \hat{\ell}_{\rm A} < \hat{\lambda}\leq 1 & {\rm[K41]}\\
      & & &\\
      M_{\rm A,0}^{\,-3/2}\,\hat{\lambda}_\perp^{\,1/2} &  \,\,\, &  \hat{\lambda}_{\perp,*}^{\rm(supA)} < \hat{\lambda}_\perp \leq \hat{\ell}_{\rm A} & {\rm[B06]}\\
      & & &\\
      S_0^{\,2/5}\,M_{\rm A,0}^{\,-3/5}\,\hat{\lambda}_\perp^{\,6/5} & \,\,\, &  \hat{\lambda}_{\perp,{\rm diss}}^{\rm(supA)} < \hat{\lambda}_\perp \leq \hat{\lambda}_{\perp,*}^{\rm(supA)} & {\rm[TMT]} 
    \end{array}
    \right.
,\end{equation}
with $\hat{\lambda}_{\|,\hat{\lambda}_\perp}=\lambda_{\|,\hat{\lambda}_\perp}/\ell_0$.
\\
See Figures \ref{fig:app:Spectra_examples} and \ref{fig:app:Anisotropy_examples} for the resulting spectra and fluctuation anisotropy versus perpendicular wavenumber $k_\perp$.


\end{appendix}

\end{document}